\AtBeginDocument{\date{April 28, 2022}} 
\documentclass[preprintnumbers,aps,prd,floatfix,nofootinbib,twocolumn,longbibliography]{revtex4-2}
\usepackage{fix-revtex-4-2-bib}  

\usepackage{epsfig}
\usepackage{amsmath}
\usepackage{graphicx}
\usepackage[normalem]{ulem} 
\usepackage[dvipsnames]{xcolor}
\usepackage[bookmarksnumbered,bookmarksopen]{hyperref}

\DeclareMathOperator*{\trace}{Tr}
\allowdisplaybreaks

\newcommand{\T}[1]{\boldsymbol{#1}_{\text{T}}}
\newcommand{\kT}{\ensuremath{k_{\rm T}}}

\newcommand{\Tsc}[2]{#1_{#2\text{T}}}
\newcommand{\Tscsq}[2]{#1^2_{#2\text{T}}}

\newcommand{\no}{\nonumber \\}
\newcommand{\parz}[1]{\ensuremath{\left(#1\right)}}
\newcommand{\order}[1]{\ensuremath{O\parz{#1}}}

\newcommand{\xbj}{\ensuremath{x_{\rm bj}}}

\DeclareRobustCommand*\diff[2][]{%
   \mathop{
        \mathrm{d}^{#1}
     \mskip-0.2\thinmuskip
   #2}\nolimits
}
\newcommand{\msbar}{\ensuremath{\overline{\rm MS}}}

\newcommand{\tarmass}{m_p}
\newcommand{\mquark}{\ensuremath{m_{q}}}
\newcommand{\mgluon}{\ensuremath{m_{s}}}

\newcommand{\eref}[1]{Eq.~(\ref{e.#1})}
\newcommand{\fref}[1]{Fig.~\ref{f.#1}}

\newcommand{\sref}[1]{Sec.~\ref{s.#1}}
\newcommand{\aref}[1]{Appendix~\ref{a.#1}}

\begin{document}

\title{Positivity and renormalization of parton densities}

\preprint{JLAB-THY-21-3507}

\author{John Collins}
\email{jcc8@psu.edu}
\affiliation{%
  Department of Physics, Penn State University, University Park PA 16802, USA}
\author{Ted~C.~Rogers}
\email{tedconantrogers@gmail.com}
\affiliation{Department of Physics, Old Dominion University, Norfolk, VA 23529, USA}
\affiliation{Jefferson Lab, 12000 Jefferson Avenue, Newport News, VA 23606, USA}
\affiliation{\\ \href{https://orcid.org/0000-0002-0762-0275}{ORCID: 0000-0002-0762-0275}}
\author{Nobuo~Sato}
\email{nsato@jlab.org}
\affiliation{Jefferson Lab, 12000 Jefferson Avenue, Newport News, VA 23606, USA}
\affiliation{\\ \href{https://orcid.org/0000-0002-1535-6208}{ORCID: 0000-0002-1535-6208}}

\begin{abstract}
  There have been recent debates about whether \msbar{} parton
    densities exactly obey positivity bounds (including the Soffer bound),
    and whether the bounds should be applied as a constraint on global fits
    to parton densities and on nonperturbative calculations. A recent paper
    (JHEP {\bf 11} (2020) 129) appears to provide a proof of positivity in
    contradiction with earlier work by other authors.  We examine their
    derivation and find that its primary failure is in the apparently
    uncontroversial statement that bare pdfs are always positive. We show
    that under the conditions used in the derivation, that statement fails.
    This is associated with the use of dimensional regularization
    for both UV divergences (space-time dimension $n<4$) and for
    collinear divergences, with $n>4$.  Collinear divergences appear in
    massless partonic quantities convoluted with bare pdfs, in the approach
    used by these and other authors, which we call ``track B''.  Divergent
    UV contributions are regulated and are positive when $n<4$, but can and
    often do become negative after analytic continuation to $n>4$.   We
    explore ramifications of this idea, and provide some elementary
    calculations in a model QFT that show how this situation can
    generically arise in reality.  We examine the connection with the
    origin of the track B  method. Our
    examination pinpoints considerable difficulties with track B that
    render it either wrong or highly problematic, and explain that a
    different approach, which appears in some literature and that we call
    track A does not suffer from this set of problems.  The issue of
    positivity highlights that track-B methods can lead to wrong results of
    phenomenological importance.  From our analysis we identify the
    restricted situations in which positivity tends to be violated.  
\end{abstract}

\maketitle

\section{Introduction}
\label{s.intro}

Central to many phenomenological applications of QCD is the concept of a
parton density (or distribution) function (pdf).  An issue that has become
particularly important recently is whether or not pdfs are always positive.
Although they generally obey positivity, there has been disagreement
on whether it is possible for some pdfs to be slightly negative under some
conditions.

In the literature, one can find cautionary statements to the effect that
negative pdfs are possible, at least at low scales~\cite{Diehl:2019fsz},
and some fitting procedures do allow for slightly negative pdfs (e.g.,
Ref.~\cite{Martin:2009iq}).  The reason given is that while the most
elementary definition of a pdf does manifestly obey positivity, it also has
ultraviolet (UV) divergences.  The necessary UV renormalization
counterterms are not guaranteed to preserve positivity, as we will
explain in Sec.\ \ref{s.examples} with explicit counterexamples to
positivity.

However, it has also been recently argued, notably by Candido, Forte and
Hekhorn \cite{Candido:2020yat}, that positivity is an automatic and general
property of pdfs defined in the $\msbar$ scheme.  Since this result is in
contradiction with explicit calculations, it creates an apparent paradox
that needs to be resolved.  We will find that the problem is that certain
simple and apparently uncontroversial assertions in
Ref.~\cite{Candido:2020yat} are in fact false, but for non-trivial reasons.
We will give more details later, but we summarize what goes wrong here.

Fundamental to the argument in Ref.~\cite{Candido:2020yat} is the
positivity of bare pdfs and of partonic cross sections in a theory
dimensionally regulated in the UV and infrared (IR).  These positivity
properties result from standard properties of quantum mechanical states,
notably the positivity of the metric on state space.  However, once a
theory is regulated, the state-space metric need not be positive.  A
classic case of such a violation of positivity is given by the
Pauli-Villars method.  In the case of Ref.~\cite{Candido:2020yat},
dimensional regularization is used for \emph{both} the UV and collinear
divergences.  (Bare pdfs have UV divergences, and massless partonic cross
sections have collinear divergences.)  In that case, positivity properties
fail, as we will show explicitly.

If instead one were to use regulators that preserved positivity, we will
show that another of the foundations of Ref.~\cite{Candido:2020yat} fails.
This is the commonly made assertion that structure functions in deep inelastic scattering (DIS) on a
target factor into unsubtracted massless partonic structure functions and bare pdfs
on the target: $F(Q,\xbj) = F^{\text{partonic}} \otimes f^{\text{bare,B}}$.

We stress that the failures just identified do not in fact affect the final
factorization into renormalized pdfs and subtracted coefficient functions.
But they do break the argument for the absolute positivity of \msbar{}
pdfs, as we will see, and this observation motivates greater scrutiny of
the properties of pdfs more generally.  Moreover, failure of a widely
asserted factorization property deserves a closer analysis, which is the
main purpose of this paper, and we use the positivity issue to motivate it.

The failure of positivity of \msbar{} pdfs in some situations occurs
despite the fact that the original concept of a pdf, within Feynman's
parton model \cite{Feynman:1972}, entailed positivity; that was simply
because a parton density was intended to be the number density of a
particular flavor of parton in a fast-moving hadron.  But, as is well-known
and as we will review below, the situation in real QCD requires
modification of the parton model.

Since factorization gives predictions for cross sections, and cross sections are
intrinsically positive, the scope for negative pdfs is severely limited.
For each parton flavor, one can construct a DIS-like process in which the
lowest-order term in the hard scattering is initiated by only the chosen
parton flavor.  (This can be done by replacing the currents in the hadronic
part of deep inelastic scattering by suitable operators containing only
fields for the chosen flavor.)  At a high scale $Q$, the effective coupling
$\alpha_s(Q)$ is small.  Therefore, the lowest-order term typically dominates,
so that positivity of a cross section (or other quantity) entails
positivity of the pdf. The only way this can be avoided is if some other
pdf is sufficiently much larger in magnitude for flavors and/or regions that do not contribute at lowest order, 
such that perturbative corrections to the cross section dominate the
lowest-order part. That is, any negative pdf must be small in magnitude relative
to other pdfs, which are necessarily positive, by the argument involving a
lowest-order approximation to a hard scattering.

This argument gradually loses its force as $Q$ gets smaller, since then
perturbative corrections are no longer so suppressed.  This leads to the
expectation that negative pdfs can occur at most at low scales.  (Later, we
will see support for this in our calculations.)  

It is desirable to have a treatment of the positivity issue in terms of
  the pdfs themselves and their definition, rather than indirectly through
  factorization and the positivity of cross sections.  We will present the
  basics of such an analysis in Sec.\ \ref{s.examples}.

One important implication of the possibility of negative pdfs arises in
phenomenological fits of pdfs, since often the scale $Q_0$ used for the
initial scale for evolution is rather low, and may even be below scales at
which it is reasonable to use factorization.  One aim of a low $Q_0$ is to
ensure that the fitted pdfs are provided at all scales where factorization
could conceivably be usefully applied.  Therefore, if a positivity
constraint were applied to fitted pdfs, especially at a low initial scale
$Q_0$, it is likely to introduce excessive theoretical bias.

Note that positivity constraints, if they are valid, not only apply
directly to unpolarized pdfs, but also give constraints on polarized pdfs,
with a particularly non-trivial case being the Soffer
bound~\cite{Soffer:1994ww,Goldstein:1995ek,Barone:1997fh}.

Another important situation where the same issue arises is when one is
making calculations of pdfs from QCD by non-perturbative methods.  A common
method is to calculate a quasi-pdf or a pseudo-pdf by lattice Monte-Carlo methods, and to
infer the pdfs themselves by a factorization property
\cite{Ji:2013dva,Xiong:2013bka,Ma:2014jla,Constantinou:2020pek,Bhat:2020ktg,Joo:2019jct},
similarly 
to the way global fits of pdfs are made to experimental scattering data.
Such calculations typically give results for a low value of $Q_0$, and in
some lattice QCD calculations positivity is imposed as a
constraint~\cite{Constantinou:2020pek,Bhat:2020ktg,Joo:2019jct} on
parametrizations, especially in the limit of large momentum fractions, $\xi \to
1$.  It is critical to know whether it is correct to apply positivity
constraints in this situation.  Another similar example is in calculations
based on the Dyson-Schwinger equation, where a scale as low as $\mu_0 =
0.78$~GeV~\cite{Bednar:2018mtf} is used.

As regards applications that use perturbative calculations, a
circumstance where pdfs do definitely become negative is in the treatment
of heavy quarks~\cite{Thorne:2006qt,Buza:1996wv}. In such treatments, one uses
perturbative calculations to match the versions of pdfs with different
numbers of active quark flavors.  The pdf for a heavy quark which is active
is perturbatively related to pdfs defined with a lower number of active
quarks.  The lowest order calculation expresses the heavy quark pdf in
terms of the gluon pdf, with a coefficient that is just the \msbar{}
pdf of the heavy quark in a massless on-shell gluon at perturbative order
$\alpha_s$. It contains a factor of $\ln(\mu/m_h)$, where $m_h$ is the
heavy-quark mass and $\mu$ is the \msbar{} renormalization scale; in
this particular instance there is no non-logarithmic
term. The calculation of a pdf for an active heavy quark can be applied 
where the $\msbar$ scale $\mu$ is somewhat less than the heavy-quark's
mass. Because of the factor of $\ln(\mu/m_h)$, the result is a negative
heavy-quark pdf in this region; the 
negative pdf is essential to preserving the momentum sum rule.  

The
smallness of the effective coupling at the heavy quark scale implies that
higher-order corrections are generally minor corrections to the
  leading-order result.\footnote{In contrast, light-quark pdfs in QCD
    cannot be usefully computed by low order perturbation theory.}
  In particular, they only slightly shift the scale
  where the heavy quark distribution becomes negative; they do not change
  the fact the pdf becomes negative for $\mu$ a bit less than $m_h$.  These
  statements rely on the specific \msbar{} definition of pdfs.
  Indeed, the corrections shift the zero to a higher value of $\mu$ --- e.g.,
  Fig.\ 1 of \cite{Thorne:2006qt} --- so that the heavy quark pdf is
  negative even at $\mu=m_h$, without at the same time making a DIS structure
  function negative.

Observe that when $\mu$ is comparable to $m_h$, the size of the 
heavy-quark pdf is substantially less than the gluon pdf, since to
a first approximation it is given in terms of the gluon pdf by a one-loop
calculation.  Hence in the application of factorization the $O(\alpha_S)$
gluon-induced term can be comparable to the lowest order
heavy-quark-induced term, thereby allowing preservation of positivity of the cross
section.\footnote{Note that the coefficient function for the
    gluon-induced process includes a subtraction to avoid double-counting
    heavy quark contributions.  If the heavy quark term is negative, that
    results in an increased value for the gluon term.} 

However, when using factorization it is generally only important to treat a
heavy quark as active when a process's physical scale is significantly
larger than the quark's mass.  In that case the heavy quark pdf is evolved
from its calculated value at the scale of the quark's mass to a
substantially higher scale; it is then positive.

Let us now return to examining the differing approaches to the positivity
question.  We will show that the differences originate in a long-standing
divergence in views about certain conceptual foundations for QCD factorization and
the definitions of pdfs.  We trace this to pioneering QCD literature of the
1970s, written while much of the technical framework of factorization was
being developed.  One track, which we call track-A, originated in efforts
to give the earliest parton ideas a concrete realization in quantum field
theory, with inspiration for derivations coming from those for
the operator-product expansion, which was already applied in QCD to deep
inelastic scattering.  The second track (track-B) arose early out of a
practical desire to perform partonic calculations. In the absence at that
time of a full track-A treatment, conjectures were made as to appropriate
methods.

First, we will review the basics of track A in Sec.\ \ref{s.tracka}. Then, in Sec.~\ref{s.trackb} we will explain track B and present a critique of it. We  will argue that track-A is actually the correct
one.  In~\sref{reconstruct} we will
examine what is necessary for track B expressions in dimensional
regularization to reproduce the DIS cross section.  In~\sref{dimreg} we point out pitfalls with using dimensional
regularization simultaneously for UV and IR divergences; these
  pitfalls break the positivity argument of Ref.\ \cite{Candido:2020yat}. In~\sref{pos.assess} we will combine the observations of the previous sections to summarize the reasons why the 
argument in \cite{Candido:2020yat} fails and argue that it is traceable to
the use of track B. In \sref{examples}, we illustrate that an \msbar{} pdf
can turn negative in the concrete example of a Yukawa field theory where
everything is calculable in fixed, low order perturbation
theory.\footnote{The reason for presenting an example in Yukawa theory is
  that the arguments in \cite{Candido:2020yat} are independent of the
  theory. So we choose a theory and coupling where low-order perturbative
  calculations on a massive target give a sufficiently accurate
  calculation to determine the validity of the methods for proving
  positivity.} 
We end with concluding remarks in~\sref{discussion}.

In our examination of DIS, we will use a fairly standard notation: $P$ is
the momentum of the target, $q$ is the incoming momentum at the current in
the hadronic part, and $M$ is the target mass.  We use light front
coordinates, with the inner product of two vectors being given by $a\cdot b
  = a^+b^- + a^-b^+ - \T{a}\cdot\T{b}$, and the components of a vector being
  written as $a=(a^+,a^-,\T{a})$.  
The coordinate
axes are chosen such that the components of $P$ and $q$ are
\begin{equation}
  \label{eq:P.q}
  P = \left( P^+, \frac{M^2}{2P^+}, \T{0} \right),
\quad
  q = \left( -xP^+, \frac{Q^2}{2xP^+}, \T{0} \right),
\end{equation}
where $x$ is the Nachtmann variable, which agrees with the Bjorken $x_{\rm
  Bj}$ up to power-suppressed corrections.

\section{Track-A: Renormalization and light-cone pdfs}
\label{s.tracka}

One of the motivating points of track-A was work to provide a definite
field-theoretical implementation of the original pdf concept. At the
beginning, this led to the insight that light-front quantization provides a
suitable candidate definition as the expectation of a light-front number
operator~\cite{Kogut:1969xa,Bouchiat:1971mj,Soper:1976jc}, provided that no
difficulties arise.  But difficulties do arise.  The difficulties are
particularly notable when one includes a treatment of
transverse-momentum-dependent pdfs, as in Soper~\cite{Soper:1979fq} and
Collins~\cite{Collins:1980ui}. For the case of the
transverse-momentum-integrated pdfs and fragmentation functions in full
QCD, the elementary definition of these quantities needs to be modified
\cite{Collins:1981uw} to allow for UV renormalization. At least for
collinear pdfs, this form of the definition has continued to be used to the
present, without modifications.  It is the pdfs with this definition
that one now studies using lattice QCD and other nonperturbative techniques.

A second motivation was the realization that the methods that led to
operator-product expansion (OPE) could be generalized. For DIS, the OPE
applies to certain integer moments of structure functions.  The methods
can be extended to obtain the large-$Q$ asymptotics of the structure
functions themselves. The factorization and OPE derivations have overall
structures that are very similar. In fact, when one takes the appropriate
moments of DIS structure functions, one recovers the results from the
OPE.  Although the factorization work drew on the derivation of the OPE by
Wilson and Zimmermann (e.g.~\cite{Wilson:1969zs,Zimmermann:1972tv}), there
were some important enhancements/modifications that we will make more
  explicit in \sref{reconstruct}.

For unpolarized quark pdfs, a \textit{bare} quark pdf is defined by
\begin{multline}
\label{e.pdfdef}
f_{j/H}^{\text{bare,A}}(\xi) \equiv \int \frac{\diff{w^-}{}}{2 \pi} \, e^{-i \xi p^+ w^-} 
\\
\; \langle p | \, \bar{\psi}_{j,0}(0,w^-,\T{0}{}) {\frac{\gamma^+}{2}} W[0,w^-]
\psi_{j,0}(0,0,\T{0}{}) \, | p \rangle \, .
\end{multline}
where $\psi_{j,0}$ is the bare field for a quark of flavor $j$ as appears
in the Lagrangian density that defines the theory. The factor $W[0,w^-]$
is a light-like Wilson line, also defined with bare field operators and the
bare coupling.  The label $H$ denotes the kind of target particle that
is used for the state $| p \rangle$.  This definition is actually the
expectation value of a quark number density
operator~\cite{Bouchiat:1971mj,Soper:1976jc}, \cite[Chap.~6]{Collins:2011qcdbook},
expressed in terms of bare 
fields in a gauge-invariant form.  The ``A'' superscript is to distinguish
this track-A bare pdf from a different track-B concept of the same name, as
discussed later in \sref{trackb}.

In the bare pdf there is a logarithmic UV divergence associated with the
bilocal operator in \eref{pdfdef}.  This is distinct from the UV
divergences that are canceled by counterterms in the QCD Lagrangian, and
hence the bare pdf is UV divergent.

An $\msbar$ renormalized pdf is defined in terms of the bare pdf by including a renormalization factor $Z^A$, 
\begin{equation}
f^{\text{renorm,A}}(\xi) \equiv Z^A \otimes f^{\text{bare,A}} \, , \label{e.pdfren}
\end{equation}
with the product being in the sense of a convolution in $\xi$ and a matrix in
flavor space. In the \msbar{} scheme,
$Z^A$ is defined by analogy with renormalization factors in
other cases, and in perturbation theory it has the form: 
\begin{equation}
Z_{ij}^A(\xi) = \delta(1 - \xi)\delta_{ij} + \sum_{n=1}^\infty C_{n,ij}(\xi,\alpha_s) \parz{\frac{S_\epsilon}{\epsilon}}^n \, . \label{e.Zdef}
\end{equation}
Here the $C_{n,ij}$ are the coefficients necessary to subtract only
  powers of $S_\epsilon/\epsilon$, where $S_{\epsilon} \equiv (4 \pi)^\epsilon/\Gamma(1 - \epsilon) \simeq (4 \pi e^{-\gamma_E})^\epsilon$,
  as in Ref.\ \cite[Eq.~(3.18)]{Collins:2011qcdbook}.
  The dimension of spacetime is $n=4-2\epsilon$.  
  The significance of the second formula for $S_\epsilon$ is that many authors
  use the second formula to define $S_\epsilon$, or an equivalent method. In the
  cases we are interested in, the difference does not affect physical
  quantities \cite[Eq.~(3.18)]{Collins:2011qcdbook}.

The definition of the convolution over
collinear momentum fraction is
\begin{equation}
\left[ A \otimes B \right](\xi) \equiv \sum_j \int_\xi^1 \frac{\diff{\xi'}}{\xi'} A_j(\xi/\xi') B_j(\xi') \, . 
\end{equation}
Notice that we carefully distinguish parton momentum fraction ($\xi$) from process-specific kinematic variables like Bjorken $\xbj$, although we 
will frequently drop $\xi$ arguments for brevity.
Associated with the use of the \msbar{} scheme and dimensional
regularization is the renormalization scale $\mu$ that is defined to appear
as a factor $\mu^\epsilon$ in the bare coupling in the Lagrangian density of the theory.

It is the renormalized version of the pdf in \eref{pdfren} that enters into
the derived factorization formulas for physical observables like cross sections. For a DIS structure function $F$, for example, 
\begin{align}
F(Q,\xbj) & = \mathcal{C}^A \otimes f^{\text{renorm,A}} + \mbox{error} \\
     & \hspace*{-1.7cm}= \sum_j \int_{\xbj}^1 \diff{\xi}
     \mathcal{C}^A_j(\xbj/\xi,\alpha_s(Q)) \, f_j^{\text{renorm,A}}(Q,\xi) + \mbox{error}.
\label{e.structfunct}
\end{align}
Here, $\mathcal{C}^A$ is a perturbatively calculable coefficient
function, and the error is suppressed by a power of $1/Q$. 

A convenient technique for perturbative calculations of $\mathcal{C}^A$
arises from recognizing that it is independent of which target is used for
the structure function $F$; all the target dependence is in the parton
density $f$.  So we can work with perturbative calculations of pdfs and
structure functions on partonic targets.  The results all follow from
definite Feynman rules.  Since the coefficients $\mathcal{C}^A$ are
independent of light-parton masses, it is sufficient to simplify the
calculations by setting the mass parameters for
all fields to zero.  Then the hard coefficient is effectively an inverse
of \eref{structfunct} when a partonic target is used, i.e.,
\begin{equation}
\label{inverse}
\mathcal{C}^A  =  \frac{F^{\rm partonic}(Q,\xbj)}{f^{\text{partonic,renorm,A}}} \, ,
\end{equation}
and this gives a perturbative calculation of $\mathcal{C}^A$.  Here
division is in the sense of an inverted convolution integral.  Although
there are collinear divergences in both the structure function and pdfs
with a massless partonic target, they necessarily cancel in the hard
scattering coefficient $\mathcal{C}^A$.  When the $\mathcal{C}^A$
coefficients are used phenomenologically, the renormalization group scale
$\mu$ is generally fixed numerically to be proportional to a physical hard
scale (e.g., $Q$), thereby ensuring that $\mathcal{C}^A$ is
perturbatively well-behaved, i.e., that useful calculations can be made
by expanding it to low orders in the effective coupling $\alpha_s(Q)$.

Equation (\ref{inverse}) can be regarded as equivalent to a version of the
$R*$ operation \cite{Chetyrkin:1982nn,Chetyrkin:1984xa,Smirnov:1985yck} that was devised to subtract both IR and UV divergences in Feynman
  graphs.
As to the situation with hadronic targets, the definition of a pdf in Eqs.\
\eqref{e.pdfdef} and \eqref{e.pdfren} is complete enough to be used for
calculations from first principles QCD with nonperturbative techniques like
lattice QCD, at least given sufficiently advanced methods, and to some
approximation.

In view of later discussions, it is important to observe that because all
actual hadrons in QCD have mass, 
there are no actual soft or collinear divergences in structure functions
and pdfs for a hadronic target.  There is sensitivity to the collinear
region in these quantities, but no actual divergence. When collinear
divergences do appear in calculations, it is at intermediate 
stages of a calculation of a hard scattering; they are artifacts of having
perturbatively calculated structure functions and pdfs with massless, on-shell partonic
targets before applying Eq.~(\ref{inverse}). 

It is perfectly possible to do the calculations for the right-hand side of
(\ref{inverse}) with quark masses kept non-zero. Then some of the collinear
divergences\footnote{The masslessness of the gluon in perturbation
    theory in QCD continues to
    provide some collinear divergences.} in the all-massless calculation correspond to logarithms of
$\text{mass}/Q$ in perturbative calculations in the limit of zero mass(es).
Given the common situation where all the masses are small compared with
$Q$, one would then take the limit of zero mass to obtain $\mathcal{C}^A$,
an operation which would need to be inserted on the right-hand side of
(\ref{inverse}).

However, considerable simplifications in Feynman graph calculations occur
in the massless case, especially with dimensional regularization, so it is
normal to use only massless calculations.  One of the simplifications is
that perturbative corrections to bare pdfs on partonic targets are
zero to all orders of \emph{massless} perturbation theory.  That is simply
because all the integrals are scale free, and therefore vanish in
dimensional regularization; there is an exact numerical cancellation of the quantified
IR and UV divergences with no remaining finite part.
Therefore, a bare track-A pdf for a parton in a massless parton is just
\begin{equation}
\label{bare.partonic.pdf}
    f^{\text{partonic,bare,A}}_{i/j}(\xi) = \delta(\xi-1) \delta_{ij},
~ ~ \mbox{(massless, dim.\ reg.)}
\end{equation}
where $i$ and $j$ label parton flavors. Hence,
the renormalized pdf on a
massless partonic target is exactly equal to the \msbar{} renormalization
factor:
\begin{equation}
\label{renormalized.partonic.pdf}
    f^{\text{partonic,renorm,A}}_{i/j} = Z^A_{i/j}.
~ ~ \mbox{(massless, dim.\ reg.)}
\end{equation}
It is important that the conceptual status of the divergences in $Z^A$ as
$\epsilon\to0$ has changed dramatically, between Eq.\ (\ref{e.pdfren}) and Eq.\
(\ref{renormalized.partonic.pdf}).  The poles in $Z^A$ in Eq.\ (\ref{e.pdfren}) are all UV poles, to cancel UV divergences in the bare pdf.  On a
hadronic target, this results in finite renormalized pdfs, of course.  But
in (\ref{renormalized.partonic.pdf}), the numerically identical poles are
actually collinear divergences in a UV-finite pdf on a partonic target.

Although working with massless partonic pdfs is a useful
 technique for calculating quantities like hard factors, what phenomenology ultimately needs is the set of pdfs for hadrons. It is \eref{pdfren} that is relevant for these pdfs.
There are situations where non-zero quark masses are
needed in perturbative calculations.  An
important one is where one deals with heavy quarks whose masses are
comparable to or bigger than $Q$.  Then the heavy-quark masses need to be
retained in the calculations, and equations like
(\ref{bare.partonic.pdf}) and (\ref{renormalized.partonic.pdf}) are no
longer true.

\section{Track-B: Collinear absorption}
\label{s.trackb}

Next we contrast the above with an alternative way that factorization is often described and 
used to derive properties of pdfs and other parton correlation functions. We will ultimately critique this approach, which we will call track B.

\subsection{Content of track B}

The starting point of track B, is the assertion that a structure function
on a hadronic target is the convolution of the corresponding massless on-shell
partonic structure function with bare pdfs on the same target:
\begin{equation}
\label{e.bare.factn}
F(Q,\xbj) = F^{\text{partonic}} \otimes f^{\text{bare,B}} \, .
\end{equation}
In contrast with the similar-looking factorization formula
(\ref{e.structfunct}) in track A, the first factor is an unsubtracted
partonic structure function and has collinear divergences, unlike the
corresponding quantity $\mathcal{C}^A$ in (\ref{e.structfunct}).
Although the pdf factor $f^{\text{bare,B}}$ is called a ``bare'' pdf, it
must in general be different from the bare pdf in track A, as we will see,
and we have therefore distinguished it by a label ``B''.

To deal with the collinear divergences in $F^{\text{partonic}}$, it is then
proved \cite{Ellis:1978ty,Curci:1980uw} that the partonic structure
function can be written as a convolution of a finite coefficient function
$\mathcal{C}^B$ with a factor $Z^B$ containing the collinear
divergences,
\begin{equation}
F^{\text{partonic}} = \mathcal{C}^B \otimes Z^B \, .  \label{e.abs1}
\end{equation}
When the collinear divergences are quantified as poles in dimensional
regularization, $Z^B$ can be defined to be of the \msbar{} form, similarly
to the UV renormalization factor in (\ref{e.Zdef}).  Commonly this is
modified by the use of a factorization scale $\mu_F$ which is distinct from
the renormalization scale $\mu$.  The exact \msbar{} form is obtained when
$\mu_F=\mu$. The pole structure can be modified by extra finite contributions,
the choice of which defines the scheme.  In all cases, the
collinear-divergence factor is independent of which hard process is
considered, e.g., which DIS structure function is treated, or whether DIS
or Drell-Yan is treated.

Process-independence of $Z^B$ permits the final step, which is the
absorption of the collinear divergences into a redefinition of the pdfs:
\begin{align}
\label{e.trackb_rearrange}
F(Q,\xbj) &{} = \parz{\mathcal{C}^B \otimes Z^B} \otimes f^{\text{bare, b}} \no
&{} = \mathcal{C}^B \otimes \parz{Z^B \otimes f^{\text{bare,B}}} \no
&{}= \mathcal{C}^B \otimes f^{\text{renorm,B}} \, ,
\end{align}
where the renormalized pdf is defined to be
\begin{equation}
f^{\text{renorm,B}} = Z^B \otimes f^{\text{bare,B}} \, .
\label{e.fbrenorm}
\end{equation}
The final line of (\ref{e.trackb_rearrange}) has the same form and nature
as the factorization formula (\ref{e.structfunct}) in track A.

In standard phenomenological applications to scattering processes, the
coefficients $\mathcal{C}^B$ or $\mathcal{C}^A$ and the corresponding
quantities for other processes are computed perturbatively, while the pdfs
at some initial scale are obtained from fits to data. Scale-evolution of
renormalized pdfs is implemented by the DGLAP equations, with their
perturbatively calculable kernels.

\subsection{Equality of coefficient functions and renormalized pdfs between
tracks A and B}

Although, as we will see shortly, there are important reasons to at least
question the starting point (\ref{e.bare.factn}) of track B, nevertheless
the structure of the final factorization formula (last line of
(\ref{e.trackb_rearrange})) for the standard applications agrees with that
of track A and is correct.

In fact, when the \msbar{} prescription is used and $\mu_F=\mu$, as is common,
the coefficient functions in the two tracks are equal:
$\mathcal{C}^A=\mathcal{C}^B$. This is because in both cases, the partonic
structure function and the coefficient function differ by a factor of an
\msbar{} form, and the poles can be uniquely determined by the requirement
that the coefficient function be finite.  In track A, this follows from
Eq.\ (\ref{inverse}) and the form for the renormalized pdf on a massless
target given by Eqs.\ (\ref{renormalized.partonic.pdf}) and (\ref{e.Zdef}).

It follows that the \emph{collinear}-divergence factor $Z^B$ in track B
equals the \emph{UV-renormalization} factor $Z^A$ in track A.  The
renormalized pdfs also have to agree, since they can be fit to the same
cross sections with hadronic targets, and the coefficient functions are
equal.

From $Z^A=Z^B$ it follows that the bare pdfs in both schemes are equal.
However, this result relies on the use of dimensional regularization,
massless partonic calculations, and the consequent vanishing of scale free
integrals.  It is these properties that led to Eq.\
(\ref{renormalized.partonic.pdf}) for the values of the massless partonic
pdfs in track A.

\subsection{Critique}

Completely essential to track B is the statement (\ref{e.bare.factn}) that
a structure function on a hadron is the convolution of an unsubtracted
partonic structure
function and bare pdfs.  Let us call this statement ``bare factorization''.
However, as far as we can see, bare factorization is merely asserted and
never actually derived.  In addition, the bare pdfs are commonly not
defined.  Especially in the early literature, the assertion of bare
factorization appears with a reference to Feynman's parton model --- e.g.,
see Refs.\ \cite{Politzer:1977fi,Ellis:1978ty} --- perhaps as the natural
generalization of the parton model to QCD.

But when the statement of bare factorization is examined in more detail, it
becomes highly implausible.  The parton model itself \cite{Feynman:1972}
can be motivated by examining relevant space-time scales for DIS in the
Breit frame with a hadronic target of high energy.  The hadron is time
dilated from its rest frame, and therefore the natural scales for internal
processes in the hadron are the large ranges of time and longitudinal
position that arise from the boost to a high energy.  The scattering with
the virtual photon involves much smaller scales, of order $1/Q$.  To obtain
the parton model, it was hypothesized that the transverse momenta and
virtualities of constituents in the hadron are limited.  Then a
factorization formula arises in which the virtual-photon-quark interaction
is restricted to \emph{lowest-order} in strong interactions.

However this motivation does not extend to the generalization from the
parton model to bare factorization in QCD.  This can be seen from the
collinear divergences in massless partonic cross sections.  The divergences
involve infinitely long times and distances (in the longitudinal direction,
the same direction as the hadron).  Such scales are much longer than the
scales for hadrons, since hadrons are massive and their actual interactions
therefore do not have collinear divergences.  This indicates that the
collinear divergences in the partonic structure function are a property of
intermediate results in a method of calculation, rather than a property of
full QCD.

Another way to see the problems is to examine the nature of the divergences
in the three quantities in (\ref{e.bare.factn}). The hadronic structure
function on the left-hand side is measurable and finite.  In particular it
has no collinear divergences because all true particles in QCD are massive.
Possible UV divergences are canceled by renormalization counterterms in the
Lagrangian.

On the right-hand side, the massless partonic structure function also has
no UV divergences, for the same reason.  But it does have
perturbative collinear divergences because of the masslessness of the
partons.

As to the bare pdf, let us copy Candido, Forte and Hekhorn
\cite{Candido:2020yat} and say that the track-B bare pdf is given by the
standard operator formula, as in their Eq.\ (2.2), essentially the same as
our (\ref{e.pdfdef}) for track A.  When a hadronic target is used, the pdf
has no collinear divergences, because of the massiveness of hadrons.  But
it does have UV divergences associated with the operator; these are beyond
those canceled by counterterms in the Lagrangian.

So we have a mismatch of divergences: The right-hand side of
(\ref{e.bare.factn}) has both collinear and UV divergences, whereas the
left-hand side has none.  The obvious conclusion is that
(\ref{e.bare.factn}) is wrong, despite the fact that it is so widely quoted
in the literature.

Table \ref{tab:IR.UV.divs} summarizes the divergence properties
of the various quantities we have been discussing.

\begin{table*}
  \centering
\begin{tabular}{cc}
\includegraphics[scale=0.5]{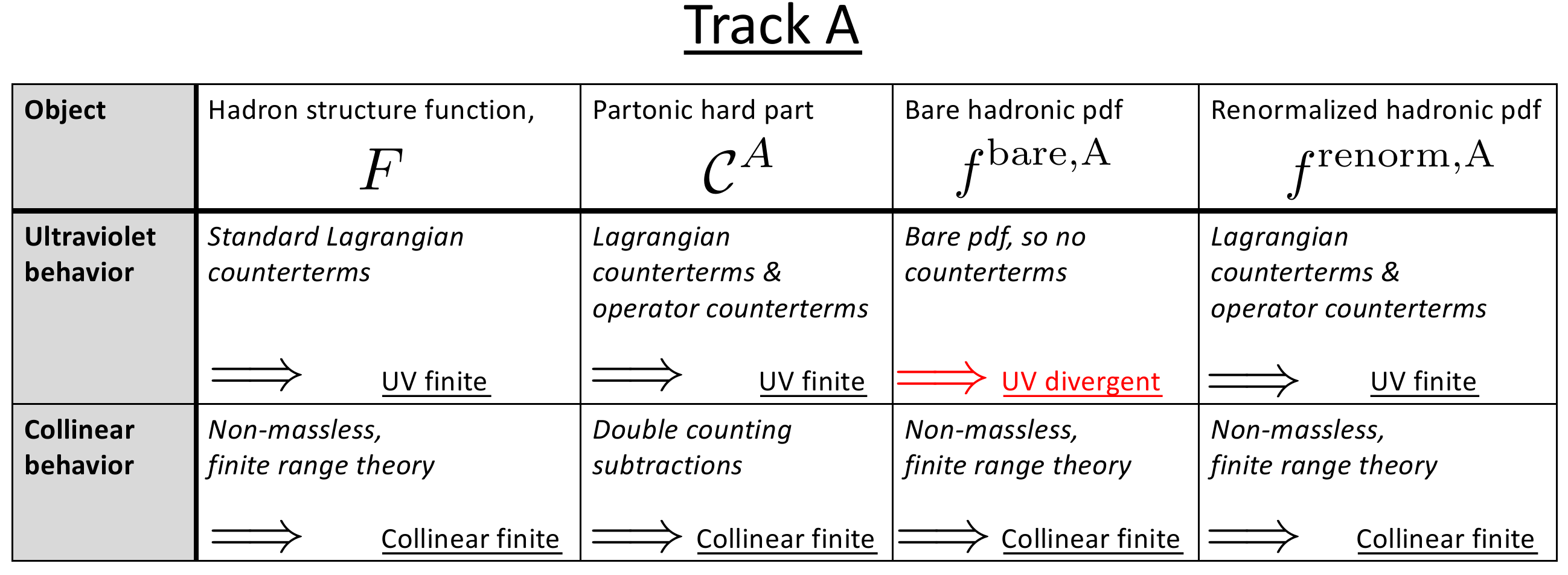}
\vspace{.25in}
\\
\includegraphics[scale=0.4]{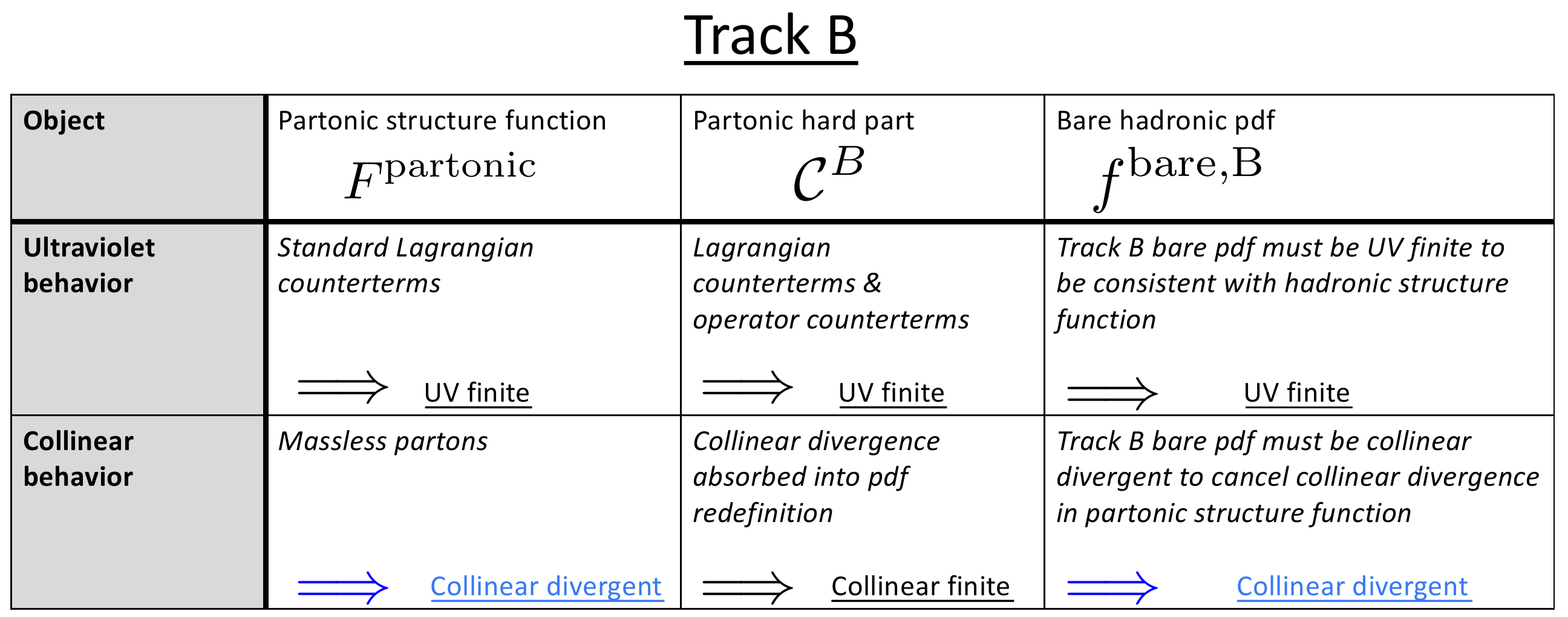}
\end{tabular}
\caption{A summary of the divergences that appear with hadron targets in track A and track B treatments. (See text for more details.) 
In the track A table, the only divergences (apart from those involved in renormalizing the QCD Lagrangian) are the UV divergences in the bare pdf. These get removed by operator counterterms when the pdf is renormalized.  $F^\text{partonic}$ and $f^\text{bare,B}$ appear in \eref{bare.factn}, and both contain collinear divergences that must cancel once all factors are combined to reproduce the physical structure function. (So $f^\text{bare,B}$ is not actually the bare pdf in \eref{pdfdef}.)
}
\label{tab:IR.UV.divs}
\end{table*}

However, there is in fact a loophole in the argument for the mismatch of
divergences.  This is that it might happen that the two kinds of divergence
cancel.  Within the context of dimensional regularization and massless
on-shell partonic calculations this does happen.  Therefore it is useful to
examine more carefully the situations in which bare factorization holds,
which we will do in \sref{reconstruct}.

But, as we will show in Sec.\ \ref{s.pos.assess} an almost immediate
consequence of relying on the cancellation of UV and collinear divergences
is that the positivity arguments in Ref.\ \cite{Candido:2020yat} fail.

\subsection{An alternative definition of a bare pdf}

A rather different definition of a bare pdf in track B was
given by Curci, Furmanski, and Petronzio in Ref.\ \cite{Curci:1980uw}, in their Eq.\ (2.46).  This
quantity is represented diagrammatically by the bottom-most object in their
Fig.\ 3, labeled ``$q_{B.H}$''.  Their definition is
obtained by modifying (\ref{e.pdfdef}) so that the quark-antiquark Green
function in a hadron is restricted to its two-particle-irreducible (2PI)
part in light-cone gauge.  The two particle irreducibility implies, by
standard power-counting arguments, that this bare pdf has no UV
divergences.  The massiveness of hadrons ensures that this kind of bare pdf
also has no collinear divergences.  

We now have a real contradiction, since the sole remaining divergences on
the right-hand side of (\ref{e.bare.factn}) are the collinear divergences
in the partonic structure functions, and there is nothing to cancel them to
make a finite left-hand side.

In Sec.\ \ref{s.Curci}, we will explain the rather trivial reasons that
Curci et al.'s assertion of bare factorization (at the start of their Sec.\
2.7) cannot be true with their definition.  Their remaining derivations are
rather clear, and it is quite simple to modify their arguments to make a
correct derivation.  For the result, see Sec.\ 8.9 of
\cite{Collins:2011qcdbook}, which itself is based on an earlier paper
\cite{Collins:1998rz}.  The announced focus of Ref.\ \cite{Collins:1998rz} was heavy
quark effects, but its argument is not so restricted.  The derivation is
definitely of the track-A kind, and the definition of a bare pdf is that of
track A, not that of Ref.\ \cite{Curci:1980uw}.

\section{Reconstructing track B parton densities}
\label{s.reconstruct}

We have questioned the validity of bare factorization, \eref{bare.factn},
which is the starting point of track B.  The problematic issue was that an
unsubtracted partonic structure function is used.  In this section, we
start from the observation that our argument in Sec.\ \ref{s.trackb}
against the validity of bare 
factorization appears to be undermined by the formulation and proof of the
OPE that was given by Wilson and Zimmermann
\cite{Wilson:1969zs,Zimmermann:1972tv}.  Their form of the OPE is rather
like bare factorization, in that their coefficient function, the analog of
$\mathcal{C}^B$ in \eref{bare.factn}, also has no subtractions for
what in this case are low-momentum regions (instead of collinear regions).  In this it differs from $\mathcal{C}^B$ only in that
all parton masses are preserved instead of being set to zero, and so there
are no actual collinear divergences.  The Wilson-Zimmermann derivation
relies on the use of a particular subtraction scheme.

Now the OPE applies in a short-distance asymptote: $q^\mu\to\infty$ at fixed hadron
momentum $P$; the operators in the analog of pdfs are local.  In contrast,
factorization applies in the Bjorken asymptote, where $Q\to\infty$ at fixed
{$Q^2/P\cdot q$}.  A generalization of the Wilson-Zimmermann method should
apply.  This would suggest that one can in fact derive bare factorization,
as used in track B.  However, it is important that in the OPE the local
operators used are UV-renormalized, not bare.  Correspondingly, the pdf in
bare factorization should be a renormalized quantity, contrary to
assertions in track-B literature.

The purpose of this section is therefore to reverse engineer what
definition of a pdf is needed in order for bare factorization,
\eref{bare.factn}, to be correct.

For the following discussion, we will find that we need to modify the
notation for bare factorization, \eref{bare.factn}, to allow for modified
definitions:
\begin{equation}
\label{e.bare.factn_subs}
F(Q,\xbj) = \left[ F^{\text{partonic}} \right]_{{{\rm R}_1,}\atop{{\rm IRR}}} \otimes \left[ f^{\text{bare,B}} \right]_{{{\rm R}_2,}\atop{{\rm IRR}}} \, .
\end{equation}
Here ${\rm R}_1$ and ${\rm R}_2$ are the UV renormalization schemes for
$F^{\text{partonic}}$ and $f^{\text{bare,B}}$ respectively, and ${\rm IRR}$
is an IR regulator scheme. ${\rm R}_1$ is simply the renormalization in the
QCD Lagrangian because there are no other UV divergences to deal with in
the partonic structure function. A separate UV scheme is allowed for
$f^\text{bare,B}$.  In the general case, some such scheme must be present
in $f^\text{bare,B}$ because for the nature of the divergences on the left
and right of \eref{bare.factn} or (\ref{e.bare.factn_subs}) to match, the
pdf cannot contain UV divergences. ${\rm IRR}$ needs to be defined such
that collinear divergences cancel between $F^{\text{partonic}}$ and
$f^{\text{bare,B}}$ in \eref{bare.factn_subs} in order to recover the
physical structure function on the left side of the equation. In
general, the choices of ${\rm IRR}$ and ${\rm R}_2$ need to be carefully
adjusted to maintain the overall correctness of \eref{bare.factn_subs}.
In~\sref{trackbpdfs} we will show an explicit example of how this works
for the specific case of dimensional regularization and $\msbar$.

The next three subsections form a rather technical detour, but they are important because they 
will allow us to state very rigorously how each factor in \eref{bare.factn_subs} must be defined for a track B approach to be consistent. 
This in turn will allow us to make a truly apples-to-apples comparison with the corresponding factors defined in the track A approach.
Once this is done, the origin of any differences between the two approaches regarding questions like positivity will be clear and 
easy to diagnose.

\subsection{The Wilson-Zimmermann treatment of the OPE, generalized to DIS}
\label{s.WZ}

To understand the relation between the Wilson-Zimmermann approach to the
OPE and the track-B treatment of factorization, it is useful to summarize
the Wilson-Zimmermann approach as it would apply to a DIS structure
function, with the aid of some of the methods of Curci, Furmanski and
Petronzio \cite{Curci:1980uw}.  

The methods of Wilson and Zimmermann can be characterized by the
observation that there is a close similarity between the operations needed
to extract the large $Q$ asymptotics of some Green function and those
needed to extract UV divergences and thereby obtain renormalization
counterterms.  Moreover, when zero-momentum subtraction is used, as they
do, the operations are identical except for the characterization of what
subgraphs they are applied to.  The proofs, as written, work to all orders
of perturbation theory.  Zero-momentum subtractions can be applied to the
integrands of Feynman graphs.  Then no regulator is needed and the scheme
is labeled ``BPHZ''.

The limit involved for the OPE is the short-distance limit where $q\to\infty$ at
fixed $P$, or, equivalently, $Q\to\infty$ with $x \propto Q$.  It applies to the
uncut amplitude for DIS, which is the expectation value in the target state
of the time-ordered product of two currents.  The short-distance limit
entails $x\to\infty$, which is not in the physical region for actual physical DIS,
but is related to it by a dispersion relation, giving results for
certain integer moments of DIS structure functions.  But the structure of
the derivation of the OPE itself applies equally to DIS in the physical
region, and that is what we will present here. 

One begins by examining situations where all the momenta $k$ inside a
subgraph are large while the momenta $l$ attaching it to the rest of the
graph are small.  The UV divergences or the leading power of $Q$ can be
quantified by expanding to the relevant order in powers of $l$ relative to
$k$. In the renormalization of UV divergences, using the first term in this
expansion to construct counterterms amounts to defining the counterterms by
zero momentum subtraction.

For the arguments in their simplest form to work in a gauge-theory,
light-cone gauge is used.\footnote{There are certain problems with the
    use of light-cone gauge, which we will describe later, but it is
    sufficient to ignore them for our present purposes.}  In this gauge,
the Wilson line in the operator in the definition of the bare pdf equals
unity and can therefore be omitted.  Most importantly, in the leading power
for the large $Q$ asymptotics of a structure function, the relevant regions
of loop momentum space are as denoted in
\begin{equation}
\label{e:DIS.region}
  \includegraphics[scale=0.5]{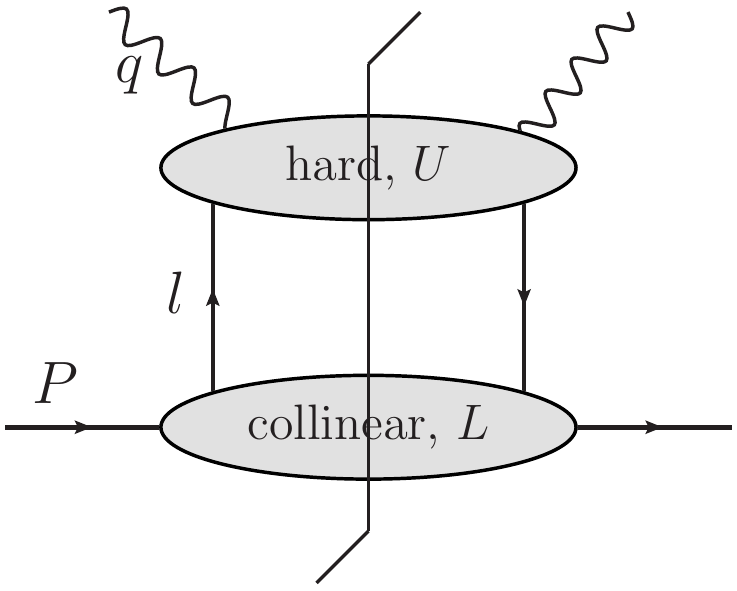}  
\end{equation}
At the top, there is a subgraph $U$ (the ``hard'' subgraph) with large
transverse momenta; it has two parton lines at its lower end.  The lower
part $L$ (the ``collinear'' subgraph) has low transverse momentum.  Each
graph typically has multiple possible regions of this form, and
for the purposes of this discussion 
we omit details of how intermediate regions are accounted for by a suitable
recursive subtraction scheme.  

Since all the possible hard subgraphs are nested with respect to each other, the
treatment can be simplified compared with the general treatment by
Zimmermann.  We then have the algebraic structures that were found in
\cite{Curci:1980uw} and that we will treat below.  In a more general case,
there can be non-trivial overlaps between different possible hard
subgraphs, and the full Zimmermann forest formula, or some equivalent,
would be
needed.  (The treatment of UV divergences renormalized by counterterms in
the Lagrangian similarly does not break the algebraic structure, and need not
be treated explicitly.)

A similar graphical structure applies for the regions that give UV
divergences in the pdfs:
\begin{equation}
\label{e:pdf.UV.div}
  \raisebox{3mm}{\includegraphics[scale=0.5]{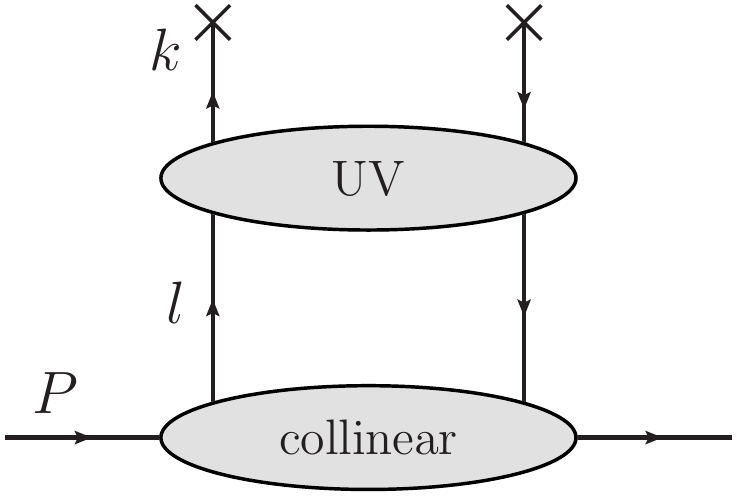}}
\end{equation}
where UV divergences arise when the transverse momenta in the upper
subgraph go to infinity, and the crosses denote the factor
  corresponding to the operator in the definition of the pdf.  Again, any single graph can have many different
regions of this form.

\begin{widetext}
Therefore to extract the large $Q$ asymptotics of DIS, we use an
expansion in two-particle-irreducible (2PI) subgraphs, as was done by Curci,
Furmanski and Petronzio \cite{Curci:1980uw}. For the DIS structure
functions on a hadron this gives
\begin{align}
\label{eq:DIS.2PI.exp}
  \raisebox{-7mm}{\includegraphics[scale=0.4]{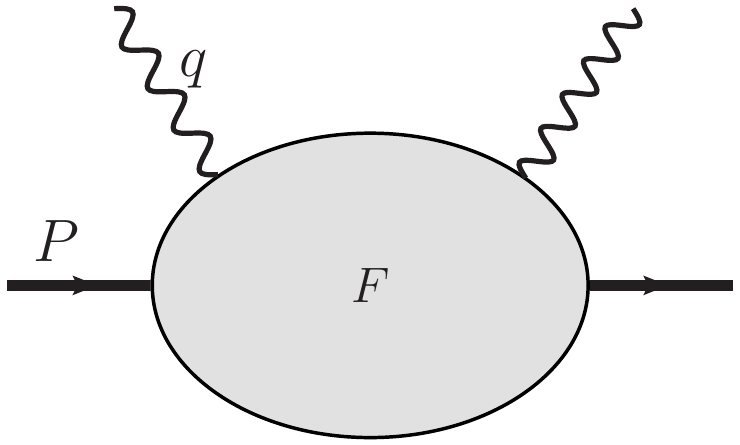}}
 ~~ ={}& ~~
   \raisebox{-2mm}{\includegraphics[scale=0.4]{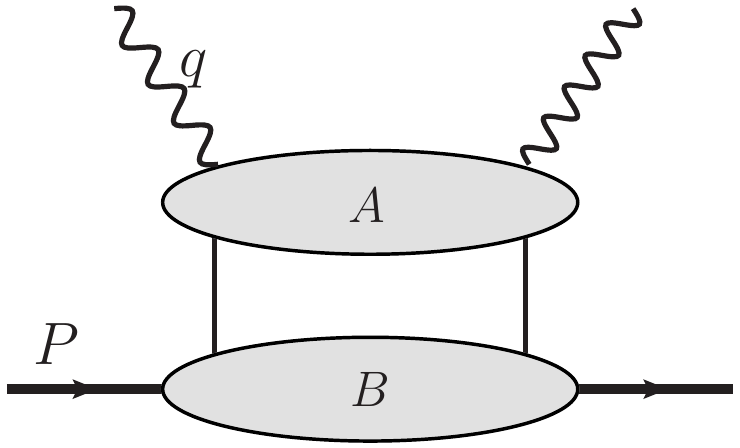}}
  ~+~ \raisebox{-2mm}{\includegraphics[scale=0.4]{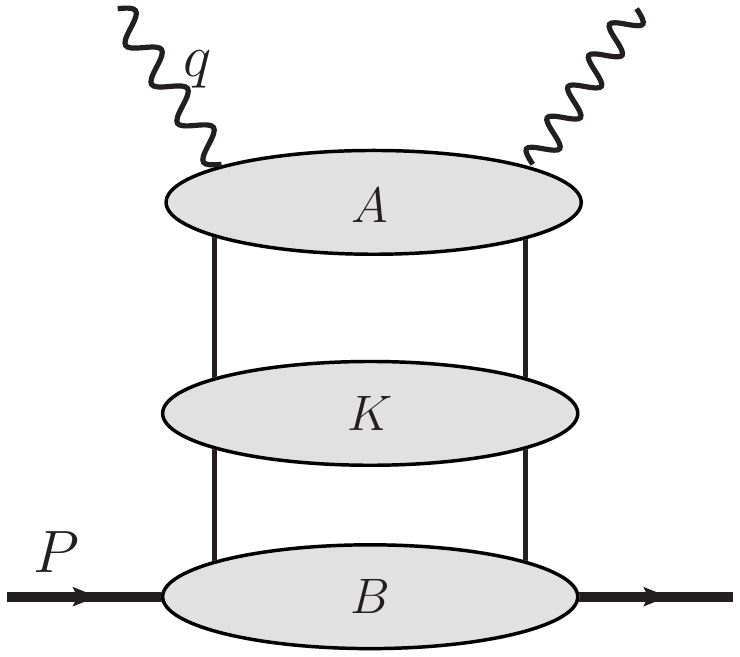}}
  ~+~ \raisebox{-2mm}{\includegraphics[scale=0.4]{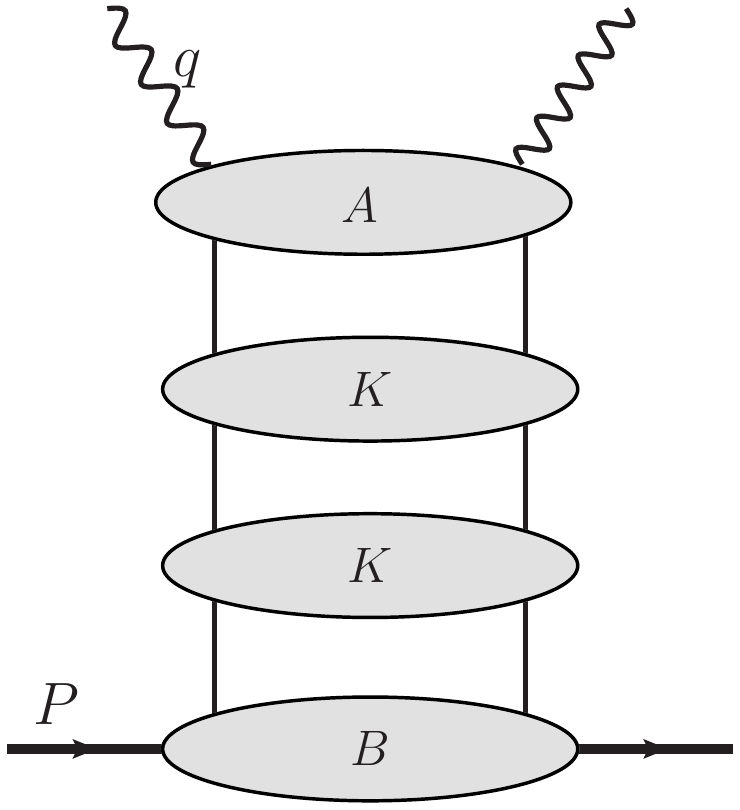}}
  ~+~ \dots . ~+~
\nonumber\\
 & ~+~
   \raisebox{-2mm}{\includegraphics[scale=0.4]{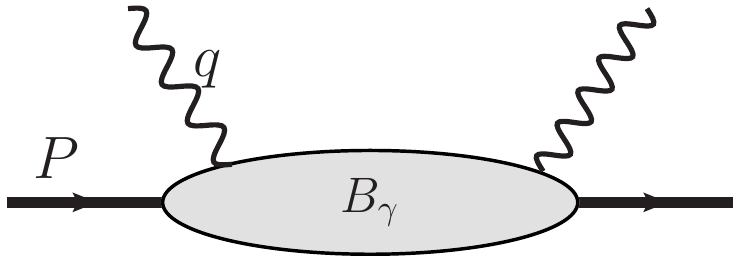}}
\nonumber\\ ={}& ~ ~
   \sum_{n=0}^\infty A K^n B + B_\gamma
 ~ ~ = ~ ~
   A \frac{1}{1-K} B + B_\gamma
.
\end{align}
Here, $F$ denotes the full matrix element of two currents in a state of a
target of momentum $P$. The subgraphs $A$, $K$, and $B$ are two-parton
irreducible in the vertical channel, with $K$ and $B$ including full parton
propagators on their top two lines, but excluding the propagators on the
lower lines.  Finally the quantity $B_\gamma$ is completely two-parton irreducible
in the vertical channel; it turns out to be power suppressed compared
with the contributions from the 2PI graphs in the top line. Generally, a
hadronic target state will entail the use of some kind of bound-state wave
function.  The definition that the lower parton lines of $K$ are amputated
can be notated by short lower lines:
\begin{equation}
  \includegraphics[scale=0.4]{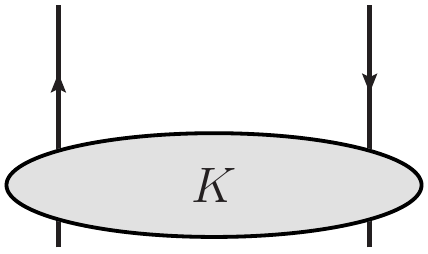}.
\end{equation}
In Eq.\ (\ref{eq:DIS.2PI.exp}), there is a sum over the number of $K$ rungs
from zero to infinity, and the products of the different factors, as in
$AK^nB$, are defined to entail integration over the loop momenta connecting
the factors and the appropriate sums over any spin indices. We define each 2PI subgraph to include all the appropriate counterterms
from the Lagrangian for UV renormalization.  

For the actual DIS structure functions, a final-state cut should be
inserted in the graphical structures in Eq.\ (\ref{eq:DIS.2PI.exp}).  But
essentially all the analysis and factorization apply equally to the
corresponding matrix elements in a target state of a time-ordered
  product of two currents, as well as to
the corresponding structure-function-like objects. 

With a hadronic target, the bottom 2PI rung $B$ in Eq.\ (\ref{eq:DIS.2PI.exp})
never participates in the hard part.
Similarly, in the 2PI expansion, (\ref{eq:pdf.2PI.exp}) below, for a
pdf, $B$ never participates in the UV divergence of the pdf.

Similar but simpler expansions in 2PI graphs apply for a bare pdf (in the
track-A sense) on the same target:
\begin{align}
\label{eq:pdf.2PI.exp}
  \raisebox{-7mm}{\includegraphics[scale=0.4]{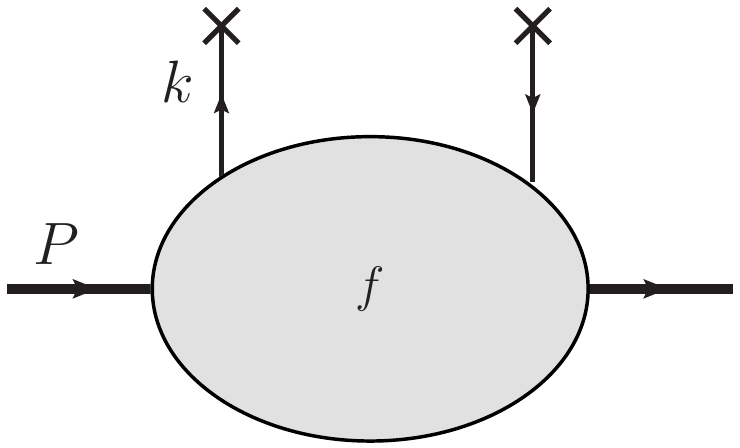}}
 ~~ = {}& ~~
   \raisebox{-2mm}{\includegraphics[scale=0.4]{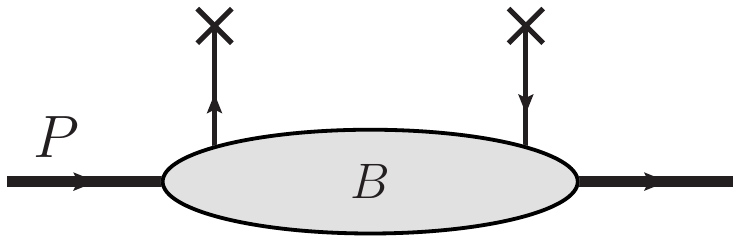}}
  ~+~ \raisebox{-2mm}{\includegraphics[scale=0.4]{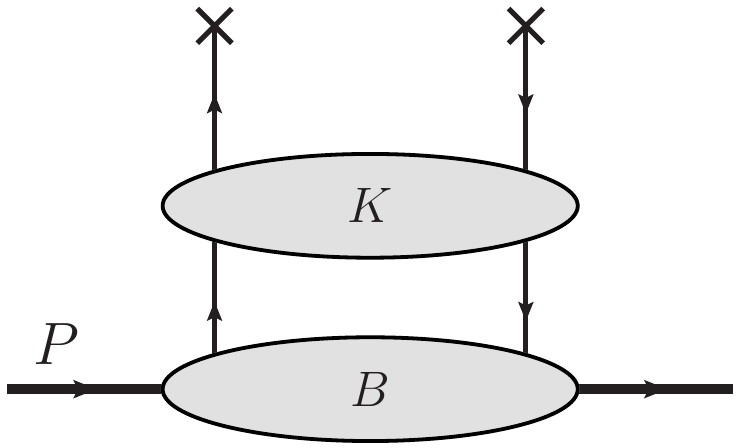}}
  ~+~ \raisebox{-2mm}{\includegraphics[scale=0.4]{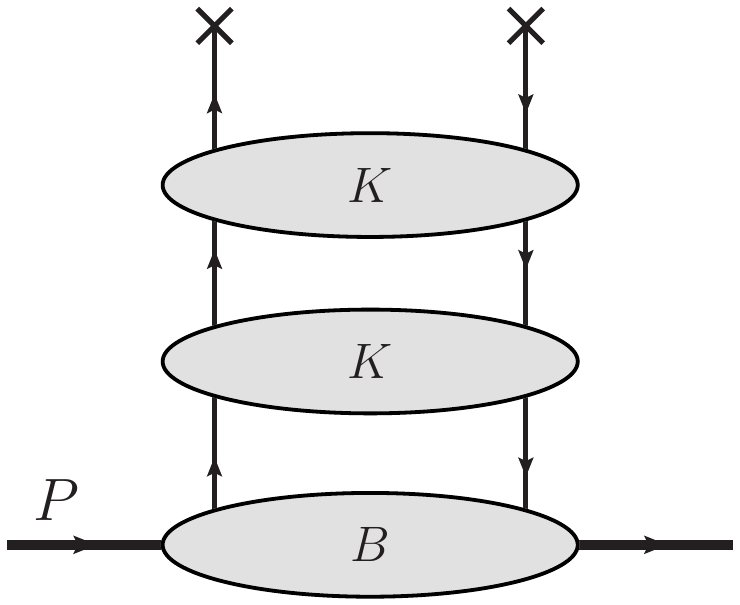}}
  ~+~ \dots
\nonumber\\ ={}& ~ ~
   T \frac{1}{1-K} B .
\end{align}
Here the crosses and $T$ correspond to the operator in the defining
matrix element (\ref{e.pdfdef}) of a pdf 
(where the case of the pdf of a quark is shown). Given that we are working in the
light-cone gauge here, the Wilson line is simply unity.  But
note that because we constructed $K$ and $B$ to be UV finite, the quark
fields in (\ref{e.pdfdef}) must now be renormalized fields, not bare
fields.

The explicit definition of $T$, in the case of an unpolarized
quark pdf, is
\begin{equation}
\label{e.T.def1}
Tf ~ ~ = ~ ~
  \raisebox{-7mm}{\includegraphics[scale=0.4]{gallery/pdf-total}}
 ~~ = ~~
  \int  \frac{ \diff{k^-} \diff[2-2\epsilon]{\T{k}} }{ (2\pi)^{4-2\epsilon}  } 
  ~ ~ \trace_{\text{Dirac, color}} \frac{\gamma^+}{2}
~ ~
  \raisebox{-7mm}{\includegraphics[scale=0.35]{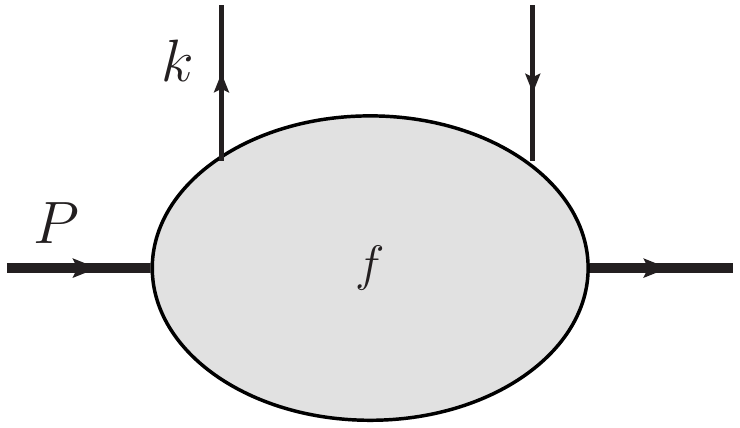}},
\end{equation}
with $k^+\ \xi P^+$.

Compared with the expansion of a DIS matrix element, the main change in
(\ref{eq:pdf.2PI.exp}) is that the two currents at the top are replaced by
partonic fields and a suitable integral over the parton momentum $k$. In
addition, there is no special 2PI subgraph like $B_\gamma$ containing both pdf
vertex and the target.

Next we write the corresponding expansions when the target is a parton
  instead of a hadron. Since the target state is elementary, the 2PI
graphs $B$ and $B_\gamma$ are no longer needed.  Instead we just need an
external line factor for each of the two lines for the incoming parton
target. Then we have for DIS:
\begin{align}
\label{eq:DIS.2PI.exp.parton}
  \raisebox{-7mm}{\includegraphics[scale=0.4]{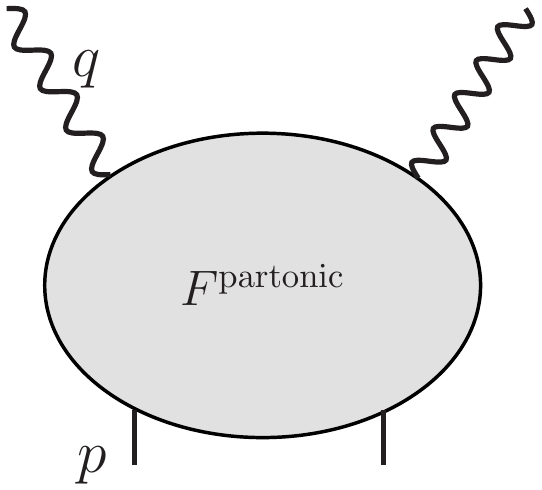}}
 ~~ = {}& ~~
   \raisebox{-7mm}{\includegraphics[scale=0.4]{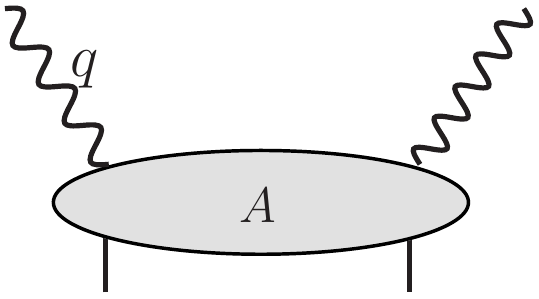}}
  ~+~ \raisebox{-7mm}{\includegraphics[scale=0.4]{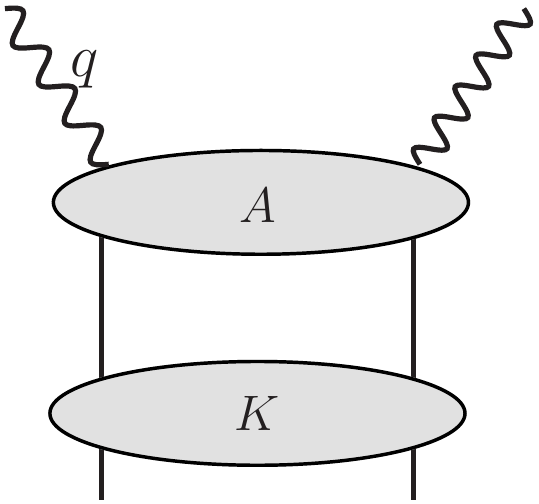}}
  ~+~ \raisebox{-7mm}{\includegraphics[scale=0.4]{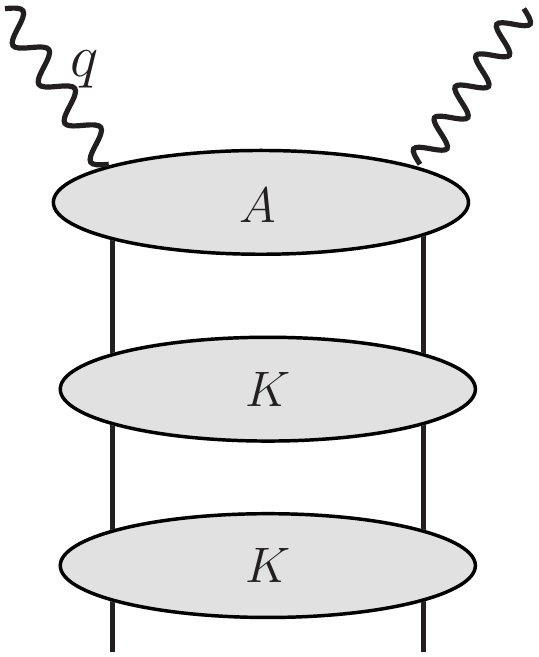}}
  ~+~ \dots 
\nonumber\\ ={}& ~ ~
   \sum_{n=0}^\infty A K^n
 ~ ~ = ~ ~
   A \frac{1}{1-K},
\end{align}
where each term has the external parton propagators amputated, as in the
definitions of $A$ and $K$.  The external-line factors are the same for all
the terms; they play no role in the rest of our treatment, so we have
omitted them.

The similar expansion for a pdf is
\begin{align}
\label{eq:pdf.2PI.exp.parton}
  \raisebox{-7mm}{\includegraphics[scale=0.4]{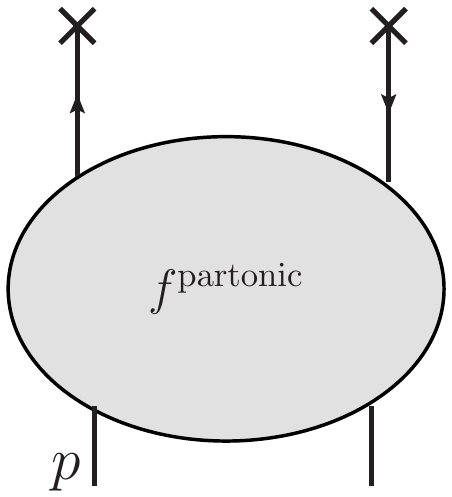}}
 ~~ = {}& ~~
   \raisebox{-2mm}{\includegraphics[scale=0.4]{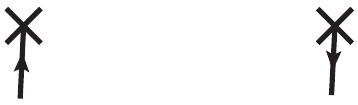}}
  ~+~ \raisebox{-2mm}{\includegraphics[scale=0.4]{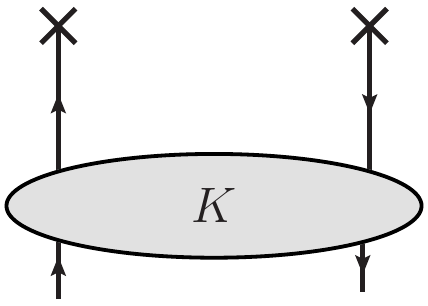}}
  ~+~ \raisebox{-2mm}{\includegraphics[scale=0.4]{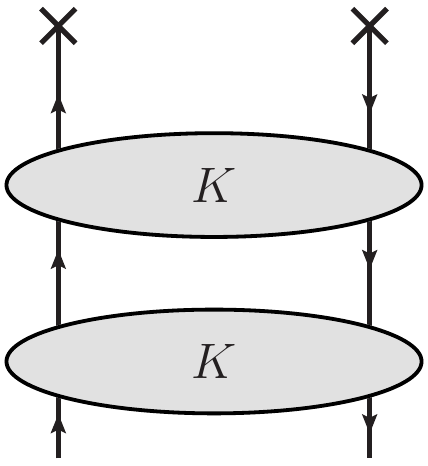}}
  ~+~ \dots 
\nonumber\\ ={}& ~ ~
   \sum_{n=0}^\infty T K^n
 ~ ~ = ~ ~
   T \frac{1}{1-K}.
\end{align}

\end{widetext}

In a gauge theory, like QCD, when the Feynman gauge is used, the graphical
specification of the leading regions is more complicated than given above~\cite{Collins:2011qcdbook}.  The use of light-cone gauge gives the simpler results stated
above.  However it comes with the penalty that the $1/n \cdot k$ term in the
gluon propagator gives what are now known as rapidity divergences.  These
lead to considerable complications in the case of
transverse-momentum-dependent pdfs and of the cross sections for which they
are used~\cite{Collins:2003fm}.  But for the case we consider here, the rapidity
divergences cancel, although general proofs, as opposed to examples, are
hard to find.  Although these problems are very non-trivial, they
are essentially orthogonal to the issues we discuss here, and so we will
ignore them. 

In the generalization of the Wilson-Zimmermann argument, the first step
is to construct for each graph $\Gamma$ its remainder
$\mbox{rem}(\Gamma)$, which is $\Gamma$ with the subtraction of both UV divergences
and of the behavior at large $Q$ to some power, which for us is the
leading power.  

In the original case, the OPE, both the counterterms for UV divergent subgraphs and
the subtractions for the large $Q$ behavior of hard subgraphs are
constructed by zero momentum subtraction, i.e., of an appropriate
polynomial in the external momenta of the subgraph in question.  A slightly
different expansion is needed in the DIS case, with a generic leading
region shown in (\ref{e:DIS.region}).  The expansion for the leading power
of $Q$ in the hard subgraph involves neglecting the relatively small minus
and transverse components of the momentum $l$ connecting the two subgraphs. In contrast, for the OPE, all components of $l$ would
be neglected in the hard subgraph.

For the DIS structure functions, we use a modification of the notation of
Ref.\ \cite{Curci:1980uw}, and obtain
\begin{align}
\label{e:rem.F}
  \mbox{rem}(F)
=
   \sum_{n=0}^\infty A (1-T) \left[ K (1-T) \right]^n B + B_\gamma.  
\end{align}
Here $T$ is what is in fact a generalization of the object of the same
name defined in (\ref{e.T.def1}), that corresponded to a pdf operator.
In (\ref{e:rem.F}), $T$ is defined to be an operation that extracts the
leading asymptotics when the factor on its left has large transverse
momenta relative to the factor on its right. Given that a product like
$AK$ means
\begin{equation}
  AK = \int \frac{ \diff[n]{l} }{ (2\pi)^n } A_{ab}(Q,l;m) K_{ab}(l, \ldots; m), 
\end{equation}
$ATK$ is defined to be
\begin{multline}
\label{e.T.def2}
  ATK =
\\
 \int \diff{l^+}  A_{ab}(Q,\hat{l};m) T_{ab;cd} 
   \frac{ \diff{l^-} \diff[n-1]{\T{l}} }{ (2\pi)^n } K_{cd}(l, \ldots; m), 
\end{multline}
where
\begin{equation}
  \hat{l} = (l^+,0,\T{0})
\end{equation}
and $T_{ab;cd}$ is a matrix that projects out the terms in the sum over
spin indices that are needed for the leading power.  In the case of
unpolarized DIS with quark lines connecting the two subgraphs, we have
\begin{equation}
  T_{ab,cd} = \frac{ \gamma^-_{ab} }{2} \frac{ \gamma^+_{cd} }{2}.
\end{equation}
The factor $\gamma^+/2$ here corresponds to the same factor in the
definition of a quark pdf. Similarly, the integrals over $l^-$ and
$\T{l}$ in (\ref{e.T.def2}), correspond to the integrals in the
definition of a pdf. Thus the result of inserting $T$ is the convolution
product of an approximation to the factor on its left with some kind of
pdf vertex applied to the factor on its right.  Hence it is useful to
overload the semantics of $T$: If there is no factor to its left, it
denotes simply the operator for the pdf.  If instead there is a factor to
its left, an approximation is applied to that factor. In its uses in
factorization, a pdf is always multiplied by a coefficient function that
is obtained by an approximant applied to some graph.

Observe that, as in the Wilson-Zimmermann treatment of the OPE, masses are
left unchanged in all quantities.

In a term like
\begin{equation}
  A (1-T) K B,
\end{equation}
there is a UV divergence where transverse momenta in $K$ go to infinity;
these correspond to UV divergences in a pdf.  But the corresponding term in
$\mbox{rem}(F)$ is
\begin{equation}
  A (1-T) K (1-T) B. 
\end{equation}
Then the second factor of $1-T$ removes not only the leading large $Q$
asymptotics of $A (1-T) K$, but also the UV divergence in $-ATKB$. 

Subtracting the remainder Eq.~\eqref{e:rem.F} from the original 
structure function in Eq.~\eqref{eq:DIS.2PI.exp}, and performing some algebraic manipulations gives 
\begin{equation}
\label{e:BPHZ.fact}
  F =
  \left( \sum_{n=0}^\infty AK^n\right)
  \left\{ T \sum_{n'=0}^\infty [K(1-T)]^{n'} B\right\}
  + \mbox{rem}(F).
\end{equation}
Let us define a renormalized pdf by 
\begin{equation}
\label{e:BPHZ.pdf}
  f^{\text{ren., BPHZ}}
  =
   T \sum_{n'=0}^\infty [K(1-T)]^{n'} B \, .
\end{equation}
Then, since $\mbox{rem}(F)$ is suppressed by a power $Q$, Eq.\
(\ref{e:BPHZ.fact}) has the form of a factorization property, with the
coefficient function being
\begin{equation}
\label{e:BPHZ.C}
  C^{\rm BPHZ}(q,x/\xi) =
  \left( \sum_{n=0}^\infty AK^n\right)_{l \to \hat{l} = (\xi P^+,0,\T{0})}.
\end{equation}
That is, it is an \emph{unsubtracted} parton DIS structure function, with
the external parton lines amputated, and with a light-like external parton
momentum. It is simply Eq.~\eqref{eq:DIS.2PI.exp.parton}.  Note that it can be shown that the renormalized pdf defined
above is a renormalization factor convoluted with a bare pdf, just as in
track A. That is, it is a fully renormalized version of Eq.~\eqref{eq:pdf.2PI.exp}.

Thus we have a bare factorization just like that in track B, \emph{except
  that}
\begin{itemize}
\item The masses of internal lines of the (unsubtracted or ``bare'') coefficient function are
  unchanged instead of being set to zero.
\item The pdf is definitely a UV renormalized quantity with a particular
  scheme, not a bare quantity like that in definition (\ref{e.pdfdef}).
\end{itemize}
In this approach, the renormalization scheme for the pdfs is implemented by
counterterms that are obtained by the same operation $T$ as for extraction
of large-$Q$ asymptotics.  It is in fact the same as the BPHZ scheme used
by Wilson and Zimmermann, except for being extended from the pure
zero-momentum renormalization scheme for local operators to a version
suitable for the renormalization of the bilocal operators in pdfs.  The
subtractions are at zero values of only the minus and transverse components
of momentum.

In a renormalizable non-gauge theory with non-vanishing masses, the above
procedure works as is, but in a gauge theory modifications of the
counterterms are liable to be needed (a) to preserve gauge-invariance, and
(b) to avoid IR and collinear divergences associated with a massless gluon
when external momenta are zero or light-like.

If one wanted to use \msbar{} renormalization for the pdfs, then the above
treatment needs to be modified so as to decouple the subtraction operation
for UV divergences from that for extraction of large $Q$ asymptotics.  This
was done in \cite{Collins:2011qcdbook,Collins:1998rz}, and leads directly
to the track-A formulation with its subtracted coefficient function.

\subsection{Relation of track B to the Wilson-Zimmermann treatment of OPE}
\label{s.ope}

It would be natural to expect that the coefficient function in the OPE or
factorization is a short-distance quantity, so that in QCD asymptotic
freedom implies that useful perturbative calculations can be made, since
the effective coupling $\alpha_s(Q)$ is small.

However, in the Wilson-Zimmermann form of the OPE the construction of the coefficient function does not include subtractions 
\emph{for the collinear region}, and so it does not obey the purely short-distance property. Thus it is a
non-perturbative quantity in QCD.

As regards the validity of the OPE itself, Wilson and Zimmermann point out
that it is sufficient that all dependence on $Q$ is in the coefficient
function.\footnote{As they point out, essentially the same observation for
  a similar purpose had been made much earlier by Valatin
  \cite{Valatin.1954a,Valatin.1954b,Valatin.1954c,Valatin.1954d}.}  Thus
the short-distance part is correctly contained solely in the coefficient
function.

But for the OPE to be valid a suitable definition of the operators must be
made. An important finding of Wilson and Zimmermann was that zero momentum
subtractions (with unchanged masses) accomplish this, in the BPHZ
scheme. In effect, the OPE can be regarded as giving a definition of
the composite local operators used in the OPE.

Bare factorization in track B also has an unsubtracted coefficient
function, but with masses set to zero.  So it is natural to expect that a
variation on the Wilson-Zimmermann approach would lead to bare
factorization together with a suitable definition of the pdfs.  In the
next section, we will implement this idea.  It will require a change of
scheme to what we will call the BPHZ$'$ scheme.

It is far from obvious that the initial papers for track B intended to use
the Wilson-Zimmermann method.  For example, Politzer in
Ref.~\cite{Politzer:1977fi} does not mention it. His motivation seems to be
entirely different, arising from an attempt to generalize the parton
model.

Much of the early work that applied the OPE in QCD appears not to use the
actual Wilson-Zimmermann method with its non-perturbative coefficient
functions, even when the Wilson-Zimmermann papers are referred to.  
Instead the composite operators were
often defined by \msbar{} renormalization for UV divergences.  This is
exactly like track-A factorization.  The corresponding coefficient function
is then perturbative.  The evolution equations are then standard
renormalization-group equations, rather than the Callan-Symanzik equations
that apply to the coefficient functions in the Wilson-Zimmermann approach,
where the pdfs (or their analogs in the OPE) do not evolve at all.

\subsection{The BPHZ$'$ scheme}
\label{s.trackbpdfs}

With inspiration from the Wilson-Zimmermann papers, we will now show how to
define the track B ``bare'' pdfs in a way that ensures that (a) the track B
equations are correct, (b) the pdfs have a definite relationship to
standard operator matrix elements such as given in (\ref{e.pdfdef}), and
(c) the definition applies independently of the choice of dimensional
regularization for both UV and IR divergences.

Given how Wilson and Zimmermann derive the OPE, as summarized in Sec.\
\ref{s.WZ}, this amounts to recognizing that the track-B ``bare'' pdfs are
actually UV-renormalized, and to determining which renormalization scheme
is needed.  Recall that, in general, from the UV finiteness of structure
functions, both partonic and hadronic, it follows that $f^{\text{bare,B}}$
is also UV finite in order for \eref{bare.factn} to be true. 

Renormalization counterterms for pdfs are used for subgraphs of the form of
the hard subgraphs specified in (\ref{e:pdf.UV.div}). So we can
infer the UV renormalization scheme for the pdfs, by examining DIS on an
on-shell partonic target, with all parton masses set to zero.  Then
$F(Q,\xbj)$ in Eq.\ (\ref{e.bare.factn}) is the same as
$F^{\text{partonic}}$.  Hence the parton densities for massless partonic
targets must be exactly the lowest order, or free-field values:
\begin{equation}
\label{e.bphz0_pdf}
    f^{\text{massless partonic,bare,B}}_{i/j}(\xi) = \delta(\xi-1) \delta_{ij}.
\end{equation}
The unique renormalization scheme that achieves this is therefore the one
where renormalization counterterms exactly remove all perturbative
contributions to the massless partonic pdf.  We call it ``BPHZ$'$''.

To see the nature of this scheme, it is useful to examine the structure of
one-loop calculations of pdfs on a partonic target, but with non-zero
masses, and with the external parton permitted to be
off-shell\footnote{I.e., we are really treating a Green function with the
  pdf operator and two parton fields; a pdf-like Green function.}.  As seen
in many examples, e.g., in Sec.\ \ref{s.examples}, the basic form can be
written as an integral over transverse momentum:
\begin{equation}
  \label{eq:pdf.one.loop.int}
  I^{\rm bare}(m^2,p^2,x) = 
  \int \frac{\diff{^{2}\T{k}}}{(2 \pi)^{2}} \frac{1}{\kT^2 + C(x)} ,
\end{equation}
multiplied by an overall factor that depends on $x$ but not on $\T{k}$.  This
integral will need a regulator to cutoff its UV divergence. The quantity
$C(x)$ summarizes the dependence on parton mass and external virtuality:
\begin{equation}
  C(x) = m^2 \times A(x)
  ~ + ~
  p^2 \times B(x), 
\end{equation}
with the coefficients $A$ and $B$ depending on $x$ but not $\T{k}$.  We
could, of course, use dimensional regularization for the UV divergence, but
that would obscure certain conceptual issues.  Instead for a one-loop
integral we can simply use an upper cutoff $\Lambda$, so that
\begin{equation}
  \label{eq:pdf.one.loop.int.reg}
  I^{\rm bare}(m^2,p^2,x) = 
  \int_0^{\Lambda^2} \frac{\diff{\kT^2}}{4\pi} \frac{1}{\kT^2 + C(x)}.
\end{equation}

With purely a zero-momentum subtraction, i.e., the BPHZ scheme, the
renormalized value is
\begin{equation}
  \label{eq:pdf.one.loop.BPHZ}
  I^{\text{ren.\ BPHZ}} = 
  \int_0^\infty \frac{\diff{\kT^2{}}}{4\pi}
  \left( \frac{1}{\kT^2 + C(x)} -  \frac{1}{\kT^2+m^2A(x)} \right),
\end{equation}
in which the counterterm is applied in the integrand, so that the UV
regulator can be removed to give a finite result.

But for the BPHZ$'$ scheme we need for track B, the subtraction is of the
value of the integrand when both $p^2$ and $m^2$ are zero,
\begin{equation}
  \label{eq:pdf.one.loop.ren}
  I^{\text{ren.\ BPHZ$'$}} = 
  \int \frac{\diff{^{2-2 \epsilon}\T{k}{}}}{(2 \pi)^{2-2 \epsilon}}
  \left( \frac{1}{\kT^2 + C(x)} -  \frac{1}{\kT^2} \right).
\end{equation}
Although the integral is UV convergent, it has a collinear divergence at
$\T{k}=0$. As such, the BPHZ$'$ scheme is not really completely defined until 
an IR regulator scheme is chosen. (This is in contrast to the standard BPHZ scheme.)
We have chosen dimensional regularization as the IR regulator. 

The UV divergence is from the asymptote $1/\kT^2$ of the integrand, which
can be characterized by saying that it is obtained by setting to zero both of the
quantities $m^2$ and $p^2$ that are negligible with respect to $\kT^2$
when it goes to infinity. Hence BPHZ$'$ is actually a very natural scheme,
in a sense more so than BPHZ.  It is a kind of minimal subtraction.  By its
motivation, this scheme is exactly and uniquely what is needed to give a
renormalized value of zero when the parton mass is zero and the external
parton is on-shell.

The penalty for this subtraction is the introduction of a collinear
divergence that was not at all present in the original integral, but that
is present when we restore the parton mass and/or the external parton is
off shell.  Recall the ``IRR'' subscript in \eref{bare.factn_subs}. The off-shell and massive case applies to a graph that
appears as a subgraph in the pdf on a hadronic target.  

A simple lowest-order example with the application of renormalization is 
\begin{equation}
\label{e.LO.ladder}
    \raisebox{-10mm}{\includegraphics[scale=0.46]{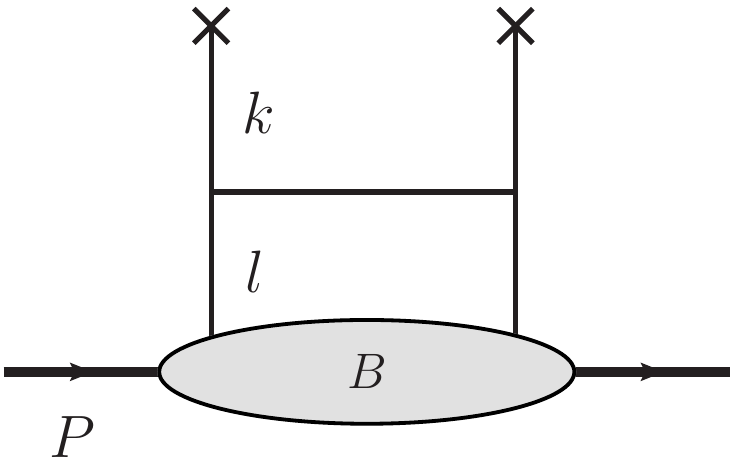}}
  ~ ~ ~ - ~ ~ ~ 
  \raisebox{-10mm}{\includegraphics[scale=0.46]{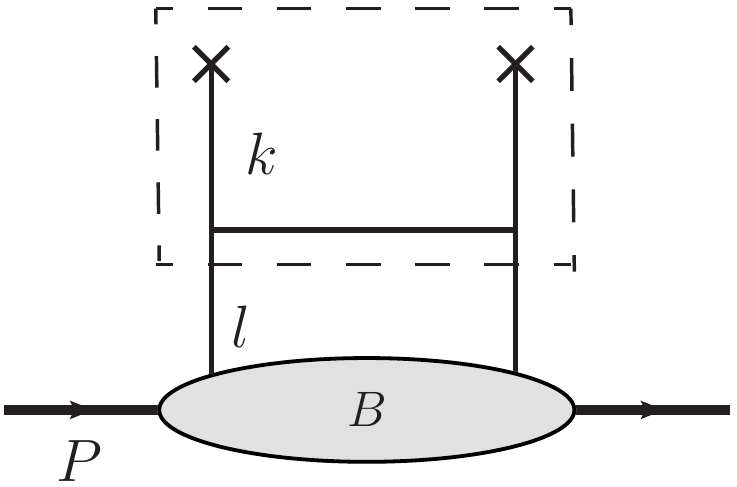}}
.
\end{equation}
The sole UV divergence is in the upper loop of the left-hand graph.  The
box denotes the operation that replaces the integrand of the upper loop by
a quantity like the $1/\T{k}^2$ term in Eq.\
(\ref{eq:pdf.one.loop.ren}), whose negative gives the counterterm for the
subdivergence.

Since the renormalization counterterm for the upper loop has a collinear
divergence, the full hadronic pdf also has this collinear divergence, even
though there is no collinear divergence in the bare pdf.  Here ``bare'' is
in the track-A sense that the graphs for the pdf are obtained purely from
graphs for the quantity defined in Eq.\ \eqref{e.pdfdef} without extra
counterterms or a renormalization factor.

However, it could be argued that the counterterm is zero, because of the
vanishing of the dimensionally regulated integral over the counterterm's
integrand:
\begin{equation}
\label{e:scale.free}
  - \int \frac{\diff{^{2-2 \epsilon}\T{k}{}}}{(2 \pi)^{2-2 \epsilon}} \frac{1}{\kT^2} 
  = 0. 
\end{equation}
However, this is quite misleading.  Suppose we used a cutoff $\Lambda$ to
regulate the UV divergence, and then used dimensional regularization with $\epsilon<0$
only to regulate the collinear divergence.  Then the integral is not only
nonzero, but is power-law divergent as $\Lambda\to\infty$:
\begin{align}
  - \int_{\kT<\Lambda} \frac{\diff{^{2-2 \epsilon}\T{k}}}{(2 \pi)^{2-2 \epsilon}} \frac{1}{\kT^2} 
  ={}& - \int_0^{\Lambda^2} \frac{\diff{\kT^2} (\kT^2)^{-\epsilon}}{\Gamma(1-\epsilon)\,(4 \pi)^{1-\epsilon}}
  \frac{1}{\kT^2} 
\nonumber\\
  ={}& \frac{ \Lambda^{-2\epsilon} }{ \epsilon\,\Gamma(1-\epsilon)\,(4 \pi)^{1-\epsilon} } \, . \label{e.eq41}
\end{align}
Of course, this UV divergence cancels the corresponding UV divergence
in the integral over the first term in Eq.\ (\ref{eq:pdf.one.loop.ren}).
As we will discuss in Sec.\ \ref{s.dimreg}, the construction of a
dimensionally regulated integral with a UV divergence has effectively and
implicitly introduced a counterterm localized at $\kT=\infty$ to give the result
in Eq.\ (\ref{e:scale.free}).  

Given that $\epsilon<0$ to regulate the collinear divergence, the collinear
divergence in the counterterm is actually negative.  Therefore the
supposed positivity of the ``bare'' track-B pdfs is actually violated.

We summarize our results in this section, together with immediate implications;
\begin{enumerate}
\item The so-called bare pdf $f_{j/H}^{\text{bare,B}}$ in track-B is
  actually a pdf renormalized to remove its UV divergence, but in the
  BPHZ$'$ scheme: $f_{j/H}^{\text{bare,B}} = f_{j/H}^{\text{ren, BPHZ$'$}}$
\item The renormalization counterterms entail collinear divergences in all
  pdfs on a hadronic target. 
\item This choice of scheme and the use of dimensional regularization for
  collinear divergences amounts to a particular choice of the schemes
  labeled R$_2$ and IRR in Eq.\ (\ref{e.bare.factn_subs}),
\item In the bare factorization formula (\ref{e.bare.factn}), these
  collinear divergences cancel against collinear divergences in the
  massless partonic structure function, so that there are no divergences in
  the hadronic structure function on the left-hand side.  
\end{enumerate}

\section{Dimensional regularization and positivity}
\label{s.dimreg}

Under conditions such as superrenormalizable theories, where it is possible
to construct pdfs as literal (light-front) number densities, the
normal properties of positivity follow automatically from positivity of
  the metric of quantum-mechanical state space.  But this is a
  property that is not necessarily true when there is a regulator.
The Pauli-Villars regulator for UV divergences is the classic
case. Here, we will explain how dimensional
regularization, when simultaneously applied to UV and IR divergences,
violates positivity.

\subsection{Dimensional regularization}
\label{s:dim.reg}

In Wilson's original argument \cite{Wilson:1972cf} for defining integration
  in $n=4-2\epsilon$ dimensions for arbitrary continuous $\epsilon$, integrals are
uniquely determined (aside from normalization) by (a) linearity in the
integrand, (b) scaling behavior, (c) invariance under translations. In
   addition, applications require an extension to the definition: (d)
Analytic continuation in $n$ is applied to extend the range of $n$ from
where an integral is convergent by 
normal mathematical criteria. (It is not even necessary to require
agreement with ordinary integrals in integer dimension when those are
convergent; that follows from the postulates and a choice of
normalization.)

One can then give a construction of the dimensionally regulated integrals we
need for Feynman graphs in terms of ordinary integration and analytic
continuation in $n$.  It is unique given the natural choice of
normalization, which is that the integral of a Gaussian in a \emph{Euclidean}
space obeys
\begin{equation}
  \label{eq:gaussian}
  \int \diff[n]{x}\, e^{-\sum_jx_j^2}
  = \left( \int \diff{y} \, e^{-y^2} \right)^n .
\end{equation}

However, dimensional regularization does not preserve all properties of standard integration. For
example, in general it is not allowed to exchange the order of a limit and
integration, e.g., for the massless limit of the integral for a massive
Feynman graph. Most importantly for us, standard integrals obey positivity, of
which a trivial example is that the integral of a positive function is
positive: That is, if $f(x)$ is strictly positive, then so is $I[f] = \int
\diff[n]{x} f(x)$.  If the integrand is merely required to be non-negative,
the integral is also non-negative; moreover, it is zero if and only the
integrand is zero everywhere.

In dimensional regularization, those properties do apply if the integral is
convergent by the standard mathematical criteria.  Otherwise, it is often
violated whenever continuation in $n$ is used in the construction of the
integral. 

The vanishing of scale free integrals violates positivity very
much. Thus, in dimensional regularization, the Euclidean integral
\begin{equation}
\label{e:scale.free.1}
  \int \diff[n]{k} \frac{1}{k^2}
\end{equation}
is zero but has an integrand that is positive everywhere.  The integrand is
rotationally invariant, so that vanishing of the $n$-dimensional integral
is equivalent to vanishing of the following one-dimensional integral:
\begin{equation}
\label{e:scale.free.2}
  \int_0^\infty \diff{k} k^{n-1} \frac{1}{k^2}  = 0.
\end{equation}
With the standard mathematical definition this integral is unambiguously
positive infinite. Technically, we could say that in dimensional
regularization the integration measure is not positive everywhere, unlike
standard integration.  Whenever the degree of UV divergence of the
  integral is non-negative, i.e., $n\geq2$, there is a negative contribution
  which we can treat as localized at $k=\infty$.  Similarly, whenever the degree
  of IR divergence is non-negative, i.e., $n\leq2$, there is a negative
  contribution localized at $k=0$.  These two ranges of $n$ overlap at
  $n=2$, thereby preventing the integral from existing at any $n$, with the
  ordinary mathematical definition.\footnote{The proof that a
      scale-free integral is defined (and zero) relies on defining the
      integral as a sum of a term with no UV divergence and one with no IR
      divergence.  Each term is defined by ordinary integration for values
      of $n$ where it is convergent by the ordinary criterion, and is then
      analytically continued to all $n$ except for $n=2$. In the sum, the
      poles at $n=2$ cancel, so the sum is also defined at $n=2$, by
      analytic continuation.}

By themselves, the above statements about non-positivity of the
  integration measure can prevent a naive application of the standard
  positivity argument for pdfs whenever we use dimensional regularization
  for both UV and IR/collinear divergences.  The positivity argument for
  pdfs involves sums and integrals over final states of the absolute square
  of matrix elements.

An interesting mathematical example is the integral 
\begin{equation}
\int \diff[2-2 \epsilon]{\T{k}}{} \frac{(\kT^2 - Q^2)^2}{\kT^2 (\kT^2 + Q^2)^2} \, .
\end{equation}
The integrand is positive definite everywhere, but the integral
evaluated in dimensional regularization is $-4 \pi$ in the limit that
$\epsilon\to0$.  For general $\epsilon$, the integral equals $-4\pi \Gamma(1+\epsilon)(\pi Q^2)^{-\epsilon}$,
which is negative for all $\epsilon>-1$.

None of above is to deny that dimensional regularization is an extremely
useful and elegant method for doing many calculations.  But one has to be
careful in going beyond those properties and manipulations that follow from
its definition and construction.


\subsection{Application to pdfs}
\label{s.pos.pdf}

In light of these results for dimensional regularization, we examine
the basic argument that pdfs are positive.

Now the operator definition of a bare pdf is equivalent to the expectation
value of a light-front number operator integrated over transverse momentum
\cite{Bouchiat:1971mj,Soper:1976jc}, \cite[Chap.~6]{Collins:2011qcdbook}.
Specifically, the pdf is an integral over transverse momentum of the
following expectation value:\footnote{For the purposes of this section, we
  ignore the added complications in giving a fully correct definition of a
  transverse-momentum-dependent pdf in QCD.}
\begin{equation}
\label{e.tmd.pdf}
  f(x,k_T) = \lim_{|\psi\rangle\to\text{state of definite $P$}}
            \frac{ \langle \psi| a_k^\dag a_k |\psi\rangle }{ \langle \psi|\psi\rangle }.
\end{equation}
A minor complication is that the expectation value of an operator in a
state requires the state to be normalizable, and so the state cannot be a
momentum eigenstate.  So we define the pdf in terms of a limit as the
target's state becomes a momentum eigenstate.  Elementary manipulations
\cite[Chap.~6]{Collins:2011qcdbook} convert that to a standard form like
\eref{pdfdef}.

Now the numerator in (\ref{e.tmd.pdf}) is given by a
sum/integral over intermediate states:
\begin{align}
  \langle \psi| a_k^\dag a_k |\psi\rangle =& \sum_X \langle \psi| a_k^\dag{} |X\rangle \, \langle X| a_k |\psi\rangle
\nonumber\\
  =& \sum_X \left| \langle X| a_k |\psi\rangle \right|^2 .
\label{e:state.contrib}
\end{align}
This is non-negative, provided that the sums and integrals have their
standard meanings.  Hence the TMD pdf is also non-negative. 

The definition of a collinear pdf has an insertion of an integral over
$k_T$, and the result is similarly non-negative:
\begin{multline}
  f^{\text{bare}}(x) =
\\
 \lim_{|\psi\rangle\to\text{state of definite $P$}}
           \int d^2k_T \, \sum_X
            \frac{ \left| \langle X| a_k |\psi\rangle \right|^2 }{ \langle \psi|\psi\rangle }
      \geq 0.
\label{e:state.integral}
\end{multline}

The UV divergence arises from $\T{k}\to\infty$, and hence where also the final
state $X$ has large transverse momentum.  When we consider a pdf in the
massless partonic case, divergences similarly arise when $\T{k}\to0$, with
corresponding final states.  The divergences are not in the integrand
itself, $\left| \langle X| a_k |\psi\rangle \right|^2 / \langle \psi|\psi\rangle$, but in the integral over
certain limits of it.  

Then when we perform the integral and use dimensional regularization to
construct a bare pdf in a massless partonic target, our analysis of
  Eqs.\ (\ref{e:scale.free.1}) and (\ref{e:scale.free.2}) shows that the
integration measure acquires negative terms in any limit of the integration
variables that would otherwise give a divergence.  For a massless
  integral for a pdf, divergences are present for all $n$, and hence the
  dimensional regulated integral also violates positivity for all $n$.

Now positivity of the integrand in Eq.\ (\ref{e:state.integral})
  results from positivity of the metric on a normal quantum-theoretic state
  space. So we could interpret a violation of positivity of the
  dimensionally regulated integral as corresponding to a non-negative
  metric in some kind of extended state space.

\section{Failure of an argument for positivity of \msbar{} pdfs}
\label{s.pos.assess}

We now show how Ref.~\cite{Candido:2020yat}'s argument for positivity of
\msbar{} pdfs breaks down.

A number of critical steps are in the first part of their Sec.\ 2. It
begins with a statement of track B bare factorization (to use our
terminology) in Eq.\ (2.1). It has the same form as our \eref{bare.factn},
except for changed notation and normalizations.

One factor is an unsubtracted structure function ($F^{\rm partonic}$ in our
notation) for DIS on an on-shell partonic target with all masses set to
zero.  The other factor is of a pdf, whose operator definition was given in
Eq.\ (2.2) of \cite{Candido:2020yat}.  That definition agrees with the
definition that we gave in \eref{pdfdef} for a bare pdf $f^{\rm bare, A}$
in track A.  (The fields and coupling in \cite{Candido:2020yat} must be
bare quantities in order that (a) the pdf is a number density in the
light-front sense, and (b) the operator is gauge-invariant.)

In our formula for bare factorization, the bare parton density is notated
$f^{\rm bare, B}$ rather than $f^{\rm bare, A}$. Hence Eqs.\ (2.1), (2.2)
and (2.6) of \cite{Candido:2020yat} are equivalent to an assertion of our
\eref{bare.factn} together with an assertion that $f^{\rm bare, B}=f^{\rm
  bare, A}$.  We have shown in previous sections that these assertions are
valid provided that dimensional regularization is used for both UV and
collinear/IR divergences, but not in general.  In Ref.\
\cite{Candido:2020yat}, only dimensional regularization is used. 

The derivation of positivity of \msbar{} pdfs relies on positivity both of
bare pdfs and of partonic cross sections.  The derivation is indirect, with
the definition and use of subtraction schemes named DPOS and POS, followed
by a scheme change to \msbar.  Primarily the argument is given in terms of
the results of one-loop calculations.

The intermediate subtraction schemes were motivated by the fact that the
standard \msbar{} subtraction of a collinear divergence from a partonic
structure function is actually an oversubtraction.  It results in negative
subtracted partonic structure functions.  The definitions of the DPOS and
POS scheme remove the oversubtraction, so that the subtracted partonic
structure functions remain positive.  This was needed because one part of
the argument---in the early part of Sec.\ 3 of
\cite{Candido:2020yat}---required positivity of all four quantities in the
first line of our Eq.\ (\ref{e.trackb_rearrange}), including the subtracted
partonic structure function $\mathcal{C}^B$.  But to show that the change
of scheme to \msbar{} preserves positivity of the pdf did not need any
properties of $\mathcal{C}^B$.

We can gain an overall view of the derivation from the statement
\cite[p.~5]{Candido:2020yat}: ``If all contributions which are factored
away from the partonic cross section and into the PDF remain positive, then
the latter also stays positive.''  In our notation, what is factored away
from the partonic cross section is $Z^B$ in Eq.\ (\ref{e.abs1}).  We
absorbed it into a redefinition of the pdf in Eq.\
(\ref{e.trackb_rearrange}).  

Now dimensional regularization with space-time dimension $n=4-2\epsilon$ was used,
so we apply the results of our discussion in Sec.\ \ref{s:dim.reg}.

To regulate collinear divergences, a space-time dimension \emph{above} 4 is
needed, i.e., $\epsilon<0$. So, when $\epsilon\to0$ from below, the collinear
divergences are positive, Hence the collinear-divergence factor $Z^B$
  in Eqs.\ (\ref{e.abs1}) and (\ref{e.fbrenorm}) only obeys positivity
  in space-time dimensions above 4. Positivity of \msbar{} pdfs would
  then follow were the bare pdfs also positive for negative $\epsilon$, i.e.,
  for space-time dimensions above 4.

Positivity of bare pdfs appears to be almost trivial to prove---e.g., Sec.\
\ref{s.pos.pdf}---and as such it seems that it should be an uncontroversial
statement.  However, we also saw that the argument only applies if the
integrals giving the pdfs are convergent; failures can occur 
when dimensional regularization is used for both
UV and IR/collinear divergences.  Now a pdf defined by Eq.\
(\ref{e.pdfdef}) has UV divergences, As we have seen in Sec.\
\ref{s.dimreg} that implies that the contribution from the UV region is
necessarily positive only if the degree of UV divergence is negative, i.e., in
space-time dimensions \emph{below} 4, i.e., $\epsilon>0$.  But that is where the
collinear divergence factor does not obey positivity. 

  If we go in the opposite direction in dimension, i.e., $\epsilon<0$, to make the
  collinear contribution positive, then the UV contribution obtained by
  analytic continuation from positive $\epsilon$ does not obey positivity.

  Hence there is no value of $\epsilon$ for which positivity is obeyed by both the
  factors in the track-B formula for renormalized pdfs, Eq.\
  (\ref{e.fbrenorm}).  A proof of positivity of renormalized pdfs from
  positivity of the collinear divergence factor $Z^B$ and the bare pdfs
  fails.

Alternatively one could use methods of regulation or cutoff other than
dimensional regularization, at the price of losing the simplicity that goes
with its use.  We have seen that bare factorization then generally fails if
the pdf is still the bare pdf defined by its standard formula.

The only way of recovering the validity of the formula for bare
factorization is to replace the bare pdf by a renormalized pdf with the UV
divergences removed by the BPHZ$'$ scheme.  As we have seen, these pdfs
acquire collinear divergences not present in the bare pdf itself -- recall Eq.~\eqref{eq:pdf.one.loop.ren}. The
subtractions violate positivity of the resulting pdfs.

To summarize, there are three assertions that need to be true
simultaneously for the derivation of strict positivity in
Ref.~\cite{Candido:2020yat} to be valid
\begin{enumerate}
\item Bare factorization Eq.~\eqref{e.bare.factn} is valid.
\item Each bare pdf, as given by the standard operator definition
  (\ref{e.pdfdef}), obeys positivity.
\item Partonic cross sections $F^{\rm partonic}$ obey positivity.
\end{enumerate}
As we have shown, at least one of items 1--3 must be false. As a result,
negative pdfs are not excluded in the $\msbar$ scheme.


\section{Curci et al.}
\label{s.Curci}

The treatment in Sec.\ \ref{s.WZ} now enables us to critique the
derivation by Curci et al.\ \cite{Curci:1980uw}.  They combine a form of
bare factorization and a version of the derivation in Sec.\ \ref{s.WZ}, but
applied to a massless partonic structure function.   They use light-cone
gauge just as we did in Sec.\ \ref{s.WZ}. 

Their definition of a bare pdf is modified from the one given in Eq.\
(\ref{e.pdfdef}).  The matrix element is restricted to 2PI graphs:
\begin{equation}
  \label{e.pdf.bare.CFP}
  f^{\text{bare, CFP}}
=
     \raisebox{-2mm}{\includegraphics[scale=0.4]{gallery/pdf-ladder_0}}
= T B.
\end{equation}
It has no UV divergences, and so is a purely collinear object.  Moreover,
the factor $B$ is in full QCD, with massive hadrons, so there are also no
collinear divergences in this pdf.  Its definition is clearly different
from the standard one that is given by Eq.\ (\ref{eq:pdf.2PI.exp}), which
transcribes Eq.\ (\ref{e.pdfdef}) in light-cone gauge.

In their Fig.\ 3, they assert (but do not prove) a bare factorization
formula, which is exactly the same as our Eq.\ (\ref{e.bare.factn}), except
that the bare pdf is not given by Eq.\ (\ref{eq:pdf.2PI.exp}) but by Eq.\
(\ref{e.pdf.bare.CFP}).  In the notation of Sec.\ \ref{s.WZ}, this
factorization is
\begin{align}
  F = {}&
  \left( A \frac{1}{1-K} \right)_{m=0} T ~ B
     ~ ~ + ~ ~ \mbox{power-suppressed}
\nonumber\\
   ={}& F^{\text{massless, partonic}} \otimes f^{\text{bare, CFP}}
     + \mbox{power-suppressed}.
\end{align}
This equation cannot be correct: When a hadronic target is used, both
the bare pdf used here and the hadronic structure function have no
divergences, but the unsubtracted massless partonic structure function
does have collinear divergences.

Despite this problem, it is interesting that, as we saw in Sec.\
\ref{s.WZ}, their methods can be used very easily to provide a correct
derivation, either in the BPHZ version, the \msbar{} version (track-A) or
even the BPHZ$'$ version.

\section{Examples}
\label{s.examples}

So far, the discussion has been general but abstract. Concrete examples illustrate the issues very clearly.

The most direct way to test whether pdfs obey properties like
positivity would be to simply calculate a suitable sample of them
directly from \eref{pdfren} from first principles in QCD. But this requires a
calculation of their nonperturbative behavior at a level which is
  beyond current abilities. 

However, neither the derivation of factorization nor the derivation of
positivity of \msbar{} pdfs in Ref.~\cite{Candido:2020yat} is specific to
QCD.  Instead they apply generally to all theories with the standard
desirable properties like renormalizability, etc. Therefore, it is
convenient to stress-test proposed general features by examining them
in a theory 
where it is straightforward to perform appropriate reliable calculations, i.e., a model QFT in a parameter region where low-order perturbative
  calculations are accurate for pdfs and structure functions.

We will do this in a Yukawa theory, with all particles massive, and
  with weak coupling.  Of course, even though factorization is still valid,
  its utility is much less than in QCD, which is asymptotically free and
  where we always have substantial non-perturbative contributions to pdfs
  and structure functions.

  Perturbative results in this theory provide a counter-example to any
  general theorem that \msbar{} pdfs are always positive.  Examining the
  details of the calculation will also indicate that there is the limited
  range of low scales over which positivity can be violated.  A primary
  impact on QCD is then that it is incorrect to impose a priori positivity
  constraints on \msbar{} pdfs at a low initial scale when making
  phenomenological fits to data. Equally, it is incorrect to impose
  positivity on fits to the results of non-perturbative calculations at low
  scales.  One caution, though, is that the \msbar{} scheme is defined
  within perturbation theory, so that it is not at all clear how to ensure
  it is sufficiently well defined at low scales that are too close to where
  QCD is clearly non-perturbative.

  In addition, a careful examination in the model theory of both how the
  results for pdfs arise and for where factorization is valid will suggest
  some conclusions for QCD itself.

\subsection{Calculation of pdf}

We will use a scalar Yukawa field theory with two separate fermion fields
and the following interaction term: 
\begin{align}
\mathcal{L}_{\rm int}
  = -\lambda\, \overline{\Psi}_N\, \psi_q\, \phi\ +\ {\rm H. C.}\, .
\label{e.lagrangian}
\end{align}
Here, $\Psi_N$ is a field that we will refer to as corresponding to a
  ``nucleon'' or a ``proton'', of spin-$1/2$ and mass $\tarmass$; we will
  use the particle as a target in our calculations.  In addition, there is
a spin-1/2 ``quark'' field $\psi_q$ with mass $\mquark$, and a zero charge
scalar ``diquark'' field $\phi$ with a mass $\mgluon$.\footnote{The
    sole purpose of these names is to indicate how we will use the model
    theory to construct analogs of what in QCD are the standard pdfs on
    hadronic targets.} We will use the notation $a_\lambda(\mu) \equiv \lambda^2/(16 \pi^2) $
in analogy with a similar notation, $a_s = g_s^2/(16 \pi^2)$ from
perturbative QCD.  Keeping all masses nonzero ensures that the theory is
finite range in coordinate space, like full QCD but not massless
  perturbative QCD. We may choose the coupling $\lambda$ small enough
that low order graphs in perturbation theory approximate DIS structure
functions across a wide range of scales to sufficient accuracy for any given $\xbj < 1$, with controllable sizes of error.

The bare fermion pdf is~\eref{pdfdef}, but without the Wilson line, i.e.
\begin{multline}
f^\text{bare,A}_{i/p}(\xi) =
 \int \frac{\diff{w^-}{}}{2 \pi} \, e^{-i \xi p^+ w^-} 
\\
\; \langle p | \, \bar{\psi}_{0,i}(0,w^-,\T{0}{}) {\frac{\gamma^+}{2}} 
\psi_{0,i}(0,0,\T{0}{}) \, | p \rangle \, , \label{e.pdfdef2}
\end{multline}
where the $i$ label indicates either the $\psi_q$ or the $\Psi_N$ field.
The renormalized collinear parton density has the form
\begin{align}
{f^\text{renorm,A}_{i/p}(\xi;\mu)} &{}=  Z^A_{i/i'} \otimes {f^\text{bare,A}_{0,i'/p}} \no
&{}\equiv \sum_{i'} \int \frac{\diff{z}}{z} 
\,  Z^A(z,a_\lambda)_{i/i'} \, f^\text{bare,A}_{0,i'/p}(\xi/z)  \, . \label{e.pdf}
\end{align}
\begin{figure}
\centering 
  \includegraphics[width=0.3\textwidth]{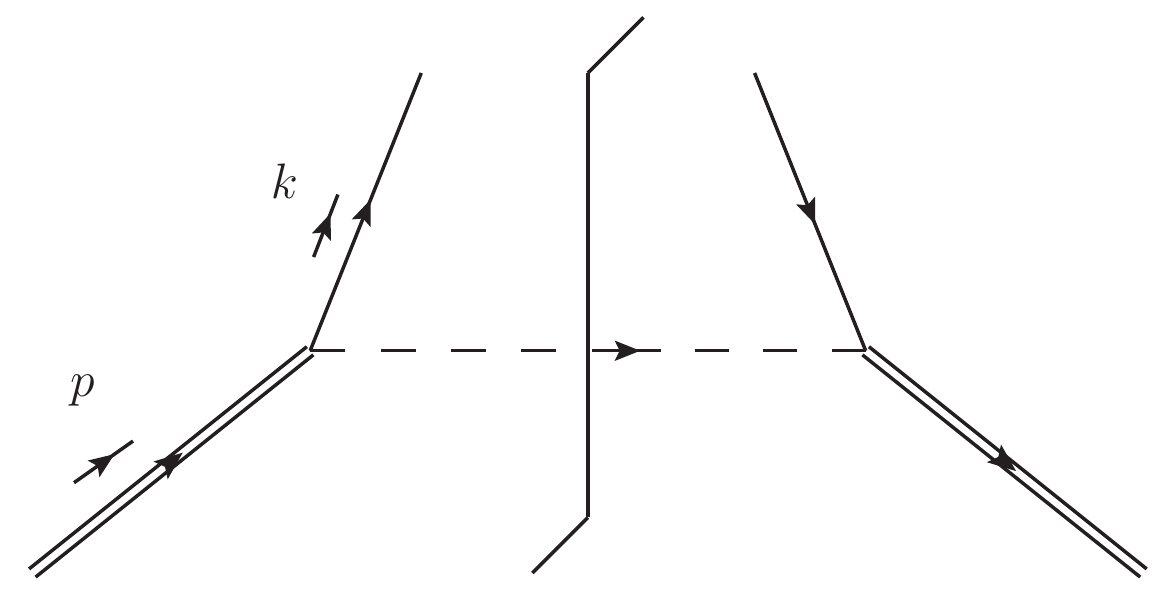} 
\caption{Lowest order contribution to $f^\text{renorm,A}_{q/p}(\xi;\mu)$.}
\label{f.fqinP}
\end{figure}
\begin{figure*}
\centering
\includegraphics[width=\textwidth]{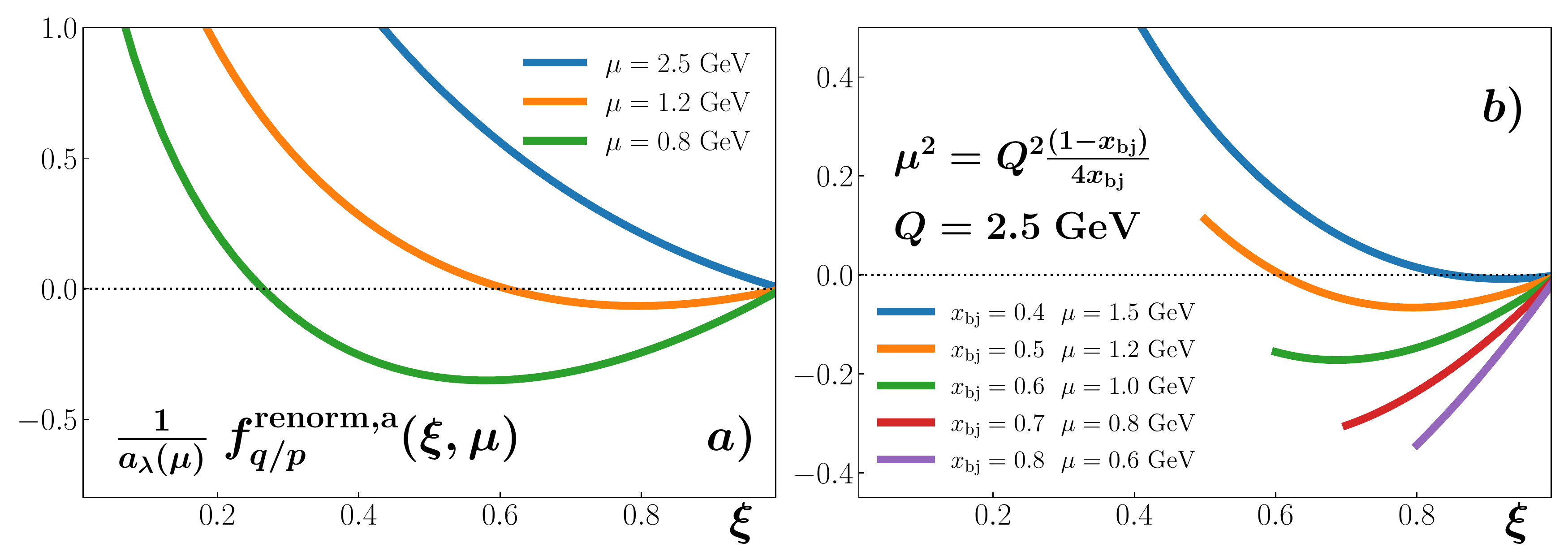}
\caption{(a) An example of the $\msbar$ quark-in-proton pdf in \eref{pdf1}
  for several values of $\mu$, 
  with $\mquark = 0.3$~GeV, $\tarmass = 1.0$~GeV, and $\mgluon = 1.5$~GeV.
  (b) Similar curves but in a form applicable to use in factorization
  for DIS with $\mu^2 = \hat{s}/4$, given $Q=2.5~$GeV and several values
  for  $\xbj$.  The pdfs are only used in the range $\xbj \leq \xi \leq 1$,
  so the curves are restricted to this region.
}
\label{f.pdfplot}
\end{figure*}

We will work with the quark-in-proton pdf, for which the lowest order
  value is at order $a_\lambda$, from the graph in \fref{fqinP}.  A direct
computation gives
\begin{align}
 f^\text{renorm,A}_{q/p}(\xi;\mu) 
\hspace*{-10mm}&\no 
={}& a_\lambda(\mu) (1 - \xi) 
 \parz{\frac{(\mquark + \xi \tarmass)^2}{\Delta(\xi)} 
 + \ln \left[ \frac{\mu^2}{\Delta(\xi)} \right] - 1} \no 
 & + \order{a_\lambda^2} \, , \label{e.pdf1}
\end{align}
where
\begin{equation}
\Delta(\xi) \equiv \xi \mgluon^2 + (1 - \xi) \mquark^2 - \xi (1-\xi) \tarmass^2 \, .
\end{equation}
(Note that $\Delta(\xi)$ is positive if the target state is stable, i.e.,
  $m_p < m_q+m_s$.)
The $\msbar$ counterterm used to obtain Eq.\ \eqref{e.pdf1} is
\begin{equation}
\msbar \;\; \text{C.T.} 
= -a_\lambda(\mu) (1 - \xi) \frac{S_\epsilon}{\epsilon} \, .
\label{e.msct}
\end{equation}
This gives $Z^A_{q/p} = -a_\lambda(\mu) (1 - \xi) S_\epsilon/\epsilon + O(a_\lambda^2)$, thereby matching
the general form of \eref{pdfren}.

By choosing $a_\lambda(\mu_0)$ small enough at some reference scale $\mu_0$, we
  ensure that the one-loop renormalized pdf in \eref{pdf1} is a good
  approximation to the exact pdf to some given accuracy  over a range of
  $\mu$.  Since the effective coupling does not increase out of the
  perturbative range at small scales, unlike QCD, the calculation retains
  its accuracy when $\mu$ is of order particle masses.  It only loses
  accuracy when $\mu$ is so large\footnote{The calculation also loses
      accuracy when $\mu$ is very small.  But that is irrelevant to the uses
      of pdfs, which are in factorization for hard processes where $\mu$ is
      chosen to be proportional to a large scale $Q$.}
  that the logarithms of $\mu/\mbox{mass}$ in 
  higher orders of perturbation theory compensate the smallness of the
  coupling, and use of DGLAP evolution becomes necessary; that is not a
  concern here.

  So that the results of calculations give suggestions as to what happens
  in QCD, we choose mass parameters to be in a range reminiscent of masses
  in QCD: $\mquark = 0.3$~GeV, $\tarmass = 1.0$~GeV and $\mgluon =
  1.5$~GeV.  Thus the quark mass is similar to the ``constituent mass''
  \cite{DeRujula:1975qlm} of
  a light quark in QCD, and similarly the hadron mass is similar to a
  nucleon mass.  But we choose the diquark mass to be somewhat larger than
  might be expected were we to treat the calculation as an actual model for
  a pdf in non-perturbative QCD; this diquark mass allows us to illustrate
  that more than one mass scale could be relevant in a $\xi$-dependent way.

  From \eref{pdf1}, it immediately follows that for any given value of $\xi$,
  the pdf is negative for low enough $\mu$ and positive for large $\mu$.  This
  is illustrated in Fig,~\ref{f.pdfplot}(a) which shows the $\xi$-dependence
  of the quark pdf for three different values of $\mu$.  The values are chosen to
  be representative of the low end of the range of $\mu$ used in QCD fits. At
  fairly low values of $\mu$, there is a range of moderately large $\xi$
  where the pdf is negative.  As $\mu$ increases, the range of negativity
  shrinks and eventually disappears.  Later we will interpret these results
  in terms of scales in the shape of the transverse momentum distribution.

  One might worry that the strong negativity might be incompatible with the
  momentum and flavor sum rules, which entail that some of the pdfs are
  sufficiently positive.  This issue is resolved by observing that the
  nucleon in our Yukawa model is a possible parton, and that a first
  approximation to the corresponding pdf $f_{p/p}$ (diagonal in
  parton/particle labels) is the free-field value $\delta(\xi-1)$, which is
  positive, and does not involve any UV renormalization. Thus the
  quark-in-nucleon pdf that we have calculated is a minority contribution,
  i.e., much smaller than the other pdf at large $\xi$.

\subsection{Systematics of why the pdf becomes negative}
\label{s.why.neg.pdf}

To understand how and where the \msbar{} pdf becomes negative, we
  relate it to an integral over transverse momentum of the corresponding
  transverse momentum dependent (TMD) pdf. Calculating \eref{pdf1}
involves calculating the following integral in dimensional regularization:
\begin{equation}
  a_\lambda(\mu) \int_0^\infty \diff{\Tscsq{k}{}} \Tsc{k}{}^{-2\epsilon}
  \frac{(1-\xi) \big[\kT^2 + (\mquark +\xi \tarmass )^2\big]}
       {\big[\kT^2 + \Delta(\xi) \big]^2} \, .
\label{e.TMDpdf} 
\end{equation}
Suppose that instead of using dimensional regularization and
  subtracting the \msbar{} pole to define a scale-dependent pdf, we simply
  applied a cutoff on transverse momentum in the unregulated
  integral:\footnote{Such a definition is used by Brodsky and
      collaborators \cite{Lepage:1980fj,Brodsky:2000ii}. }
\begin{equation}
  a_\lambda(\mu) \int_0^{\Tscsq{k}{\rm cut,}} \diff{\Tscsq{k}{}}
  \frac{(1-\xi) \big[\kT^2 + (\mquark +\xi \tarmass )^2\big]}
       {\big[\kT^2 + \Delta(\xi) \big]^2} \, .
\label{e.TMDpdf2} 
\end{equation}
The new pdf is an ordinary integral with a positive integrand
  everywhere when $0 < \xi < 1$, with no further subtraction term.  So, this
  pdf is guaranteed to be positive.  It can be verified that when
  $\Tscsq{k}{\rm cut,}$ is set to equal $\mu^2$, the cut-off integral matches
  the $\msbar$ result in \eref{pdf1} except for corrections by a power of
  $m^2/\mu^2$, which are small at high scales.  Its evolution is of the DGLAP
  form, but with a power-suppressed inhomogeneous term.  In a gauge theory,
  there are problems with a naive use of a cutoff in transverse momentum,
  since the most natural definition of a TMD pdf suffers from rapidity
  divergences~\cite{Collins:2003fm} associated with light-like Wilson
  lines.  These necessitate changes in any method of working with TMD
  pdfs\footnote{Note that the rapidity divergences cancel in \msbar{} pdfs
    \cite{Collins:1981uw}.} in QCD. But the same problem does not occur in
  a non-gauge theory.

  Equation~\eqref{e.TMDpdf2} is a very intuitive way to represent a
  scale-dependent pdf in terms of hadron structure: It contains the
  contributions from transverse momenta up to the cutoff.  In particular,
  when $\Tscsq{k}{\rm cut,}$ is large, it includes, among other things, all
  the physics associated with intrinsic hadron structure. Then the
  coefficient function in factorization for DIS at high $Q$ can simply be
  characterized as whatever else contributes to the correct structure
  function.  When $\Tscsq{k}{\rm cut,}$ is of order $Q^2$, the coefficient
  function is only concerned with physics at the scale $Q$.

  Up to terms of order $m^2/\Tscsq{k}{\rm cut,}$, the pdf in
  Eq.~\eqref{e.TMDpdf2} is related to the $\msbar$ pdf at scale $\mu$ by
  subtracting the term
\begin{equation}
a_\lambda(\mu)  (1-\xi) \int_{\mu^2}^{\Tscsq{k}{\rm cut,}}  \frac{\diff{\Tscsq{k}{}}}{{\Tscsq{k}{}}} \, ,
\end{equation}
as can be verified by explicit calculation.  The integrand is the large
$\Tsc{k}{}$ asymptote of the integrand in Eq.~\eqref{e.TMDpdf2}, as is
appropriate to the minimal subtraction used in the \msbar{} scheme.

By taking $\Tscsq{k}{\rm cut,}$ to infinity, we find that the \msbar{} pdf
equals 
\begin{multline}
  a_\lambda(\mu)(1-\xi) \int_0^\infty \diff{\Tscsq{k}{}}
\\
  \left\{ 
  \frac{ \kT^2 + (\mquark +\xi \tarmass )^2 }
       {\big[\kT^2 + \Delta(\xi) \big]^2}
   -\frac{\theta(\Tsc{k}{}-\mu)}{\kT^2}
   \right\}
\,,
\label{e.TMDpdf.msbar} 
\end{multline}
i.e., there is a subtraction of the asymptote of the integrand, with a
lower cutoff. 

At high $\Tsc{k}{}$, the subtraction term closely matches the first term.
When $\mu$ is large, there is a logarithmic contribution from $\Tsc{k}{}$ in
the range $\Delta \lesssim \Tsc{k}{} < \mu$, since the subtraction term vanishes there.
The positive logarithmic contribution overwhelms any negativity in the
remaining non-logarithmic contributions to the integral.

Now when $\Tsc{k}{}$ goes to zero the $1/\Tscsq{k}{}$ term goes to
infinity, unlike the first term in \eqref{e.TMDpdf.msbar}, giving a strong
mismatch.  Hence when $\mu$ is small, there is a large negative logarithmic
contribution to the subtracted integral.  Therefore, where the \msbar{} pdf
becomes negative is governed by the mass scale(s) below which the mismatch
is strong.

When $\xi\to0$, the first term in the integrand becomes
$1/(\Tscsq{k}{}+\mquark^2)$.  Then the pdf becomes negative when
$\mu<\mquark$, which is a low scale, well under a GeV.

When  $\xi\to1$, the term is 
\begin{equation}
  \frac{ \kT^2 + (\mquark + \tarmass )^2 }
       { ( \kT^2 + \mgluon^2 )^2 }\,.
\end{equation}
Where the integral becomes negative is now governed by a combination of the
scales $\mquark+\tarmass$ and $\mgluon$, which are around a GeV or larger. 

The existence of these two scales differing by a modest factor
qualitatively explains Fig.~\ref{f.pdfplot}.  For $\mu$ in the middle of the
range between the quark mass and about a GeV, the pdf is positive at low
$\xi$ but not at high $\xi$.  Only when $\mu$ is well above a GeV does the pdf
become positive everywhere.  The effect is enhanced by our choice of a
somewhat large value of $\mgluon$, but it illustrates the effects of
different scales of transverse momentum in different ranges of $\xi$ in a way
that can easily happen in QCD.  Indeed recent fits do provide results with negative
values in some ranges of $\xi$ and $\mu$; we will discuss this below.

\subsection{Relation to factorization}
\label{s.pdf.v.fact}

As to the impact of the negative values of pdfs on possibly predicting a
negative cross section from a factorized form, we recall first that the
factorized cross section is independent of $\mu$.  But in finite order
approximations, $\mu$-independence is valid only up to errors of the order of
uncalculated higher order corrections. One should choose $\mu$ to avoid large
logarithms in the perturbative expansion of the coefficient functions.

Furthermore, the factorization theorem itself is only valid up to errors of
size $\mquark^2/Q^2$, $\mgluon^2/Q^2$ and $\tarmass^2/Q^2$, with 
$\xbj$-dependent coefficients.  These
arise from mismatches between the exact integrands for graphs for a
structure function and the approximations used in obtaining the factorized
form.  Once the errors are too large compared with the factorized value of
a cross section, negative values from the factorized cross section are
irrelevant.  In addition, when we work in QCD, the coupling becomes too
large at low scales to allow low-order perturbative calculations of the
coefficient functions to be useful.

In our model the second issue does not arise, and we examine the sizes of
the power-law errors and their impact on the effect of negative pdfs,
especially in relation to the mass scales involved. Because of our use of a
weak coupling, it is sufficient to work at order $a_\lambda$.  Some details of
our calculations are given in \aref{full}.

\begin{figure}
\centering 
  \includegraphics[width=0.48\textwidth]{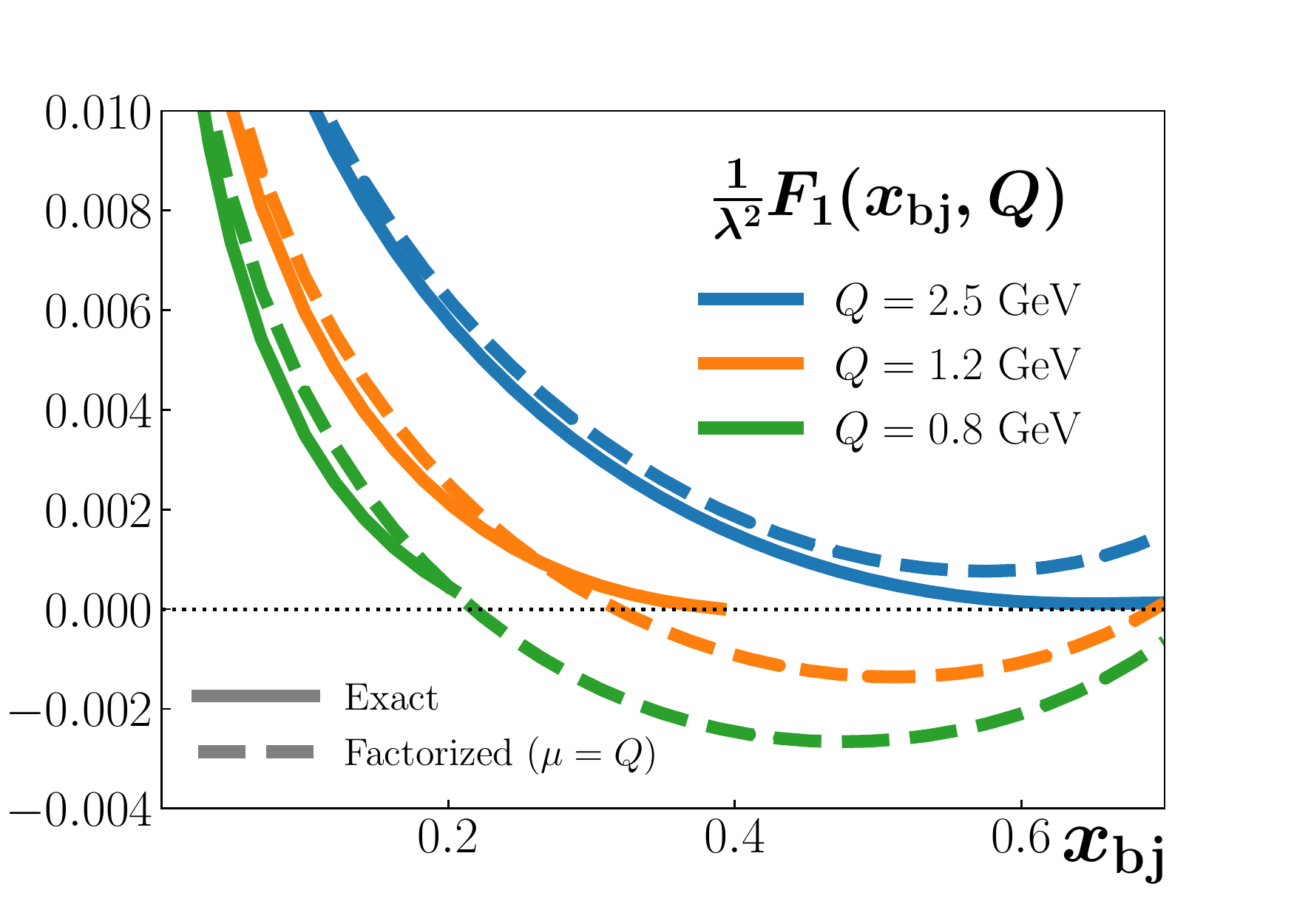} 
  \caption{A comparison between \eref{finalfactorized} and the unfactorized
    graphs in \fref{basicmodel} for the $F_1$ structure function (dropping
    $\order{a_\lambda^2}$ terms), for three values of $Q$. The graphs are
    independent of the choice of $\mu$. The zero in the blue curve just above
    $\xbj = 0.5$ is the kinematical upper bound on Bjorken-$\xbj$.}
\label{f.F1plot}
\end{figure}

In Fig.\ \ref{f.F1plot}, we compare an exact calculation of $F_1$ at order
$a_\lambda$ with the factorized approximation, for three quite low values of $Q$.
These are within the range of a number of experiments, e.g.,
Ref.~\cite{COMPASS:2013bfs}.  The lowest value is $Q=0.8~$GeV, which is
below where one normally uses factorization in QCD.  But even there, there
is a range of $\xbj$, viz.\ $\lesssim0.2$, where the factorized approximation is
reasonably good.  Recall that as $\xbj$ gets small, the invariant mass of
the final state gets large, so that the collision is quite inelastic and
there is another larger scale in the problem than $Q$.

In contrast, at the larger $\xbj$ values and fairly low $Q$, the factorized approximation is
not merely a poor approximation, but in some places gives an unphysical
negative value to $F_1$, while the true value is zero because the final
state mass is below threshold.

As $Q$ increases, the range of $\xbj$ where the factorized approximation is
good increases, while the range of negative values for the approximation
decreases. But the range of negative values does not closely match the
range of negative values for the pdf. This is to be expected, since
  the calculation contains contributions from the one-loop coefficient
  function as well as directly from the negative pdf. At the
highest value $Q=2.5~$GeV, the factorized approximation is positive
everywhere.  But at large $\xbj$ it increases. Much of the increase is
where the true value $F_1$ is zero because the final state is below the
quark-diquark threshold, so that the factorized approximation is very
incorrect. 

From our calculation of the pdf in Sec.~\ref{s.why.neg.pdf}, we saw that at
small values of parton momentum fraction $\xi$, the width of the internal
transverse momentum integration in the pdf is
governed by the quark mass, which is $0.3~$GeV, a typical value for a
constituent light quark mass, whereas at large $\xi$ the width is governed by
the larger diquark mass.  The approximation that leads to factorization
involves neglecting quark transverse momentum in the hard scattering; in
addition kinematic approximations involve neglecting $\xbj m_p$ with
respect to $Q$.

This suggests that at small enough $\xbj$, the power error in factorization
involves a relatively low mass, while at larger $\xbj$ it involves a
relatively high mass.  

Now a standard choice of scale is $\mu=Q$.  But, as observed in
\cite{Candido:2020yat} for example, this is quite inappropriate at large
$\xbj$. The partonic transverse phase space is limited by $\hat{s}/4 = Q^2
(1 - \xbj)/4 \xbj$, which produces large logarithms of $\hat{s}/Q^2$ at
larger values of $\xbj$ in the coefficient function.  A consequence
\cite{Candido:2020yat} is that choosing $\mu=Q$ gives considerable
oversubtraction in the coefficient function.  A more sensible value would be $\mu^2
= \hat{s}/4$, at least at large $\xbj$. For a given value of $Q$, this gets
very small as $\xbj\to1$.  Although this choice of $\mu$ should improve the
perturbative treatment of hard parts, it substantially exacerbates the
possibility that the value of a pdf in a calculation will turn negative, as
illustrated in Fig.~\ref{f.pdfplot}(b).  This is the same as
\fref{pdfplot}(a) but with $\mu^2 = \hat{s}/4$ and a sequence of large values
for $\xbj$, with $Q^2=2$\,GeV. But with such low values of the final
state's invariant mass, we must also expect that factorization also gives a
poor approximation to the actual cross section.  

Regardless of the details, we see that, given the mass scales $\tarmass$,
$\mquark$, and/or $\mgluon$, low values of $\mu$ result in a pdf that turns
negative.  The example is enough, therefore, to show that a pdf defined in
the $\msbar$ scheme is not constrained by general principles to be positive
everywhere.  As mentioned in the Introduction, the result of
  calculations of heavy-quark pdfs provides one example of this
  phenomenon in QCD.

We have also seen that a negative pdf often occurs in a region of momentum fraction $\xi$ where power corrections cause factorization to be a poor
approximation if $\xi = \xbj$. Recall, however, that pdfs in the range $\xbj
\leq \xi \leq 1$ enter calculations at higher orders. So $\xi \sim 1$ pdfs are relevant
even for small $\xbj$ calculations.

\subsection{``Bare'' pdf of track B}

We now find the corresponding bare pdf in track B. From the treatment
  in \sref{reconstruct}, we know that $f^\text{bare,B}$
is just
\eref{pdf1}, but before the $\msbar$ pole in \eref{msct} has been
subtracted:
\begin{align}
& \hspace*{-5mm}
f^\text{bare,B}_{q/p}(\xi;\mu) = f^{\text{renorm,A}}_{\text{BPHZ}'}\no 
={}& a_\lambda(\mu) (1 - \xi) 
 \parz{\frac{(\mquark + \xi \tarmass)^2}{\Delta(\xi)} 
 + \ln \left[ \frac{\mu^2}{\Delta(\xi)} \right] - 1} \no 
 & + \left. a_\lambda(\mu) (1 - \xi) \frac{S_\epsilon}{\epsilon} \right|_{\text{IR}} +
 \order{a_\lambda^2} \,,
\label{e.pdf_bare_b}
\end{align}
with terms of $O(\epsilon)$ ignored, and with the pole identified as
collinear rather than UV. In the methods of track B, dimensional
  regularization is used to regulate collinear (and IR) divergences, so
  that $\epsilon<0$. Then the bare pdf in \eref{pdf_bare_b} is negative for small
  negative $\epsilon$. 

\section{Valence v.\ non-valence pdfs}

The argument in the previous section about \msbar{} pdfs becoming
negative at small $\mu$ might 
appear to be quite general.  It may seem that \emph{all} pdfs should become entirely negative at small enough $\mu$ for all $\xi$. But this would be in contradiction with the
sum rules obeyed by the pdfs, notably the momentum sum rule.  These entail
that the negativity at low $\mu$ cannot be a general property.

That worry is removed by observing that not all graphs for pdfs have UV
divergences. In the model, the simplest case is for the nucleon-in-nucleon
pdf, for which the lowest order value, including transverse momentum
dependence is a simple delta function: $\delta(\xi-1) \delta^{(2)}(\T{k}{})$; this
obviously has no UV divergence when integrated over $\T{k}{}$. 

We characterize the situation by saying that in the model, the nucleon is
itself the analog of a valence quark in QCD, while the quark in the
  model is an analog of a sea quark in QCD, given that the target particle
is a nucleon.  (For our rough purpose here, we characterize a valence
  quark, or, more generally, a valence parton as one that at large and
  moderately large momentum fraction corresponds at low scales to the
  largest pdfs and corresponds to the basic structure of the target
  state.)

The generalization to other theories, including QCD, can be explained by
using the expansion (\ref{eq:pdf.2PI.exp}) of a pdf in terms of 2PI graphs.
The part without a UV divergence is the first term in the expansion, i.e.,
the term without any rungs in the ladder.  We notated it as $TB$.  All the
remaining terms in (\ref{eq:pdf.2PI.exp}) have UV divergences.  Those
statements follow from the standard power counting for UV divergences in
pdfs.  (Our Yukawa model provides a degenerate case where the lowest order
2PI graph is just a lowest order disconnected graph.)

The TB term is exactly the bare pdf $f^{\text{bare, CFP}}$ that was
defined by Curci et al.\ \cite{Curci:1980uw}.  This term is necessarily
positive, by a version of the argument given in Sec.\ \ref{s.pos.pdf}, but
without any complications arising due to the need for a
regulator.\footnote{Here, ``regulator'' means ``UV regulator''. As
    mentioned earlier, there are rapidity divergences due to the use of
    light-cone gauge. In principle, a regulator needs to be applied to
    rapidity divergences.  but those divergences cancel and they concern
    orthogonal issues to those we are concerned with here.}

It is natural to suppose that $TB$ is dominated by the valence quark terms,
which essentially correspond to simple quark models of hadrons.  Any other
terms, for non-valence partons, are presumably substantially smaller.

The argument leading to possible negative \msbar{} pdfs applies to the
graphs with rungs.  We then get a two-component picture: A positive valence
contribution from the TB term for the relevant pdfs, and then potentially
negative contributions from the graphs with rungs.  For non-valence
partons, the first component is small, and so the potentially negative
component from the renormalization of the UV divergent terms can dominate.
Of course in QCD the coupling is not particularly small in the low $\mu$
region.  So there is the potential for higher-order terms to modify the
results.

As we observed, the negative contribution to a pdf goes away once $\mu$ is
enough larger than the scale setting the width of the $\T{k}{}$
distribution at small $\T{k}{}$. But this is a soft constraint.  As
  regards the potential for negative cross sections, the primary point is
  that negative values for pdfs arise in pdfs that are small compared with
  others, so higher-order perturbative terms in the hard scattering induced
  by valence quarks are not suppressed with respect to the lowest-order
  term induced by the minority partons.

Motivated by the calculations,  a simple idea to see that some non-valence
  pdfs are likely to become negative at low enough scales is as follows.  We examine
  consequences of DGLAP evolution:
  $df_{\rm min.}/d\ln\mu \simeq \text{kernel} \otimes f_{\rm maj}$.
  We first observe
  that at the \emph{lowest} order those DGLAP kernels $P_{ij}(z,\alpha_s)$ that are off-diagonal in 
  parton flavor  are positive.  The diagonal kernels are positive when $z<1$, but have a 
  negative delta-function at $z=1$.\footnote{As is well-known, in a gauge theory there is a plus
  distribution at $z=1$, so that the coefficient of the delta function is effectively infinitely negative, with the
  divergence cancelled by a positive divergence in the integral over $z<1$.  But that doesn't
  really affect our argument.}  Suppose that at large $x$ we have a valence pdf, notably the $u$
  quark in QCD,  that is substantially
  larger than the others, and that is positive.  Then the positivity of the off diagonal kernel 
  implies that small pdfs increase with scale, but only for those partons with a non-zero LO
  DGLAP kernel with the valence parton.  This is simply because a sufficient smallness of a pdf
  implies that the diagonal term in its evolution is smaller than the off-diagonal term. 
  
  Hence going to a lower scale takes such small pdfs to negative values.  Direct examples
  of this are given not only by our model calculations, but also by the evolution of a heavy-quark
  pdf in QCD, when the scale $\mu$ is close to the heavy quark's mass.
   
  The increase with $\mu$ for a minority pdf contrasts with the
  well-known decrease of valence pdfs at the larger values of $x$.  That
  decrease arises from the dominance of the delta-function terms in the DGLAP kernels
  that are diagonal in flavor.  At sufficiently smaller $x$, the positive continuum part becomes
  more important and then the valence pdfs increase.

\begin{figure}
  \centering
  \includegraphics[width=0.5\textwidth]{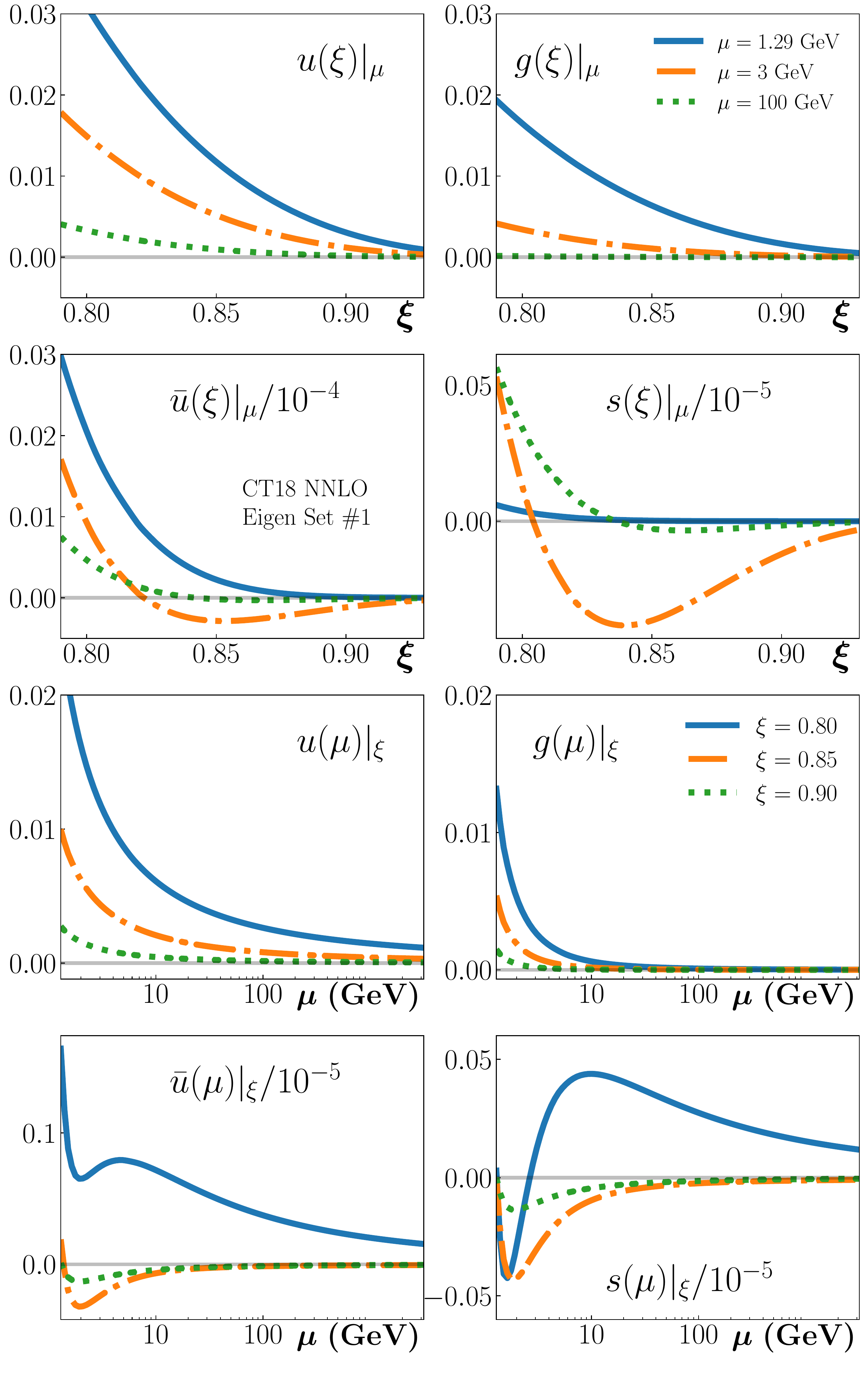}
  \caption{Pdfs from the CT18 fit \cite{Hou:2019efy} restricted to eigen-direction \texttt{\#1}
    at several values of scale as a function of $\xi$ (upper two rows) and at several large
    values of $\xi$ as a function of scale (lower two rows). To keep
    the number of plots lower, we have restricted the sea quark plots to
    $s$ and $\bar{u}$. Behavior of the $\bar{d}$ pdf is similar to that
      of the $\bar{u}$. }
  \label{fig:CT18.plots}
\end{figure}


However, the expectations just summarized are somewhat in contrast
  with the results of actual fits, as illustrated  in Fig.\ \ref{fig:CT18.plots}. 
   In these plots we have focussed on $x$ above 0.8 using a particular 
   eigen-direction pdf set from CT18 that displays negativity as an example.
  There is a clear hierarchy of sizes of pdf, with the pdf 
  of the $u$ quark being the largest. We also notice that there are large
  discrepancies between the different fits for the pdfs of the sea quarks
  ($s$, $\bar{u}$)
  and of the gluon.  This simply reflects the large uncertainties on these small pdfs in
  a region where they are weakly probed by the currently available data. 
  
  Moreover, the evolution for the minority pdfs, at scales of a GeV or two 
  makes them decrease instead of increase with scale, initially.  Some of the
  minority pdfs even stay negative up to $\mu \sim 100\,\textrm{GeV}$, which
  is not at all in agreement with the elementary prediction.

  It is beyond the scope of this article to perform a detailed analysis. We make the
  following comments:
  \begin{enumerate}

  \item None of the other quarks has a lowest-order DGLAP kernel that connects 
      it to the quark with the largest pdf, the $u$ quark.  So the effect of the $u$ pdf 
      on the evolution of the other quarks is indirect, both through an NLO kernel, and
      via the gluon pdf.  Given the use of subtractions in defining the higher-order
      terms, the positivity properties are not at all clear compared with simple situations.

  \item The coefficient functions in factorization, Eq.\ \ref{e.structfunct}, have singularities
      at $x/\xi=1$, as do the DGLAP kernels.  These generate large logarithms whenever pdfs
      are rapidly decreasing, as they are at large $\xi$. So fixed-order calculations of the kernels
      and of coefficient functions can be insufficient in this region.
      Then the errors due to omitted higher-order  
      terms may not be small relative to lower-order terms, and
      large-$x$ resummation is needed to get  accurate results. This was 
      already observed by Candido et al.\ \cite{Candido:2020yat}; their argument for positivity 
      of pdfs fails unless large-$x$ resummation is used.  (That failure is beyond the other issues
      we discussed earlier.)

  \item The negative pdfs are a few orders of magnitude below the
     largest pdf, of the $u$ quark. Thus the negativity has no direct
     consequences for negativity of cross sections. 
     
  \item They are also in a region where they are badly determined, and the
      uncertainty range \cite{Hou:2019efy} includes positive values.  So one can always say that
      the negative values are not significant.  The quoted values of pdfs in the
      unmeasured region are effectively extrapolations from regions where they are
      well measured.
      
  \item Therefore there are liable to be implicit or explicit assumptions
      about the functional form of pdfs at their starting scale. For example, when an
      explicit parameterized form is used a restricted parameterization may be used
      which may not in reality be accurate enough at large $x$.  
      
  \end{enumerate}

Another situation where negativity of pdfs is sometimes
    encountered is for gluons at small $x$ and fairly low $\mu$.  There is
    a strong increase with scale of the gluon pdf, and depending on the
    size of the singlet and the gluon pdfs, reverse evolution to lower
    scales can give a negative gluon. However at small-$x$
    there is loss of accuracy in fixed order calculations that requires
    small-$x$ resummation.  For instance in \cite{Ball:2017otu} a global
    analysis with small-$x$ evolution was reported with the inferred gluon
    replicas being all positive in the small-$x$ in contrast to their
    baseline analysis based on fixed order calculations.  As with the case
    of pdfs at large $x$, the possibly negative pdfs are in a region where
    they are weakly constrained by data.


\section{Conclusions}
\label{s.discussion}

One of our main goals with this article has been to draw attention to gaps 
in one particularly common approach to QCD factorization. However, 
the issues are abstract and it is tempting to view them as only formal, with no impact on practical phenomenology. We have focused on the positivity question, therefore, to illustrate how those gaps can influence practical considerations. The positivity example shows that the track-A/track-B dichotomy is especially relevant to questions about the properties of the pdfs themselves. Other interesting examples likely exist. 

Past approaches to pdf phenomenology have mainly focused on the universal
nature of pdfs to constrain them from experimental scattering data.  But
increasingly sophisticated methods are being used to study pdfs and their
properties directly using nonperturbative techniques like lattice QCD, and
to combine those approaches with more traditional phenomenological
approaches.  It is important to check the derivation of properties that
originally relied on a track-B view before they are adopted unchecked into
larger phenomenological programs. More generally, the pitfalls of the
track-B approach need to be taken into consideration in nonperturbative QCD
approaches. Many of these calculations are performed at rather low scales
where, as illustrated by the positivity example, the problems with track B
are most prominent.

As regards the positivity issue itself, there are several points to make. First, we emphasize that we have not argued that $\msbar$ pdfs \emph{must} be negative for any particular choice of scales or $\mu_{\msbar}$.  Rather we proved that nothing in the definition of pdfs or in the factorization theorems themselves excludes negativity as a possibility, especially at low or moderate input scales. But we did show arguments that indicate that certain generic situations do tend to lead to negative
pdfs of partons with small pdfs, notably for non-valence quarks.
Giving a full theoretical answer to
the question of whether a particular pdf turns negative depends on its large distance/low energy nonperturbative properties, as the sensitivity to mass scales in the example of \sref{examples} illustrates. Also, the failure of absolute positivity in the $\msbar$ scheme does not necessarily imply that other schemes do not exist which exactly preserve positivity. 

Instead, we argue that imposing strict positivity on $\msbar$ pdfs as an absolute phenomenological constraint is an excessive theoretical bias. 
As our examples show, this is especially relevant to the question of how
low $Q$ may be before factorization theorems become
unreliable. Applications of factorization to low or moderate $Q$ are often
important for studies of hadron structure.  Note, for example, that past
Jefferson Lab DIS measurements are in the range of $Q \sim
\tarmass$~\cite{Gaskell:2010zz}, with $Q > 1$~GeV identified as the
``canonical'' DIS range. It is possible that for factorization theorems to
continue to hold at these scales and with the desired precision, exact
positivity constraints need to be relaxed. Indeed, if positivity
constraints are relaxed, it is possible that DIS
factorization theorems can be extended to significantly lower $Q$ than
might otherwise be expected.  

As we have seen, once the $\msbar$ scale $\mu$ is large enough compared with
intrinsic transverse momentum scales, the pdfs indeed become positive.  

Further refinements in knowledge about transverse momentum mass scales will
likely help to sharpen estimates of where in kinematics a positivity
assumption begins to be warranted.  We leave such considerations to future
work.

Given that positivity of pdfs is not absolutely guaranteed, there is the
possibility that cross sections predicted by factorization can be negative.  Indeed, we illustrated by
calculation that a fixed-order factorized cross section can be unambiguously negative,
without contradicting positivity of physical cross sections:  A factorized
cross section is only an approximation to the true cross section.  If the
difference between a measured and calculated cross section is within the expected error in factorization, then there is no
contradiction. Two notable sources of error in particular kinematic
  regions are higher-order perturbative terms with logarithmic
  enhancements, and power corrections to factorization.

We conclude that positivity should not be imposed as an absolute constraint
in fits for pdfs, neither on the pdfs nor on the calculated cross
sections. Instead, any finding of a negative pdf or especially a negative
cross section should be simply treated as a focus of attention.  Does a
negative pdf occur in a region where it can be expected?  For a negative
cross section, what are the expected errors and uncertainties?  Which data
are critical to obtaining the negative results? 

\vskip 0.3in
\acknowledgments
We thank Markus Diehl, Pavel Nadolsky, Duff Neill, Jianwei Qiu, Juan Rojo and George Sterman for useful discussions. 
T.R. was supported by the U.S. Department of Energy, Office of Science, Office of Nuclear Physics, under Award Number DE-SC0018106. 
The work of N.S. was supported by the DOE, Office of Science, Office of Nuclear Physics in the Early Career Program.
This work 
was also supported by the DOE Contract No. DE- AC05-06OR23177, under which 
Jefferson Science Associates, LLC operates Jefferson Lab.


\appendix

\section{The full cross section}
\label{a.full}

Since the $\msbar$ pdf only turns negative for rather low $\mu$, it is worthwhile to consider whether the scales must be so low that factorization theorems are badly violated. 
To test this, we can work out the exact, unfactorized lowest order cross section for deep inelastic scattering in the Yukawa theory of \sref{examples} and compare the results with the factorized expression that uses \eref{pdf1}. 
The graphs contributing to the DIS cross section at lowest order and for $\xbj < 1$ are shown in 
\fref{basicmodel}.  (The Hermitian 
conjugate of \fref{basicmodel}(c) is also needed, but for brevity we do not show it.) 

The steps for deriving the factorization theorem~\cite{Collins:2011qcdbook} can be applied directly to these graphs. Namely, a sequence of approximations that neglect small,  intrinsic mass scales relative to $Q$ lead (order-by-order in $a_\lambda$) to the separation into factors in \eref{structfunct}. Errors are suppressed by powers of $\text{mass}^2/Q^2$. In the case of the $F_1$ structure function, the $\mathcal{C}$ becomes a partonic $\hat{F}_1$, and the quark-in-hadron pdf for the Yukawa theory is just the result in \eref{pdf1}, while the lowest order hadron-in-hadron pdf is a $\delta$ function.
\begin{figure*}
\centering
  \begin{tabular}{c@{\hspace*{.01mm}}c@{\hspace*{.01mm}}c}
    \includegraphics[scale=0.42]{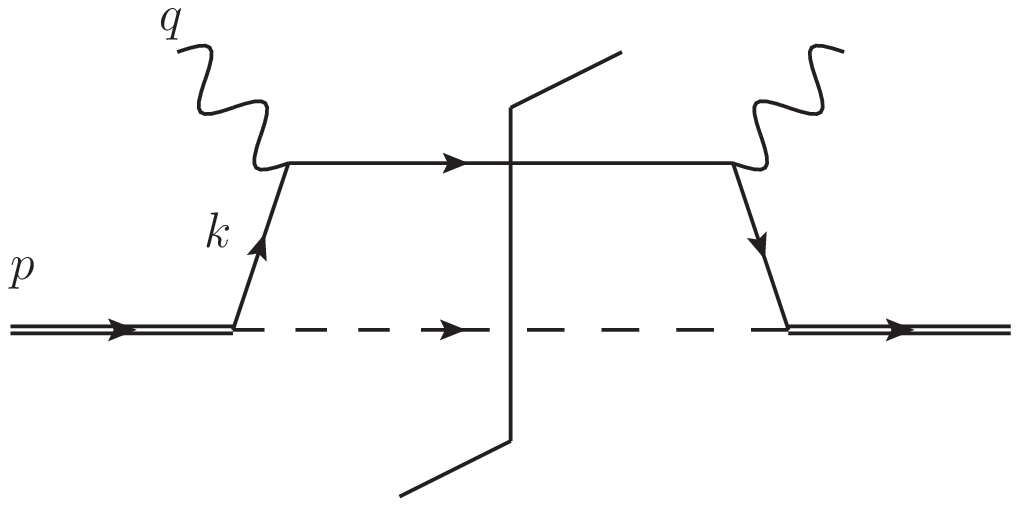}
    \hspace{0.2cm}
    &
    \hspace{0.2cm}
    \includegraphics[scale=0.42]{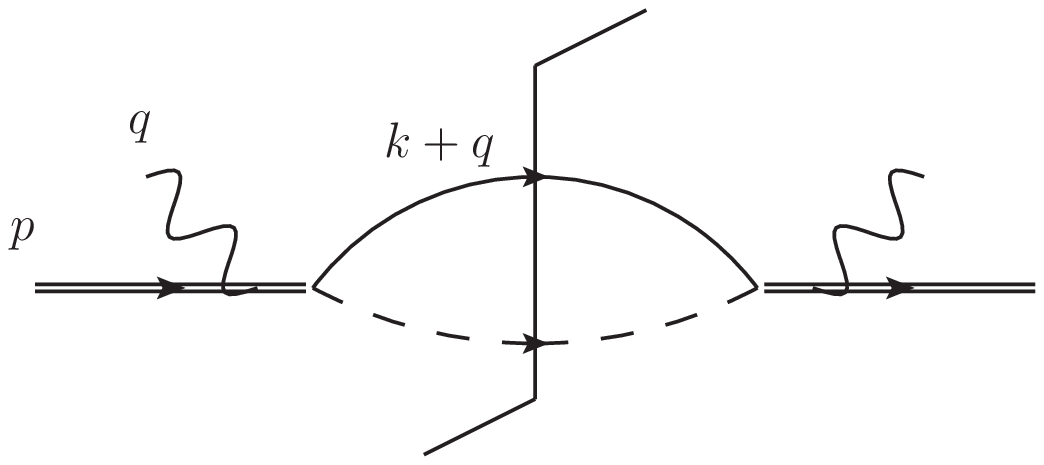}
    &
    \hspace{0.2cm}
    \includegraphics[scale=0.42]{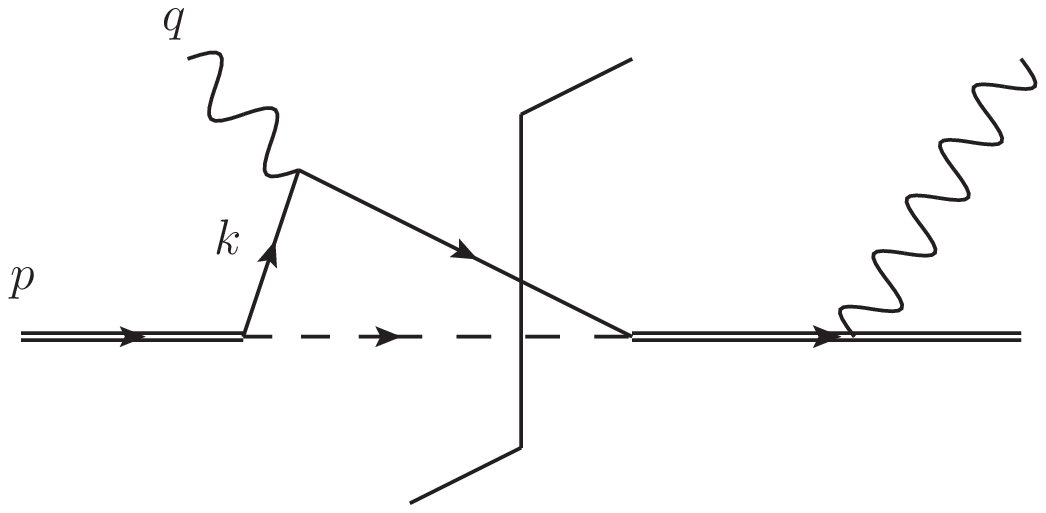}
  \\
  (a) & (b) & (c)
  \end{tabular}
\caption{Contributions to DIS structure functions from \eref{lagrangian} at $\order{a_\lambda}$.
  Graph~(a) is the ``handbag'' diagram that contributes at leading power and small 
  transverse momentum. Graphs~(b) and (c) contribute at leading power to large 
  $\Tsc{k}{}$. (the Hermitian conjugate for (c) is not shown).
}
\label{f.basicmodel}
\end{figure*}
The expression for the factorized $F_1$ structure function in DIS is 
\begin{widetext}
\begin{align}
&{}F_{1}(\xbj,Q)  \stackrel{\xbj < 1}{=} \sum_f \int_{\xbj}^1 \frac{\diff{\xi}{}}{\xi} \times \no
 {}&\times \frac{1}{2}  \left\{\delta\parz{1 - \frac{\xbj}{\xi}} \delta_{qf} 
+ a_\lambda(\mu) \parz{1 - \frac{\xbj}{\xi}} \left[ \ln \parz{4} 
- \frac{\parz{\frac{\xbj}{\xi}}^2 - 3 \frac{\xbj}{\xi} + \frac{3}{2}}{\parz{1 - \frac{\xbj}{\xi}}^2} 
-  \ln \frac{4 \xbj \mu^2}{Q^2 (\xi - \xbj)} \right] \delta_{pf} \right\} \times \no
\no
&{} \times \underbrace{\left\{ \delta\parz{1 - \xi} \delta_{fp}
+ a_\lambda(\mu) (1 - \xi) \left[  \frac{(\mquark + \xi \tarmass)^2}{\Delta(\xi)} 
+ \ln \parz{\frac{\mu^2}{\Delta(\xi)}} - 1 \right] \delta_{fq} \right\}}_{f_{f/p}(\xi;\mu)} \no
&{} \qquad + \order{a_\lambda(\mu)^2} + \order{m^2/Q^2} \, .
 \label{e.finalfactorized}
\end{align}
\end{widetext}
Both the hadron-in-hadron pdf and the order-$a_\lambda$ quark-in-hadron pdf appear 
in the braces on the third line. Note that the $\mu$-dependence cancels between the second and third lines up to order $a_\lambda^2$, as expected. The expression for the unfactorized 
structure function is straightforward but rather complex and so we do not display it here. 

Using the same numerical values for parameters as were used in \fref{pdfplot}, we compare 
\eref{finalfactorized} (dropping both the $\order{a_\lambda^2}$ terms and the
$\order{m^2/Q^2}$ terms) with the exact $\order{a_\lambda}$ $F_1$ in
\fref{F1plot} for several values of $Q$.

\bibliography{bibliography}

\providecommand{\noopsort}[1]{}
\begin{thebibliography}{45}%
\makeatletter
\providecommand \@ifxundefined [1]{%
 \@ifx{#1\undefined}
}%
\providecommand \@ifnum [1]{%
 \ifnum #1\expandafter \@firstoftwo
 \else \expandafter \@secondoftwo
 \fi
}%
\providecommand \@ifx [1]{%
 \ifx #1\expandafter \@firstoftwo
 \else \expandafter \@secondoftwo
 \fi
}%
\providecommand \natexlab [1]{#1}%
\providecommand \enquote  [1]{``#1''}%
\providecommand \bibnamefont  [1]{#1}%
\providecommand \bibfnamefont [1]{#1}%
\providecommand \citenamefont [1]{#1}%
\providecommand \href@noop [0]{\@secondoftwo}%
\providecommand \href [0]{\begingroup \@sanitize@url \@href}%
\providecommand \@href[1]{\@@startlink{#1}\@@href}%
\providecommand \@@href[1]{\endgroup#1\@@endlink}%
\providecommand \@sanitize@url [0]{\catcode `\\12\catcode `\$12\catcode
  `\&12\catcode `\#12\catcode `\^12\catcode `\_12\catcode `\%12\relax}%
\providecommand \@@startlink[1]{}%
\providecommand \@@endlink[0]{}%
\providecommand \url  [0]{\begingroup\@sanitize@url \@url }%
\providecommand \@url [1]{\endgroup\@href {#1}{\urlprefix }}%
\providecommand \urlprefix  [0]{URL }%
\providecommand \Eprint [0]{\href }%
\providecommand \doibase [0]{https://doi.org/}%
\providecommand \selectlanguage [0]{\@gobble}%
\providecommand \bibinfo  [0]{\@secondoftwo}%
\providecommand \bibfield  [0]{\@secondoftwo}%
\providecommand \translation [1]{[#1]}%
\providecommand \BibitemOpen [0]{}%
\providecommand \bibitemStop [0]{}%
\providecommand \bibitemNoStop [0]{.\EOS\space}%
\providecommand \EOS [0]{\spacefactor3000\relax}%
\providecommand \BibitemShut  [1]{\csname bibitem#1\endcsname}%
\let\auto@bib@innerbib\@empty
\bibitem [{\citenamefont {Diehl}\ and\ \citenamefont
  {Stienemeier}(2020)}]{Diehl:2019fsz}%
  \BibitemOpen
  \bibfield  {author} {\bibinfo {author} {\bibfnamefont {M.}~\bibnamefont
  {Diehl}}\ and\ \bibinfo {author} {\bibfnamefont {P.}~\bibnamefont
  {Stienemeier}},\ }\bibfield  {title} {\bibinfo {title} {{Gluons and sea
  quarks in the proton at low scales}},\ }\href
  {https://doi.org/10.1140/epjp/s13360-020-00200-6} {\bibfield  {journal}
  {\bibinfo  {journal} {Eur. Phys. J. Plus}\ }\textbf {\bibinfo {volume}
  {135}},\ \bibinfo {pages} {211} (\bibinfo {year} {2020})},\ \Eprint
  {https://arxiv.org/abs/1904.10722} {arXiv:1904.10722 [hep-ph]} \BibitemShut
  {NoStop}%
\bibitem [{\citenamefont {Martin}\ \emph {et~al.}(2009)\citenamefont {Martin},
  \citenamefont {Stirling}, \citenamefont {Thorne},\ and\ \citenamefont
  {Watt}}]{Martin:2009iq}%
  \BibitemOpen
  \bibfield  {author} {\bibinfo {author} {\bibfnamefont {A.}~\bibnamefont
  {Martin}}, \bibinfo {author} {\bibfnamefont {W.}~\bibnamefont {Stirling}},
  \bibinfo {author} {\bibfnamefont {R.}~\bibnamefont {Thorne}},\ and\ \bibinfo
  {author} {\bibfnamefont {G.}~\bibnamefont {Watt}},\ }\bibfield  {title}
  {\bibinfo {title} {Parton distributions for the {LHC}},\ }\href
  {https://doi.org/10.1140/epjc/s10052-009-1072-5} {\bibfield  {journal}
  {\bibinfo  {journal} {Eur. Phys. J.}\ }\textbf {\bibinfo {volume} {C63}},\
  \bibinfo {pages} {189} (\bibinfo {year} {2009})},\ \Eprint
  {https://arxiv.org/abs/0901.0002} {arXiv:0901.0002 [hep-ph]} \BibitemShut
  {NoStop}%
\bibitem [{\citenamefont {Candido}\ \emph {et~al.}(2020)\citenamefont
  {Candido}, \citenamefont {Forte},\ and\ \citenamefont
  {Hekhorn}}]{Candido:2020yat}%
  \BibitemOpen
  \bibfield  {author} {\bibinfo {author} {\bibfnamefont {A.}~\bibnamefont
  {Candido}}, \bibinfo {author} {\bibfnamefont {S.}~\bibnamefont {Forte}},\
  and\ \bibinfo {author} {\bibfnamefont {F.}~\bibnamefont {Hekhorn}},\
  }\bibfield  {title} {\bibinfo {title} {{Can $ \overline{\mathrm{MS}} $ parton
  distributions be negative?}},\ }\href
  {https://doi.org/10.1007/JHEP11(2020)129} {\bibfield  {journal} {\bibinfo
  {journal} {JHEP}\ }\textbf {\bibinfo {volume} {11}},\ \bibinfo {pages}
  {129}},\ \Eprint {https://arxiv.org/abs/2006.07377} {arXiv:2006.07377
  [hep-ph]} \BibitemShut {NoStop}%
\bibitem [{\citenamefont {Feynman}(1972)}]{Feynman:1972}%
  \BibitemOpen
  \bibfield  {author} {\bibinfo {author} {\bibfnamefont {R.~P.}\ \bibnamefont
  {Feynman}},\ }\href@noop {} {\emph {\bibinfo {title} {{P}hoton-{H}adron
  {I}nteractions}}}\ (\bibinfo  {publisher} {Benjamin},\ \bibinfo {address}
  {Reading, MA},\ \bibinfo {year} {1972})\BibitemShut {NoStop}%
\bibitem [{\citenamefont {Soffer}(1995)}]{Soffer:1994ww}%
  \BibitemOpen
  \bibfield  {author} {\bibinfo {author} {\bibfnamefont {J.}~\bibnamefont
  {Soffer}},\ }\bibfield  {title} {\bibinfo {title} {{Positivity constraints
  for spin dependent parton distributions}},\ }\href
  {https://doi.org/10.1103/PhysRevLett.74.1292} {\bibfield  {journal} {\bibinfo
   {journal} {Phys. Rev. Lett.}\ }\textbf {\bibinfo {volume} {74}},\ \bibinfo
  {pages} {1292} (\bibinfo {year} {1995})},\ \Eprint
  {https://arxiv.org/abs/hep-ph/9409254} {arXiv:hep-ph/9409254} \BibitemShut
  {NoStop}%
\bibitem [{\citenamefont {Goldstein}\ \emph {et~al.}(1995)\citenamefont
  {Goldstein}, \citenamefont {Jaffe},\ and\ \citenamefont
  {Ji}}]{Goldstein:1995ek}%
  \BibitemOpen
  \bibfield  {author} {\bibinfo {author} {\bibfnamefont {G.~R.}\ \bibnamefont
  {Goldstein}}, \bibinfo {author} {\bibfnamefont {R.}~\bibnamefont {Jaffe}},\
  and\ \bibinfo {author} {\bibfnamefont {X.-D.}\ \bibnamefont {Ji}},\
  }\bibfield  {title} {\bibinfo {title} {{Soffer's inequality}},\ }\href
  {https://doi.org/10.1103/PhysRevD.52.5006} {\bibfield  {journal} {\bibinfo
  {journal} {Phys. Rev. D}\ }\textbf {\bibinfo {volume} {52}},\ \bibinfo
  {pages} {5006} (\bibinfo {year} {1995})},\ \Eprint
  {https://arxiv.org/abs/hep-ph/9501297} {arXiv:hep-ph/9501297} \BibitemShut
  {NoStop}%
\bibitem [{\citenamefont {Barone}(1997)}]{Barone:1997fh}%
  \BibitemOpen
  \bibfield  {author} {\bibinfo {author} {\bibfnamefont {V.}~\bibnamefont
  {Barone}},\ }\bibfield  {title} {\bibinfo {title} {{On the QCD evolution of
  the transversity distribution}},\ }\href
  {https://doi.org/10.1016/S0370-2693(97)00875-7} {\bibfield  {journal}
  {\bibinfo  {journal} {Phys. Lett. B}\ }\textbf {\bibinfo {volume} {409}},\
  \bibinfo {pages} {499} (\bibinfo {year} {1997})},\ \Eprint
  {https://arxiv.org/abs/hep-ph/9703343} {arXiv:hep-ph/9703343} \BibitemShut
  {NoStop}%
\bibitem [{\citenamefont {Ji}(2013)}]{Ji:2013dva}%
  \BibitemOpen
  \bibfield  {author} {\bibinfo {author} {\bibfnamefont {X.}~\bibnamefont
  {Ji}},\ }\bibfield  {title} {\bibinfo {title} {{Parton Physics on a Euclidean
  Lattice}},\ }\href {https://doi.org/10.1103/PhysRevLett.110.262002}
  {\bibfield  {journal} {\bibinfo  {journal} {Phys. Rev. Lett.}\ }\textbf
  {\bibinfo {volume} {110}},\ \bibinfo {pages} {262002} (\bibinfo {year}
  {2013})},\ \Eprint {https://arxiv.org/abs/1305.1539} {arXiv:1305.1539
  [hep-ph]} \BibitemShut {NoStop}%
\bibitem [{\citenamefont {Xiong}\ \emph {et~al.}(2014)\citenamefont {Xiong},
  \citenamefont {Ji}, \citenamefont {Zhang},\ and\ \citenamefont
  {Zhao}}]{Xiong:2013bka}%
  \BibitemOpen
  \bibfield  {author} {\bibinfo {author} {\bibfnamefont {X.}~\bibnamefont
  {Xiong}}, \bibinfo {author} {\bibfnamefont {X.}~\bibnamefont {Ji}}, \bibinfo
  {author} {\bibfnamefont {J.-H.}\ \bibnamefont {Zhang}},\ and\ \bibinfo
  {author} {\bibfnamefont {Y.}~\bibnamefont {Zhao}},\ }\bibfield  {title}
  {\bibinfo {title} {{One-loop matching for parton distributions: Nonsinglet
  case}},\ }\href {https://doi.org/10.1103/PhysRevD.90.014051} {\bibfield
  {journal} {\bibinfo  {journal} {Phys. Rev. D}\ }\textbf {\bibinfo {volume}
  {90}},\ \bibinfo {pages} {014051} (\bibinfo {year} {2014})},\ \Eprint
  {https://arxiv.org/abs/1310.7471} {arXiv:1310.7471 [hep-ph]} \BibitemShut
  {NoStop}%
\bibitem [{\citenamefont {Ma}\ and\ \citenamefont {Qiu}(2018)}]{Ma:2014jla}%
  \BibitemOpen
  \bibfield  {author} {\bibinfo {author} {\bibfnamefont {Y.-Q.}\ \bibnamefont
  {Ma}}\ and\ \bibinfo {author} {\bibfnamefont {J.-W.}\ \bibnamefont {Qiu}},\
  }\bibfield  {title} {\bibinfo {title} {{Extracting Parton Distribution
  Functions from Lattice QCD Calculations}},\ }\href
  {https://doi.org/10.1103/PhysRevD.98.074021} {\bibfield  {journal} {\bibinfo
  {journal} {Phys. Rev. D}\ }\textbf {\bibinfo {volume} {98}},\ \bibinfo
  {pages} {074021} (\bibinfo {year} {2018})},\ \Eprint
  {https://arxiv.org/abs/1404.6860} {arXiv:1404.6860 [hep-ph]} \BibitemShut
  {NoStop}%
\bibitem [{\citenamefont {Constantinou}(2021)}]{Constantinou:2020pek}%
  \BibitemOpen
  \bibfield  {author} {\bibinfo {author} {\bibfnamefont {M.}~\bibnamefont
  {Constantinou}},\ }\bibfield  {title} {\bibinfo {title} {{The $x$-dependence
  of hadronic parton distributions: A review on the progress of lattice QCD}},\
  }\href {https://doi.org/10.1140/epja/s10050-021-00353-7} {\bibfield
  {journal} {\bibinfo  {journal} {Eur. Phys. J. A}\ }\textbf {\bibinfo {volume}
  {57}},\ \bibinfo {pages} {77} (\bibinfo {year} {2021})},\ \Eprint
  {https://arxiv.org/abs/2010.02445} {arXiv:2010.02445 [hep-lat]} \BibitemShut
  {NoStop}%
\bibitem [{\citenamefont {Bhat}\ \emph {et~al.}(2021)\citenamefont {Bhat},
  \citenamefont {Cichy}, \citenamefont {Constantinou},\ and\ \citenamefont
  {Scapellato}}]{Bhat:2020ktg}%
  \BibitemOpen
  \bibfield  {author} {\bibinfo {author} {\bibfnamefont {M.}~\bibnamefont
  {Bhat}}, \bibinfo {author} {\bibfnamefont {K.}~\bibnamefont {Cichy}},
  \bibinfo {author} {\bibfnamefont {M.}~\bibnamefont {Constantinou}},\ and\
  \bibinfo {author} {\bibfnamefont {A.}~\bibnamefont {Scapellato}},\ }\bibfield
   {title} {\bibinfo {title} {{Flavor nonsinglet parton distribution functions
  from lattice QCD at physical quark masses via the pseudodistribution
  approach}},\ }\href {https://doi.org/10.1103/PhysRevD.103.034510} {\bibfield
  {journal} {\bibinfo  {journal} {Phys. Rev. D}\ }\textbf {\bibinfo {volume}
  {103}},\ \bibinfo {pages} {034510} (\bibinfo {year} {2021})},\ \Eprint
  {https://arxiv.org/abs/2005.02102} {arXiv:2005.02102 [hep-lat]} \BibitemShut
  {NoStop}%
\bibitem [{\citenamefont {Jo\'o}\ \emph {et~al.}(2019)\citenamefont {Jo\'o},
  \citenamefont {Karpie}, \citenamefont {Orginos}, \citenamefont {Radyushkin},
  \citenamefont {Richards},\ and\ \citenamefont {Zafeiropoulos}}]{Joo:2019jct}%
  \BibitemOpen
  \bibfield  {author} {\bibinfo {author} {\bibfnamefont {B.}~\bibnamefont
  {Jo\'o}}, \bibinfo {author} {\bibfnamefont {J.}~\bibnamefont {Karpie}},
  \bibinfo {author} {\bibfnamefont {K.}~\bibnamefont {Orginos}}, \bibinfo
  {author} {\bibfnamefont {A.}~\bibnamefont {Radyushkin}}, \bibinfo {author}
  {\bibfnamefont {D.}~\bibnamefont {Richards}},\ and\ \bibinfo {author}
  {\bibfnamefont {S.}~\bibnamefont {Zafeiropoulos}},\ }\bibfield  {title}
  {\bibinfo {title} {{Parton Distribution Functions from Ioffe time
  pseudo-distributions}},\ }\href {https://doi.org/10.1007/JHEP12(2019)081}
  {\bibfield  {journal} {\bibinfo  {journal} {JHEP}\ }\textbf {\bibinfo
  {volume} {12}},\ \bibinfo {pages} {081}},\ \Eprint
  {https://arxiv.org/abs/1908.09771} {arXiv:1908.09771 [hep-lat]} \BibitemShut
  {NoStop}%
\bibitem [{\citenamefont {Bednar}\ \emph {et~al.}(2020)\citenamefont {Bednar},
  \citenamefont {Clo\"et},\ and\ \citenamefont {Tandy}}]{Bednar:2018mtf}%
  \BibitemOpen
  \bibfield  {author} {\bibinfo {author} {\bibfnamefont {K.~D.}\ \bibnamefont
  {Bednar}}, \bibinfo {author} {\bibfnamefont {I.~C.}\ \bibnamefont
  {Clo\"et}},\ and\ \bibinfo {author} {\bibfnamefont {P.~C.}\ \bibnamefont
  {Tandy}},\ }\bibfield  {title} {\bibinfo {title} {{Distinguishing Quarks and
  Gluons in Pion and Kaon Parton Distribution Functions}},\ }\href
  {https://doi.org/10.1103/PhysRevLett.124.042002} {\bibfield  {journal}
  {\bibinfo  {journal} {Phys. Rev. Lett.}\ }\textbf {\bibinfo {volume} {124}},\
  \bibinfo {pages} {042002} (\bibinfo {year} {2020})},\ \Eprint
  {https://arxiv.org/abs/1811.12310} {arXiv:1811.12310 [nucl-th]} \BibitemShut
  {NoStop}%
\bibitem [{\citenamefont {Thorne}(2006)}]{Thorne:2006qt}%
  \BibitemOpen
  \bibfield  {author} {\bibinfo {author} {\bibfnamefont {R.}~\bibnamefont
  {Thorne}},\ }\bibfield  {title} {\bibinfo {title} {{A Variable-flavor number
  scheme for next-to-next-to-leading order}},\ }\href
  {https://doi.org/10.1103/PhysRevD.73.054019} {\bibfield  {journal} {\bibinfo
  {journal} {Phys. Rev. D}\ }\textbf {\bibinfo {volume} {73}},\ \bibinfo
  {pages} {054019} (\bibinfo {year} {2006})},\ \Eprint
  {https://arxiv.org/abs/hep-ph/0601245} {arXiv:hep-ph/0601245} \BibitemShut
  {NoStop}%
\bibitem [{\citenamefont {Buza}\ \emph {et~al.}(1998)\citenamefont {Buza},
  \citenamefont {Matiounine}, \citenamefont {Smith},\ and\ \citenamefont {van
  Neerven}}]{Buza:1996wv}%
  \BibitemOpen
  \bibfield  {author} {\bibinfo {author} {\bibfnamefont {M.}~\bibnamefont
  {Buza}}, \bibinfo {author} {\bibfnamefont {Y.}~\bibnamefont {Matiounine}},
  \bibinfo {author} {\bibfnamefont {J.}~\bibnamefont {Smith}},\ and\ \bibinfo
  {author} {\bibfnamefont {W.~L.}\ \bibnamefont {van Neerven}},\ }\bibfield
  {title} {\bibinfo {title} {{Charm electroproduction viewed in the variable
  flavor number scheme versus fixed order perturbation theory}},\ }\href
  {https://doi.org/10.1007/BF01245820} {\bibfield  {journal} {\bibinfo
  {journal} {Eur. Phys. J. C}\ }\textbf {\bibinfo {volume} {1}},\ \bibinfo
  {pages} {301} (\bibinfo {year} {1998})},\ \Eprint
  {https://arxiv.org/abs/hep-ph/9612398} {arXiv:hep-ph/9612398} \BibitemShut
  {NoStop}%
\bibitem [{\citenamefont {Kogut}\ and\ \citenamefont
  {Soper}(1970)}]{Kogut:1969xa}%
  \BibitemOpen
  \bibfield  {author} {\bibinfo {author} {\bibfnamefont {J.~B.}\ \bibnamefont
  {Kogut}}\ and\ \bibinfo {author} {\bibfnamefont {D.~E.}\ \bibnamefont
  {Soper}},\ }\bibfield  {title} {\bibinfo {title} {Quantum electrodynamics in
  the infinite momentum frame},\ }\href
  {https://doi.org/10.1103/PhysRevD.1.2901} {\bibfield  {journal} {\bibinfo
  {journal} {Phys. Rev.}\ }\textbf {\bibinfo {volume} {D1}},\ \bibinfo {pages}
  {2901} (\bibinfo {year} {1970})}\BibitemShut {NoStop}%
\bibitem [{\citenamefont {Bouchiat}\ \emph {et~al.}(1971)\citenamefont
  {Bouchiat}, \citenamefont {Fayet},\ and\ \citenamefont
  {Meyer}}]{Bouchiat:1971mj}%
  \BibitemOpen
  \bibfield  {author} {\bibinfo {author} {\bibfnamefont {C.}~\bibnamefont
  {Bouchiat}}, \bibinfo {author} {\bibfnamefont {P.}~\bibnamefont {Fayet}},\
  and\ \bibinfo {author} {\bibfnamefont {P.}~\bibnamefont {Meyer}},\ }\bibfield
   {title} {\bibinfo {title} {Galilean invariance in the infinite momentum
  frame and the parton model},\ }\href
  {https://doi.org/10.1016/0550-3213(71)90117-9} {\bibfield  {journal}
  {\bibinfo  {journal} {Nucl. Phys.}\ }\textbf {\bibinfo {volume} {B34}},\
  \bibinfo {pages} {157} (\bibinfo {year} {1971})}\BibitemShut {NoStop}%
\bibitem [{\citenamefont {Soper}(1977)}]{Soper:1976jc}%
  \BibitemOpen
  \bibfield  {author} {\bibinfo {author} {\bibfnamefont {D.~E.}\ \bibnamefont
  {Soper}},\ }\bibfield  {title} {\bibinfo {title} {{The Parton Model and the
  Bethe-Salpeter Wave Function}},\ }\href
  {https://doi.org/10.1103/PhysRevD.15.1141} {\bibfield  {journal} {\bibinfo
  {journal} {Phys. Rev. D}\ }\textbf {\bibinfo {volume} {15}},\ \bibinfo
  {pages} {1141} (\bibinfo {year} {1977})}\BibitemShut {NoStop}%
\bibitem [{\citenamefont {Soper}(1979)}]{Soper:1979fq}%
  \BibitemOpen
  \bibfield  {author} {\bibinfo {author} {\bibfnamefont {D.~E.}\ \bibnamefont
  {Soper}},\ }\bibfield  {title} {\bibinfo {title} {Partons and their
  transverse momenta in {QCD}},\ }\href
  {https://doi.org/10.1103/PhysRevLett.43.1847} {\bibfield  {journal} {\bibinfo
   {journal} {Phys. Rev. Lett.}\ }\textbf {\bibinfo {volume} {43}},\ \bibinfo
  {pages} {1847} (\bibinfo {year} {1979})}\BibitemShut {NoStop}%
\bibitem [{\citenamefont {Collins}(1980)}]{Collins:1980ui}%
  \BibitemOpen
  \bibfield  {author} {\bibinfo {author} {\bibfnamefont {J.~C.}\ \bibnamefont
  {Collins}},\ }\bibfield  {title} {\bibinfo {title} {{Intrinsic transverse
  momentum. Nongauge theories}},\ }\href
  {https://doi.org/10.1103/PhysRevD.21.2962} {\bibfield  {journal} {\bibinfo
  {journal} {Phys. Rev. D}\ }\textbf {\bibinfo {volume} {21}},\ \bibinfo
  {pages} {2962} (\bibinfo {year} {1980})}\BibitemShut {NoStop}%
\bibitem [{\citenamefont {Collins}\ and\ \citenamefont
  {Soper}(1982)}]{Collins:1981uw}%
  \BibitemOpen
  \bibfield  {author} {\bibinfo {author} {\bibfnamefont {J.~C.}\ \bibnamefont
  {Collins}}\ and\ \bibinfo {author} {\bibfnamefont {D.~E.}\ \bibnamefont
  {Soper}},\ }\bibfield  {title} {\bibinfo {title} {Parton distribution and
  decay functions},\ }\href {https://doi.org/10.1016/0550-3213(82)90021-9}
  {\bibfield  {journal} {\bibinfo  {journal} {Nucl. Phys.}\ }\textbf {\bibinfo
  {volume} {B194}},\ \bibinfo {pages} {445} (\bibinfo {year}
  {1982})}\BibitemShut {NoStop}%
\bibitem [{\citenamefont {Wilson}(1969)}]{Wilson:1969zs}%
  \BibitemOpen
  \bibfield  {author} {\bibinfo {author} {\bibfnamefont {K.~G.}\ \bibnamefont
  {Wilson}},\ }\bibfield  {title} {\bibinfo {title} {{Nonlagrangian models of
  current algebra}},\ }\href {https://doi.org/10.1103/PhysRev.179.1499}
  {\bibfield  {journal} {\bibinfo  {journal} {Phys. Rev.}\ }\textbf {\bibinfo
  {volume} {179}},\ \bibinfo {pages} {1499} (\bibinfo {year}
  {1969})}\BibitemShut {NoStop}%
\bibitem [{\citenamefont {Zimmermann}(1973)}]{Zimmermann:1972tv}%
  \BibitemOpen
  \bibfield  {author} {\bibinfo {author} {\bibfnamefont {W.}~\bibnamefont
  {Zimmermann}},\ }\bibfield  {title} {\bibinfo {title} {Normal products and
  the short distance expansion in the perturbation theory of renormalizable
  interactions},\ }\href {https://doi.org/10.1016/0003-4916(73)90430-2}
  {\bibfield  {journal} {\bibinfo  {journal} {Ann. Phys.}\ }\textbf {\bibinfo
  {volume} {77}},\ \bibinfo {pages} {570} (\bibinfo {year} {1973})}\BibitemShut
  {NoStop}%
\bibitem [{\citenamefont {Collins}(2011)}]{Collins:2011qcdbook}%
  \BibitemOpen
  \bibfield  {author} {\bibinfo {author} {\bibfnamefont {J.~C.}\ \bibnamefont
  {Collins}},\ }\href@noop {} {\emph {\bibinfo {title} {Foundations of
  Perturbative QCD}}}\ (\bibinfo  {publisher} {Cambridge University Press},\
  \bibinfo {address} {Cambridge},\ \bibinfo {year} {2011})\BibitemShut
  {NoStop}%
\bibitem [{\citenamefont {Chetyrkin}\ and\ \citenamefont
  {Tkachov}(1982)}]{Chetyrkin:1982nn}%
  \BibitemOpen
  \bibfield  {author} {\bibinfo {author} {\bibfnamefont {K.~G.}\ \bibnamefont
  {Chetyrkin}}\ and\ \bibinfo {author} {\bibfnamefont {F.~V.}\ \bibnamefont
  {Tkachov}},\ }\bibfield  {title} {\bibinfo {title} {{Infrared $R$-operation
  and ultraviolet counterterms in the MS scheme}},\ }\href
  {https://doi.org/10.1016/0370-2693(82)90358-6} {\bibfield  {journal}
  {\bibinfo  {journal} {Phys. Lett. B}\ }\textbf {\bibinfo {volume} {114}},\
  \bibinfo {pages} {340} (\bibinfo {year} {1982})}\BibitemShut {NoStop}%
\bibitem [{\citenamefont {Chetyrkin}\ and\ \citenamefont
  {Smirnov}(1984)}]{Chetyrkin:1984xa}%
  \BibitemOpen
  \bibfield  {author} {\bibinfo {author} {\bibfnamefont {K.~G.}\ \bibnamefont
  {Chetyrkin}}\ and\ \bibinfo {author} {\bibfnamefont {V.~A.}\ \bibnamefont
  {Smirnov}},\ }\bibfield  {title} {\bibinfo {title} {{$R^*$-Operation
  corrected}},\ }\href {https://doi.org/10.1016/0370-2693(84)91291-7}
  {\bibfield  {journal} {\bibinfo  {journal} {Phys. Lett. B}\ }\textbf
  {\bibinfo {volume} {144}},\ \bibinfo {pages} {419} (\bibinfo {year}
  {1984})}\BibitemShut {NoStop}%
\bibitem [{\citenamefont {Smirnov}\ and\ \citenamefont
  {Chetyrkin}(1985)}]{Smirnov:1985yck}%
  \BibitemOpen
  \bibfield  {author} {\bibinfo {author} {\bibfnamefont {V.~A.}\ \bibnamefont
  {Smirnov}}\ and\ \bibinfo {author} {\bibfnamefont {K.~G.}\ \bibnamefont
  {Chetyrkin}},\ }\bibfield  {title} {\bibinfo {title} {{$R^*$ operation in the
  minimal subtraction scheme}},\ }\href {https://doi.org/10.1007/BF01017902}
  {\bibfield  {journal} {\bibinfo  {journal} {Theor. Math. Phys.}\ }\textbf
  {\bibinfo {volume} {63}},\ \bibinfo {pages} {462} (\bibinfo {year}
  {1985})}\BibitemShut {NoStop}%
\bibitem [{\citenamefont {Ellis}\ \emph {et~al.}(1979)\citenamefont {Ellis},
  \citenamefont {Georgi}, \citenamefont {Machacek}, \citenamefont {Politzer},\
  and\ \citenamefont {Ross}}]{Ellis:1978ty}%
  \BibitemOpen
  \bibfield  {author} {\bibinfo {author} {\bibfnamefont {R.~K.}\ \bibnamefont
  {Ellis}}, \bibinfo {author} {\bibfnamefont {H.}~\bibnamefont {Georgi}},
  \bibinfo {author} {\bibfnamefont {M.}~\bibnamefont {Machacek}}, \bibinfo
  {author} {\bibfnamefont {H.~D.}\ \bibnamefont {Politzer}},\ and\ \bibinfo
  {author} {\bibfnamefont {G.~G.}\ \bibnamefont {Ross}},\ }\bibfield  {title}
  {\bibinfo {title} {Perturbation theory and the parton model in {QCD}},\
  }\href {https://doi.org/10.1016/0550-3213(79)90105-6} {\bibfield  {journal}
  {\bibinfo  {journal} {Nucl. Phys.}\ }\textbf {\bibinfo {volume} {B152}},\
  \bibinfo {pages} {285} (\bibinfo {year} {1979})}\BibitemShut {NoStop}%
\bibitem [{\citenamefont {Curci}\ \emph {et~al.}(1980)\citenamefont {Curci},
  \citenamefont {Furmanski},\ and\ \citenamefont {Petronzio}}]{Curci:1980uw}%
  \BibitemOpen
  \bibfield  {author} {\bibinfo {author} {\bibfnamefont {G.}~\bibnamefont
  {Curci}}, \bibinfo {author} {\bibfnamefont {W.}~\bibnamefont {Furmanski}},\
  and\ \bibinfo {author} {\bibfnamefont {R.}~\bibnamefont {Petronzio}},\
  }\bibfield  {title} {\bibinfo {title} {Evolution of parton densities beyond
  leading order: {T}he nonsinglet case},\ }\href
  {https://doi.org/10.1016/0550-3213(80)90003-6} {\bibfield  {journal}
  {\bibinfo  {journal} {Nucl. Phys.}\ }\textbf {\bibinfo {volume} {B175}},\
  \bibinfo {pages} {27} (\bibinfo {year} {1980})}\BibitemShut {NoStop}%
\bibitem [{\citenamefont {Politzer}(1977)}]{Politzer:1977fi}%
  \BibitemOpen
  \bibfield  {author} {\bibinfo {author} {\bibfnamefont {H.}~\bibnamefont
  {Politzer}},\ }\bibfield  {title} {\bibinfo {title} {{Gluon Corrections to
  Drell-Yan Processes}},\ }\href {https://doi.org/10.1016/0550-3213(77)90197-3}
  {\bibfield  {journal} {\bibinfo  {journal} {Nucl. Phys. B}\ }\textbf
  {\bibinfo {volume} {129}},\ \bibinfo {pages} {301} (\bibinfo {year}
  {1977})}\BibitemShut {NoStop}%
\bibitem [{\citenamefont {Collins}(1998)}]{Collins:1998rz}%
  \BibitemOpen
  \bibfield  {author} {\bibinfo {author} {\bibfnamefont {J.~C.}\ \bibnamefont
  {Collins}},\ }\bibfield  {title} {\bibinfo {title} {Hard-scattering
  factorization with heavy quarks: {A} general treatment},\ }\href
  {https://doi.org/10.1103/PhysRevD.58.094002} {\bibfield  {journal} {\bibinfo
  {journal} {Phys. Rev.}\ }\textbf {\bibinfo {volume} {D58}},\ \bibinfo {pages}
  {094002} (\bibinfo {year} {1998})},\ \Eprint
  {https://arxiv.org/abs/hep-ph/9806259} {hep-ph/9806259} \BibitemShut
  {NoStop}%
\bibitem [{\citenamefont {Collins}(2003)}]{Collins:2003fm}%
  \BibitemOpen
  \bibfield  {author} {\bibinfo {author} {\bibfnamefont {J.~C.}\ \bibnamefont
  {Collins}},\ }\bibfield  {title} {\bibinfo {title} {What exactly is a parton
  density?},\ }\href {https://www.actaphys.uj.edu.pl/R/34/6/3103/pdf}
  {\bibfield  {journal} {\bibinfo  {journal} {Acta Phys. Polon.}\ }\textbf
  {\bibinfo {volume} {B34}},\ \bibinfo {pages} {3103} (\bibinfo {year}
  {2003})},\ \Eprint {https://arxiv.org/abs/hep-ph/0304122} {hep-ph/0304122}
  \BibitemShut {NoStop}%
\bibitem [{\citenamefont {Valatin}(1954{\natexlab{a}})}]{Valatin.1954a}%
  \BibitemOpen
  \bibfield  {author} {\bibinfo {author} {\bibfnamefont {J.}~\bibnamefont
  {Valatin}},\ }\bibfield  {title} {\bibinfo {title} {Singularities of electron
  kernel functions in an external electromagnetic field},\ }\href
  {https://doi.org/10.1098/rspa.1954.0055} {\bibfield  {journal} {\bibinfo
  {journal} {Proc. Roy. Soc. A}\ }\textbf {\bibinfo {volume} {222}},\ \bibinfo
  {pages} {93} (\bibinfo {year} {1954}{\natexlab{a}})}\BibitemShut {NoStop}%
\bibitem [{\citenamefont {Valatin}(1954{\natexlab{b}})}]{Valatin.1954b}%
  \BibitemOpen
  \bibfield  {author} {\bibinfo {author} {\bibfnamefont {J.}~\bibnamefont
  {Valatin}},\ }\bibfield  {title} {\bibinfo {title} {On the
  {D}irac-{H}eisenberg theory of vacuum polarization},\ }\href
  {https://doi.org/10.1098/rspa.1954.0065} {\bibfield  {journal} {\bibinfo
  {journal} {Proc. Roy. Soc. A}\ }\textbf {\bibinfo {volume} {222}},\ \bibinfo
  {pages} {228} (\bibinfo {year} {1954}{\natexlab{b}})}\BibitemShut {NoStop}%
\bibitem [{\citenamefont {Valatin}(1954{\natexlab{c}})}]{Valatin.1954c}%
  \BibitemOpen
  \bibfield  {author} {\bibinfo {author} {\bibfnamefont {J.}~\bibnamefont
  {Valatin}},\ }\bibfield  {title} {\bibinfo {title} {On the propagation
  functions of quantum electrodynamics},\ }\href
  {https://doi.org/10.1098/rspa.1954.0221} {\bibfield  {journal} {\bibinfo
  {journal} {Proc. Roy. Soc. A}\ }\textbf {\bibinfo {volume} {225}},\ \bibinfo
  {pages} {535} (\bibinfo {year} {1954}{\natexlab{c}})}\BibitemShut {NoStop}%
\bibitem [{\citenamefont {Valatin}(1954{\natexlab{d}})}]{Valatin.1954d}%
  \BibitemOpen
  \bibfield  {author} {\bibinfo {author} {\bibfnamefont {J.}~\bibnamefont
  {Valatin}},\ }\bibfield  {title} {\bibinfo {title} {On the definition of
  finite operator quantities in quantum electrodynamics},\ }\href
  {https://doi.org/10.1098/rspa.1954.0252} {\bibfield  {journal} {\bibinfo
  {journal} {Proc. Roy. Soc. A}\ }\textbf {\bibinfo {volume} {226}},\ \bibinfo
  {pages} {254} (\bibinfo {year} {1954}{\natexlab{d}})}\BibitemShut {NoStop}%
\bibitem [{\citenamefont {Wilson}(1973)}]{Wilson:1972cf}%
  \BibitemOpen
  \bibfield  {author} {\bibinfo {author} {\bibfnamefont {K.~G.}\ \bibnamefont
  {Wilson}},\ }\bibfield  {title} {\bibinfo {title} {Quantum field theory
  models in less than 4 dimensions},\ }\href
  {https://doi.org/10.1103/PhysRevD.7.2911} {\bibfield  {journal} {\bibinfo
  {journal} {Phys. Rev.}\ }\textbf {\bibinfo {volume} {D7}},\ \bibinfo {pages}
  {2911} (\bibinfo {year} {1973})}\BibitemShut {NoStop}%
\bibitem [{\citenamefont {De~R\'ujula}\ \emph {et~al.}(1975)\citenamefont
  {De~R\'ujula}, \citenamefont {Georgi},\ and\ \citenamefont
  {Glashow}}]{DeRujula:1975qlm}%
  \BibitemOpen
  \bibfield  {author} {\bibinfo {author} {\bibfnamefont {A.}~\bibnamefont
  {De~R\'ujula}}, \bibinfo {author} {\bibfnamefont {H.}~\bibnamefont
  {Georgi}},\ and\ \bibinfo {author} {\bibfnamefont {S.~L.}\ \bibnamefont
  {Glashow}},\ }\bibfield  {title} {\bibinfo {title} {{Hadron Masses in a Gauge
  Theory}},\ }\href {https://doi.org/10.1103/PhysRevD.12.147} {\bibfield
  {journal} {\bibinfo  {journal} {Phys. Rev. D}\ }\textbf {\bibinfo {volume}
  {12}},\ \bibinfo {pages} {147} (\bibinfo {year} {1975})}\BibitemShut
  {NoStop}%
\bibitem [{\citenamefont {Lepage}\ and\ \citenamefont
  {Brodsky}(1980)}]{Lepage:1980fj}%
  \BibitemOpen
  \bibfield  {author} {\bibinfo {author} {\bibfnamefont {G.~P.}\ \bibnamefont
  {Lepage}}\ and\ \bibinfo {author} {\bibfnamefont {S.~J.}\ \bibnamefont
  {Brodsky}},\ }\bibfield  {title} {\bibinfo {title} {Exclusive processes in
  perturbative quantum chromodynamics},\ }\href
  {https://doi.org/10.1103/PhysRevD.22.2157} {\bibfield  {journal} {\bibinfo
  {journal} {Phys. Rev.}\ }\textbf {\bibinfo {volume} {D22}},\ \bibinfo {pages}
  {2157} (\bibinfo {year} {1980})}\BibitemShut {NoStop}%
\bibitem [{\citenamefont {Brodsky}\ \emph {et~al.}(2001)\citenamefont
  {Brodsky}, \citenamefont {Hwang}, \citenamefont {Ma},\ and\ \citenamefont
  {Schmidt}}]{Brodsky:2000ii}%
  \BibitemOpen
  \bibfield  {author} {\bibinfo {author} {\bibfnamefont {S.~J.}\ \bibnamefont
  {Brodsky}}, \bibinfo {author} {\bibfnamefont {D.-S.}\ \bibnamefont {Hwang}},
  \bibinfo {author} {\bibfnamefont {B.-Q.}\ \bibnamefont {Ma}},\ and\ \bibinfo
  {author} {\bibfnamefont {I.}~\bibnamefont {Schmidt}},\ }\bibfield  {title}
  {\bibinfo {title} {Light-cone representation of the spin and orbital angular
  momentum of relativistic composite systems},\ }\href
  {https://doi.org/10.1016/S0550-3213(00)00626-X} {\bibfield  {journal}
  {\bibinfo  {journal} {Nucl. Phys.}\ }\textbf {\bibinfo {volume} {B593}},\
  \bibinfo {pages} {311} (\bibinfo {year} {2001})},\ \Eprint
  {https://arxiv.org/abs/hep-th/0003082} {hep-th/0003082} \BibitemShut
  {NoStop}%
\bibitem [{\citenamefont {Adolph}\ \emph {et~al.}(2013)\citenamefont {Adolph}
  \emph {et~al.}}]{COMPASS:2013bfs}%
  \BibitemOpen
  \bibfield  {author} {\bibinfo {author} {\bibfnamefont {C.}~\bibnamefont
  {Adolph}} \emph {et~al.} (\bibinfo {collaboration} {COMPASS}),\ }\bibfield
  {title} {\bibinfo {title} {{Hadron Transverse Momentum Distributions in Muon
  Deep Inelastic Scattering at 160 GeV/$c$}},\ }\href
  {https://doi.org/10.1140/epjc/s10052-013-2531-6} {\bibfield  {journal}
  {\bibinfo  {journal} {Eur. Phys. J. C}\ }\textbf {\bibinfo {volume} {73}},\
  \bibinfo {pages} {2531} (\bibinfo {year} {2013})},\ \bibinfo {note}
  {[Erratum: Eur.Phys.J.C 75, 94 (2015)]},\ \Eprint
  {https://arxiv.org/abs/1305.7317} {arXiv:1305.7317 [hep-ex]} \BibitemShut
  {NoStop}%
\bibitem [{\citenamefont {Hou}\ \emph {et~al.}(2021)\citenamefont {Hou} \emph
  {et~al.}}]{Hou:2019efy}%
  \BibitemOpen
  \bibfield  {author} {\bibinfo {author} {\bibfnamefont {T.-J.}\ \bibnamefont
  {Hou}} \emph {et~al.},\ }\bibfield  {title} {\bibinfo {title} {{New CTEQ
  global analysis of quantum chromodynamics with high-precision data from the
  LHC}},\ }\href {https://doi.org/10.1103/PhysRevD.103.014013} {\bibfield
  {journal} {\bibinfo  {journal} {Phys. Rev. D}\ }\textbf {\bibinfo {volume}
  {103}},\ \bibinfo {pages} {014013} (\bibinfo {year} {2021})},\ \Eprint
  {https://arxiv.org/abs/1912.10053} {arXiv:1912.10053 [hep-ph]} \BibitemShut
  {NoStop}%
\bibitem [{\citenamefont {Ball}\ \emph {et~al.}(2018)\citenamefont {Ball},
  \citenamefont {Bertone}, \citenamefont {Bonvini}, \citenamefont {Marzani},
  \citenamefont {Rojo},\ and\ \citenamefont {Rottoli}}]{Ball:2017otu}%
  \BibitemOpen
  \bibfield  {author} {\bibinfo {author} {\bibfnamefont {R.~D.}\ \bibnamefont
  {Ball}}, \bibinfo {author} {\bibfnamefont {V.}~\bibnamefont {Bertone}},
  \bibinfo {author} {\bibfnamefont {M.}~\bibnamefont {Bonvini}}, \bibinfo
  {author} {\bibfnamefont {S.}~\bibnamefont {Marzani}}, \bibinfo {author}
  {\bibfnamefont {J.}~\bibnamefont {Rojo}},\ and\ \bibinfo {author}
  {\bibfnamefont {L.}~\bibnamefont {Rottoli}},\ }\bibfield  {title} {\bibinfo
  {title} {{Parton distributions with small-x resummation: evidence for BFKL
  dynamics in HERA data}},\ }\href
  {https://doi.org/10.1140/epjc/s10052-018-5774-4} {\bibfield  {journal}
  {\bibinfo  {journal} {Eur. Phys. J. C}\ }\textbf {\bibinfo {volume} {78}},\
  \bibinfo {pages} {321} (\bibinfo {year} {2018})},\ \Eprint
  {https://arxiv.org/abs/1710.05935} {arXiv:1710.05935 [hep-ph]} \BibitemShut
  {NoStop}%
\bibitem [{\citenamefont {Gaskell}(2010)}]{Gaskell:2010zz}%
  \BibitemOpen
  \bibfield  {author} {\bibinfo {author} {\bibfnamefont {D.}~\bibnamefont
  {Gaskell}},\ }\bibfield  {title} {\bibinfo {title} {{Recent results in DIS
  from Jefferson Lab}},\ }\href {https://doi.org/10.22323/1.106.0009}
  {\bibfield  {journal} {\bibinfo  {journal} {PoS}\ }\textbf {\bibinfo {volume}
  {DIS12010}},\ \bibinfo {pages} {009} (\bibinfo {year} {2010})}\BibitemShut
  {NoStop}%
\end{thebibliography}%

\end{document}